\documentclass[twoside,12pt]{article}
\usepackage{epsfig}
\def\Journal#1#2#3#4{{#1} {#2} (#4) #3 }
\def\APJ{\em Astrophys. J.}
\def\ASJ{\em Astronom. J.}
\def\SCI{\em Science}
\def\NAT{\em Nature} 
\def\PLB{{\em Phys. Lett.} B}
\def\PRL{\em Phys. Rev. Lett.}
\def\ANR{\em Ann. Rev. Nucl.  Part. Sci.}
\def\BAS{\em Bull. Astro. Soc. Neth.}
\def\HPA{\em Helv. Phys. Acta}
\def\AA{\em Astronom. and Astrophys.}
\def\MONRAS{\em Mon. Not. R. Astron. Soc.}
\def\PRD{{\em Phys. Rev.} D}
\def\ANP{\em Anns. Phys.}
\def\IJMPD{{\em Int. J. Mod. Phys.} D}
\def\NCB{{\em Nuovo Cim.} B}
\def\PREV{\em Phys. Rev.}
\def\JPE{\em Jour. Phys. Essays}
\def\SPAW{\em Sitzsungber. Preuss. Akad. Wiss.}
\def\MZ{\em Math. Zeit.}
\def\JMP{{\em J. Math. Phys.}}
\def\GRG{{\em Gen. Rel. Gravit.}}
\def\ANM{\em Anns. Math.}
\def\AAR{\em Astronom. and Astrophys. Rev.}
\def\FP{\em Found. Phys.}
\def\PRA{{\em Phys. Rev.} A}
\def\ASS{\em Astrophys. and Space Sci.}
\def\CQG{\em Class. and Quant. Grav.}
\def\ARAA{\em Ann. Rev. Astron. and Astrophys.}

\begin{document}

\title{ \vspace{1cm} Alternatives to Dark Matter and Dark Energy}
\author{Philip D. Mannheim 
\\
Department of Physics, University of Connecticut, Storrs, CT 06269, USA\\
email: philip.mannheim@uconn.edu\\}
\date{July 28, 2005} 
\maketitle

\begin{abstract}

We review the underpinnings of the standard Newton-Einstein theory of
gravity, and identify where it could possibly go wrong. In particular,
we discuss the logical independence from each other of the general
covariance principle, the equivalence principle and the Einstein
equations, and discuss how to constrain the matter energy-momentum
tensor which serves as the source of gravity. We identify the a priori
assumption of the validity of standard gravity on all distance scales as
the root cause of the dark matter and dark energy problems, and discuss
how the freedom currently present in gravitational theory can enable us
to construct candidate alternatives to the standard theory in which the
dark matter and dark energy problems could then be resolved. We identify
three generic aspects of these alternate approaches: that it is a
universal acceleration scale which determines when a luminous Newtonian
expectation is to fail to fit data, that there is a global cosmological
effect on local galactic motions which can replace galactic dark matter,
and that to solve the cosmological constant problem it is not necessary
to quench the cosmological constant itself, but only the amount
by which it gravitates. 

\end{abstract}

\section{Introduction}

Following many years of research in cosmology and astrophysics, a
picture of the universe has emerged
\cite{Riess1998,Perlmutter1998,Bahcall2000,deBernardis2000,Tegmark2004}
which is as troubling as it is impressive. Specifically, a wide
variety of data currently support the view that the matter content of
the universe consists of two primary components, viz. dark matter and
dark energy, with ordinary luminous matter being relegated to a
decidedly minor role. The nature and composition of these dark matter
and dark energy components is not at all well-understood, and while both
present severe challenges to the standard theory, each presents a
different kind of challenge to it. As regards dark matter, there is
nothing in principle wrong with the existence of non-luminous material
per se (indeed objects such as dead stars, brown dwarfs and massive
neutrinos are well-established in nature). Rather, what is disturbing
is the ad hoc, after the fact, way in which dark matter is actually
introduced, with its presence only being inferred after known luminous
astrophysical sources are found to fail to account for any given
astrophysical observation. Dark matter thus seems to know where, and in
what amount, it is to be needed, and to know when it is not in fact
needed (dark matter has to avoid being abundant in the solar system in
order to not impair the success of standard gravity in accounting for
solar system observations using visible sources alone); and moreover,
in the cases where it is needed, what it is actually made of
(astrophysical sources (Machos) or new elementary particles
(Wimps)) is as yet totally unknown and elusive. 

Disturbing as the dark matter problem is, the dark energy problem is
even more severe, and not simply because its composition and nature is
as mysterious as that of dark matter. Rather, for dark energy there
actually is a very good, quite clear-cut candidate, viz. a cosmological
constant, and the problem here is that the value for the cosmological
constant as anticipated from fundamental theory is orders of magnitude
larger than the data can possibly permit. With dark matter then, we see
that luminous sources alone underaccount for the data, while for dark
energy, a cosmological constant overaccounts for the data. Thus, within
the standard picture, arbitrary as their introduction might be, there is
nonetheless room for dark matter candidates should they ultimately be
found, but for dark energy there is a need not to find something which
might only momentarily be missing, but rather to get rid of something which
is definitely there. And indeed, if it does not prove possible to quench
the cosmological constant by the requisite orders and orders of magnitude,
one would have to conclude that the prevailing cosmological theory
simply does not work.

In arriving at the predicament that contemporary astrophysical and
cosmological theory thus finds itself in, it is important to recognize that
the entire case for the existence of dark matter and dark energy is based on
just one thing alone, viz. on the validity on all distance scales of the
standard Newton-Einstein gravitational theory as expressed through the
Einstein equations of motion 
\begin{equation}
-\frac{c^3}{8\pi G}\left(R^{\mu\nu}-
\frac{1}{2}g^{\mu\nu}R^{\alpha}_{\phantom{\alpha}\alpha}
\right)=T^{\mu\nu}
\label{1}
\end{equation}
for the gravitational field $g_{\mu\nu}$. Specifically, the standard
approach to cosmology and astrophysics is to take the Einstein equations of
motion as given, and whenever the theory is found to encounter
observational difficulties on any particular distance scale, modifications
are to then be made to $T^{\mu\nu}$ through the introduction of new,
essentially ad hoc, gravitational sources so that agreement with
observation can then be restored. While a better understanding of dark
energy and explicit observational detection of dark matter
sources might eventually be achieved in the future, at the present time the
only apparent way to avoid the dark matter and dark energy problems is to
modify or generalize not the right-hand side of Eq. (\ref{1}) but rather
its left. In order to see how one might actually do this, it is thus
necessary to carefully go over the entire package represented by Eq.
(\ref{1}), to determine whether some of its ingredients might not be as
secure as others, and investigate whether the weaker ones could possibly be
replaced. To do this we will thus need to make a critical appraisal of both
of the two sides of Eq. (\ref{1}).

\section{The underpinnings of the standard gravitational picture}
\subsection{Massive test particle motion}

Following a year of remarkable achievement, achievement whose
centennial is currently being celebrated, at the end of 1905 Einstein
found himself in a somewhat paradoxical situation. While he
had surmounted an enormous hurdle in developing special relativity, its
very establishment created an even bigger hurdle for him. Specifically,
with the development of special relativity Einstein resolved the
conflict between the Lorentz invariance of the Maxwell equations and the
Galilean invariance of Newtonian mechanics by realizing that it was
Lorentz invariance which was the more basic of the two principles, and
that it was Newtonian mechanics which therefore had to be modified.
While special relativity thus ascribed primacy to Lorentz invariance so
that observers moving with large or small uniform velocity could then
all agree on the same physics, such observers still occupied a highly
privileged position since uniform velocity observers form only a very
small subset of all possible allowable observers, observers who could
move with arbitrarily non-uniform velocity. Additionally, if it was
special relativity that was to be the all-embracing principle, all
interactions would then need to obey it, and yet what was at the time
the accepted theory of gravity, viz. Newtonian gravity, in fact did
not. It was the simultaneous resolution of these two issues
(viz. accelerating observers and the compatibility of gravity with
relativity) at one and the same time through the spectacular development
of general relativity which not only established the Einstein theory of
gravity, but which left the impression that there was only one possible
resolution to the two issues, viz. that based on Eq. (\ref{1}). To
pinpoint what it is in the standard theory which leads us to the dark
matter and dark energy problems, we thus need to unravel the standard
gravitational package into what are in fact logically independent
components, an exercise which is actually of value in and of itself
regardless in fact of the dark matter and dark energy problems.  

In order to make such a dissection of the standard picture, we begin
with a discussion of a standard, free, spinless, relativistic, 
classical-mechanical  Newtonian particle of non-zero kinematic mass $m$
moving in flat spacetime  according to the special relativistic
generalization of Newton's second law of 
motion                                 
\begin{equation}
m\frac{d^2\xi^{\alpha}}{d \tau ^2} =0~~,~~R_{\mu \nu \sigma \tau}=0 ~~,
\label{2}
\end{equation}
where $d\tau=(-\eta_{\alpha \beta}d\xi^{\alpha} d\xi^{\beta})^{1/2}$ is the
proper time and $\eta_{\alpha \beta}$ is the flat spacetime metric, and 
where we have indicated explicitly that the Riemann tensor is (for the
moment) zero. As such, Eq. (\ref{2}) will be left invariant under
linear transformations of the coordinates $\xi^{\mu}$, but on making an
arbitrary non-linear transformation to coordinates
$x^{\mu}$ and using the definitions 
\begin{equation}
\Gamma^{\lambda}_{\mu \nu} = \frac{\partial x^{\lambda}}{\partial
\xi^{\alpha}} \frac{\partial ^2\xi^{\alpha}}{\partial x^{\mu} \partial
x^{\nu}}~~,~~g_{\mu \nu}= \frac{\partial \xi^{\alpha}}{ \partial
x^{\mu}} \frac{\partial \xi^{\beta}}{ \partial x^{\nu}}
\eta_{\alpha \beta}~~, 
\label{3}
\end{equation}
we directly find (see e.g. \cite{Weinberg1972}, a reference whose notation
we use throughout) that the invariant proper time is brought to the form
$d\tau=(-g_{\mu
\nu}dx^{\mu} dx^{\nu})^{1/2}$, and that the equation of motion of Eq.
(\ref{2}) is rewritten as
\begin{equation}
m \left( \frac{D^2x^{\lambda}}{ D\tau^2}\right) \equiv
m \left( \frac{d^2x^{\lambda} }{ d\tau^2}
+\Gamma^{\lambda}_{\mu \nu} 
\frac{dx^{\mu}}{ d\tau}\frac{dx^{\nu } }{ d\tau} \right) 
= 0~~,~~R_{\mu \nu \sigma \tau}=0~~, 
\label{4}
\end{equation}
with Eq. (\ref{4}) serving to define $D^2x^{\lambda}/D\tau^2$. As
derived, Eq. (\ref{4}) so far only holds in a strictly flat spacetime with
zero Riemann curvature tensor, and indeed Eq. (\ref{4}) is only a covariant
rewriting of the special relativistic Newtonian second law of motion, i.e.
it covariantly describes what an observer with a non-uniform velocity in
flat spacetime sees, with the $\Gamma ^{\lambda}_{\mu \nu}$ term emerging
as an inertial, coordinate-dependent force. The emergence of such a $\Gamma
^{\lambda}_{\mu \nu}$ term originates in the fact that even while the
four-velocity $dx^{\lambda}/d\tau$ transforms as a general contravariant
vector, its ordinary derivative $d^2x^{\lambda}/d\tau^2$ (which samples
adjacent points and not merely the point where the four-velocity itself is
calculated) does not, and it is only the four-acceleration
$D^2x^{\lambda}/D\tau^2$ which transforms as a general contravariant
four-vector, and it is thus only this particular four-vector on whose
meaning all (accelerating and  non-accelerating) observers can agree. The
quantity $\Gamma ^{\lambda}_{\mu\nu}$ is not itself a general coordinate
tensor, and in flat spacetime one can eliminate it everywhere by working in
Cartesian coordinates. Despite this privileged status for Cartesian
coordinate systems, in general it is Eq. (\ref{4}) rather than Eq.
(\ref{2}) which should be used (even in flat spacetime) since this is the
form of Newton's second law of motion which an accelerating flat
spacetime observer sees.

Now while all of the above remarks where developed purely for flat
spacetime, Eq. (\ref{4}) has an immediate generalization to curved
spacetime where it then takes the form 
\begin{equation}
m \left( \frac{D^2x^{\lambda}}{ D\tau^2}\right) \equiv
m \left( \frac{d^2x^{\lambda} }{ d\tau^2}
+\Gamma^{\lambda}_{\mu \nu} 
\frac{dx^{\mu}}{d\tau}\frac{dx^{\nu } }{ d\tau} \right) 
= 0~~,~~R_{\mu \nu \sigma \tau}\neq 0~~. 
\label{5}
\end{equation}
In curved spacetime, it is again only the quantity $D^2x^{\lambda}/D\tau^2$
which transforms as a general contravariant four-vector, and the 
Christoffel symbol $\Gamma ^{\lambda}_{\mu \nu}$ is again not a general
coordinate tensor. Consequently at any given point $P$ it can be made to
vanish,\footnote{ Under the transformation $x^{\prime \lambda} =
x^{\lambda}+ \frac{1}{2}x^{\mu}x^{\nu}(\Gamma^{\lambda}_{\mu \nu})_{P}$, the
primed  coordinate Christoffel symbols $(\Gamma^{\prime \lambda}_{\mu
\nu})_{P}$ will vanish at the point $P$, regardless in fact
of how large the Riemann tensor in the neighborhood of the point $P$
might actually be.} though no coordinate transformation can bring it to
zero at every point in a spacetime whose Riemann tensor is non-vanishing. 
As an equation, Eq. (\ref{5}) can also be obtained from an action principle,
since it is the stationarity condition $\delta I_T/\delta x_{\lambda}=0$
associated with functional variation of the test particle action
\begin{equation}
I_T=-mc\int d\tau                                                                               
\label{6}
\end{equation}               
with respect to the coordinate $x^{\lambda}$. To appreciate the ubiquity of
the appearance of the covariant acceleration $D^2x^{\lambda}/D\tau^2$, we
consider as action the curved space electromagnetic coupling
\begin{equation}
I_T^{(2)}=-mc\int d\tau + e\int
d\tau\frac{dx^{\lambda}}{d\tau}A_{\lambda}
~~,                                                                              
\label{7}
\end{equation}               
to find that its variation with respect to $x^{\lambda}$ leads to the
curved space Lorentz force law
\begin{equation}
mc \left( \frac{d^2x^{\lambda} }{ d\tau^2}
+\Gamma^{\lambda}_{\mu \nu} 
\frac{dx^{\mu}}{d\tau}\frac{dx^{\nu } }{ d\tau} \right) 
= eF^{\lambda}_{\phantom{\lambda}\alpha}\frac{dx^{\alpha}}{d\tau}~~,
~~R_{\mu \nu\sigma\tau}\neq 0~~. 
\label{8}
\end{equation}
Similarly, the coupling of the test particle to the Ricci scalar via 
\begin{equation}
I_T^{(3)}=-mc\int d\tau - \kappa\int d\tau
R^{\alpha}_{\phantom{\alpha}\alpha}
~~,                                                                              
\label{9}
\end{equation}               
leads to
\begin{equation}
\left(mc +\kappa R^{\alpha}_{\phantom {\alpha} \alpha}\right)\left(
\frac{d^2x^{\lambda} }{ d\tau^2} +\Gamma^{\lambda}_{\mu \nu} 
\frac{dx^{\mu}}{d\tau}\frac{dx^{\nu } }{ d\tau} \right) 
= -\kappa R^{\alpha}_{\phantom {\alpha} \alpha ;\beta} \left( g^{\lambda
\beta}+
\frac{dx^{\lambda}}{d\tau}                                                      
\frac{dx^{\beta}}{d\tau}\right)~~, 
\label{10}
\end{equation}
($R_{\mu \nu\sigma\tau}$ necessarily non-zero here), while the coupling of
the test particle to a scalar field $S(x)$ via 
\begin{equation}
I_T^{(4)}=-\hat{\kappa}\int d\tau S(x)
~~,                                                                              
\label{11}
\end{equation}               
leads to the curved space
\begin{equation}
\hat{\kappa}S\left(
\frac{d^2x^{\lambda} }{ d\tau^2} +\Gamma^{\lambda}_{\mu \nu} 
\frac{dx^{\mu}}{d\tau}\frac{dx^{\nu } }{ d\tau} \right) 
= -\hat{\kappa}S_{;\beta} \left( g^{\lambda
\beta}+
\frac{dx^{\lambda}}{d\tau}                                                      
\frac{dx^{\beta}}{d\tau}\right)~~,~~R_{\mu \nu\sigma\tau}\neq 0~~, 
\label{12}
\end{equation}
an expression incidentally which reduces to Eq. (\ref{5}) when $S(x)$
is a spacetime constant (with the mass parameter then being given by
$mc=\hat{\kappa}S$). In all of the above cases then it is the quantity
$D^2x^{\lambda}/D\tau^2$ which must appear, since in each such case the
action which is varied is a general coordinate scalar. 

\subsection{The equivalence principle}

Now as such, the analysis given above is a purely kinematic one which
discusses only the propagation of test particles in curved backgrounds.
This analysis makes no reference to gravity per se, and in particular
makes no reference to Eq. (\ref{1}) at all, though it does imply that
in any curved spacetime in which the metric $g_{\mu\nu}$ is taken to be
the gravitational field, covariant equations of motion involving
four-accelerations would have to based strictly on
$D^2x^{\lambda}/D\tau^2$, with neither $d^2x^{\lambda}/d\tau^2$ nor
$\Gamma^{\lambda}_{\mu \nu}  (dx^{\mu}/d\tau)(dx^{\nu}/d\tau)$ having any
coordinate independent significance or meaning. As such, we thus
recognize the equivalence principle as the statement that the
$d^2x^{\lambda}/d\tau^2$ and $\Gamma^{\lambda}_{\mu \nu} 
(dx^{\mu}/d\tau)(dx^{\nu}/d\tau)$ terms must appear in all of the above
propagation equations in precisely the
combination indicated,\footnote{This thereby secures the equality of the
inertial and passive gravitational masses of material particles.} and
that via a coordinate transformation it is possible to remove the
Christoffel symbols at any chosen point $P$, with it thereby being
possible to simulate the Christoffel symbol contribution to the
gravitational field at such a point $P$ by an accelerating coordinate
system in flat spacetime. We introduce this particular formulation of
the equivalence principle quite guardedly, since in an equation such as
the fully covariant Eq. (\ref{10}), one cannot remove the dependence on
the Ricci scalar by any coordinate transformation whatsoever, and we
thus define the equivalence principle not as the statement that all
gravitational effects at any point $P$ can be simulated by an 
accelerating coordinate system in flat spacetime (viz. that test
particles unambiguously move on geodesics and obey Eq. (\ref{5}) and
none other), but rather that no matter what propagation equation is to
be used for test particles, the appropriate acceleration for them is
$D^2x^{\lambda}/D\tau^2$.\footnote{This particular formulation of the
equivalence principle does no violence to observation, since Eotvos
experiment type testing of the equivalence principle is made in
Ricci-flat Schwarzschild geometries where all Ricci tensor or Ricci
scalar dependent terms (such as for instance those exhibited in Eq.
(\ref{10})) are simply absent, with such tests (and in fact any which
involve the Schwarzschild geometry) not being able to distinguish
between Eqs. (\ref{4}) and (\ref{10}).} With Eq. (\ref{4}) thus showing
the role of coordinate invariance in flat spacetime, with Eqs.
(\ref{5}), (\ref{8}), and (\ref{10}) exhibiting the equivalence
principle in curved spacetime, and with all of these equations being
independent of the Einstein equations of Eq. (\ref{1}), the logical
independence of the general covariance principle, the equivalence
principle and the Einstein equations is thus established. Hence, any
metric theory of gravity in which the action is a general coordinate
scalar and the metric is the gravitational field will thus
automatically obey both the relativity principle and the equivalence
principle, no matter whether or not the Einstein equations are to be
imposed as well.\footnote{To sharpen this point, we note that it could
have been the case that the resolution of the conflict between gravity
and special relativity could have been through the introduction of the
gravitational force not as a geometric entity at all, but rather as an
analog of the way the Lorentz force is introduced in Eq. (\ref{8}). In
such a case, in an accelerating coordinate system one would still need
to use the acceleration $D^2x^{\lambda}/D\tau^2$ and not the ordinary 
$d^2x^{\lambda}/d\tau^2$. However, if the gravitational field were to
be treated the same way as the electromagnetic field, left open would
then be the issue of whether  physics is to be conducted in flat space
or curved space, i.e. left open would be the question of what does
then fix the Riemann tensor. There would then have to be some
additional equation which would fix it, and curvature would still have
to be recognized as having true, non-coordinate artifact effects on
particles if the Riemann tensor were then found to be non-zero. Taking
such curvature to be associated with the gravitational field (rather
than with some further field) is of course the most economical,
though doing so would not oblige gravitational effects to only be felt
through $D^2x^{\lambda}/D\tau^2$, and would not preclude some
gravitational Lorentz force type term (such as the one exhibited in Eq.
(\ref{10})) from appearing as well.}

\subsection{Massless field motion}

Absent from the above discussion is the question of whether real, as
opposed to test, particles actually obey curved space propagation equations
such as Eq. (\ref{5}) at all, and whether such a discussion should apply
to massless particles as well since for them both $m$ and the proper
time $d\tau$ vanish identically, with equations such as Eq. (\ref{5})
becoming meaningless. Now it turns out that these two issues are not
actually independent, as both reduce to the question of how waves
rather than particles couple to gravity, since light is described by a
wave equation, and elementary particles are actually taken to be the
quanta associated with quantized fields which are also described by wave
equations. And since the discussion will be of relevance for the
exploration of the energy-momentum tensor to be given below, we present
it in some detail now. For simplicity we look at the standard minimally
coupled curved spacetime massless Klein-Gordon scalar field with wave
equation
\begin{equation}
S^{;\mu}_{\phantom{\mu};\mu}=0~~,~~~R_{\mu \nu \sigma \tau} \neq 0
\label{13}
\end{equation}
where $S^{;\mu}$ denotes the contravariant derivative $\partial S/\partial
x_{\mu}$. If for the scalar field we introduce a eikonal phase $T(x)$
via $S(x)={\rm exp}(iT(x))$, the scalar $T(x)$ is then found to obey the
equation 
\begin{equation}
T^{;\mu}T_{;\mu}-iT^{;\mu}_{\phantom{;\mu};\mu}=0~~,
\label{14}
\end{equation}
an equation which reduces to $T^{;\mu}T_{;\mu}=0$ in the short wavelength
limit. From the relation $T^{;\mu}T_{;\mu ;\nu}=0$ which then ensues, it
follows that in the short wavelength limit the phase $T(x)$ obeys
\begin{equation}
T^{;\mu}T_{;\mu;\nu}=
T^{;\mu}[T_{;\nu;\mu}+\partial_{\nu}T_{;\mu}-\partial_{\mu}T_{;\nu}]=
T^{;\mu}[T_{;\nu;\mu}+\partial_{\nu}\partial_{\mu}T
-\partial_{\mu}\partial_{\nu}T]=
T^{;\mu}T_{;\nu;\mu}=0~~.
\label{15}
\end{equation}
Since normals to wavefronts obey the eikonal relation
\begin{equation}
T^{;\mu}=\frac{dx^{\mu}}{dq}=k^{\mu}
\label{16}
\end{equation}
where $q$ is a convenient scalar affine parameter which measures distance
along the normals and $k^{\mu}$ is the wave vector of the wave,
on noting that $(dx^{\mu}/dq)(\partial/\partial x^{\mu})=d/dq$ we thus
obtain
\begin{equation}
k^{\mu}k^{\lambda}_{\phantom{\lambda} ;\mu}=\frac{d^2x^{\lambda} }{ dq^2}
+\Gamma^{\lambda}_{\mu \nu} 
\frac{dx^{\mu}}{ dq}\frac{dx^{\nu } }{ dq}=0~~,
\label{17}
\end{equation}
a condition which we recognize as being the massless particle
geodesic equation, with rays then  precisely being found to be geodesic in
the eikonal limit. Since the discussion given earlier of the coordinate
dependence of the  Christoffel symbols was purely geometric, we thus see
that once rays such as light rays are geodesic, they immediately obey the
the equivalence principle,\footnote{According to Eq. (\ref{17}), an observer
in  Einstein's elevator would not be able to tell if a light ray is
falling downwards under gravity or whether the elevator is  accelerating
upwards.} with phenomena such as the gravitational bending of light then
immediately following.

Now while we do obtain strict geodesic motion for rays when we eikonalize
the minimally coupled Klein-Gordon equation, the situation becomes somewhat
different if we consider a non-minimally coupled Klein-Gordon equation
instead. Thus, on replacing Eq. (\ref{13}) by
\begin{equation}
S^{;\mu}_{\phantom{\mu};\mu}
+\frac{\xi}{6}SR^{\alpha}_{\phantom{\alpha}\alpha}=0~~,~~~R_{\mu
\nu\sigma\tau}\neq 0~~,
\label{18}
\end{equation}
the curved space eikonal equation then takes the form
\begin{equation}
T^{;\mu}T_{;\mu}-\frac{\xi}{6}R^{\alpha}_{\phantom{\alpha}\alpha}=0~~,
\label{19}
\end{equation}
with Eq. (\ref{17}) being replaced by the Ricci scalar dependent
\begin{equation}
\frac{d^2x^{\lambda} }{ dq^2}
+\Gamma^{\lambda}_{\mu \nu} 
\frac{dx^{\mu}}{ dq}\frac{dx^{\nu } }{
dq}=\frac{\xi}{12}(R^{\alpha}_{\phantom{\alpha}\alpha})^{;\lambda}
\label{20}
\end{equation}
in the short wavelength limit. 

A similar dependence on the Ricci scalar or tensor is also obtained for
massless spin one-half and spin one fields. Specifically, even though
the massless Dirac equation in curved space, viz.
$i\gamma^{\mu}(x)[\partial_{\mu}+\Gamma_{\mu}]\psi(x)=0$ (where
$\Gamma_{\mu}(x)$ is the fermion spin connection) contains no explicit
direct dependence on the Ricci tensor, nonetheless the second order
differential equation which the fermion field then also obeys is found
to take the form
$[\partial_{\mu}+\Gamma_{\mu}][\partial^{\mu}+\Gamma^{\mu}]\psi(x)+
(1/4)R^{\alpha}_{\phantom {\alpha}\alpha}\psi(x)=0$. Likewise, even though
the curved space Maxwell equations, viz. $F^{\mu \nu}_{\phantom{\mu
\nu};\nu}=0$, $F_{\mu \nu ; \lambda}+ F_{\lambda \mu ; \nu}+ F_{\nu \lambda
; \mu }=0$ also possess no direct coupling to the Ricci tensor,
manipulation of the Maxwell equations leads to the second order 
$g^{\alpha \beta}F^{\mu\nu}_{\phantom {\mu\nu};\alpha; \beta}
+F^{\mu\alpha}R^{\nu}_{\phantom{\nu}\alpha}
-F^{\nu\alpha}R^{\mu}_{\phantom{\nu}\alpha}=0$, 
and thus to the second order equation $g^{\alpha \beta} A_{\mu;\alpha;
\beta}- A^{\alpha}_{\phantom{\alpha};\alpha ;\mu}+
A^{\alpha}R_{\mu \alpha}=0$ for the vector potential $A^{\mu}$.
Characteristic of all these massless field equations then is the
emergence of an explicit dependence on the Ricci
tensor, and thus of some non-geodesic motion analogous to that exhibited
in Eq. (\ref{20}) when we eikonalize.\footnote{We note that none of
these particular curved space field equations involve the Riemann
tensor, but only the Ricci tensor. This is fortunate since Schwarzschild
geometry tests of gravity would be sensitive to the Riemann tensor, a
tensor which in contrast to the Ricci tensor, does not vanish
in a Schwarzschild geometry.}  Now equations such as Eq. (\ref{20}) will
still obey the equivalence principle (as we have defined it above),
since the four-acceleration which appears is still one which contains
the non-tensor Christoffel symbols. Moreover, despite the fact that all
of these particular curved space equations are non-geodesic,
nonetheless, in the flat space limit all of them degenerate into the
flat space geodesics, with the eikonal rays travelling on straight
lines. Now in general, what is understood by covariantizing is to
replace flat spacetime expressions by their curved space counterparts,
with Eq. (\ref{4}) for instance being replaced by Eq. (\ref{5}).
However, this is a very restrictive procedure, since Eqs. (\ref{5}) and
(\ref{10}) both have the same flat space limit. The standard
covariantizing prescription will thus fail to generate any terms which
explicitly depend on curvature (terms which in some cases we see must
be there), and thus constructing a curved space energy-momentum tensor
purely by covariantizing a flat space one is not at all a general
prescription, a issue we shall return to below when we discuss the
curved space energy-momentum tensor in detail. 

\subsection{Massive field motion}

A situation analogous to the above also obtains for massive fields.
Specifically for the quantum-mechanical minimally coupled curved space
massive Klein-Gordon equation, viz. 
\begin{equation}
S^{;\mu}_{\phantom{\mu};\mu}
-\frac{m^2c^2}{\hbar^2}S=0~~,~~~R_{\mu
\nu\sigma\tau}\neq 0~~,
\label{21}
\end{equation}
the substitution $S(x)={\rm exp}(iP(x)/\hbar)$ yields
\begin{equation}
P^{;\mu}P_{;\mu}+m^2c^2=i\hbar P^{;\mu}_{\phantom{;\mu}; \mu}~~.
\label{22}
\end{equation}
In the eikonal (or the small $\hbar$) approximation the 
$i\hbar P^{\mu}_{\phantom{\mu};\mu}$ 
term can be dropped, so that the phase $P(x)$ is then seen to obey the
purely classical condition 
\begin{equation}
g_{\mu \nu}P^{;\mu}P^{;\nu}+m^2c^2=0~~.
\label{23}
\end{equation}
We immediately recognize Eq. (\ref{23}) as the covariant
Hamilton-Jacobi equation of classical mechanics, an equation whose
solution is none other than the stationary classical action $\int
p_{\mu}dx^{\mu}$ as evaluated between relevant end points. In the
eikonal approximation then we can thus identify the wave phase $P(x)$
as $\int p_{\mu}dx^{\mu}$, with the phase derivative
$P^{;\mu}=\partial^{\mu}P$ then being given as the particle momentum 
$p^{\mu}=mcdx^{\mu}/d\tau$, a four-vector momentum which accordingly
has to obey   
\begin{equation}
g_{\mu\nu}p^{\mu}p^{\nu}+m^2c^2=0~~,
\label{24}
\end{equation}
the familiar fully covariant particle energy-momentum relation.
With covariant differentiation of Eq. (\ref{24}) immediately leading to
the classical massive particle geodesic equation
\begin{equation}
p^{\mu}p^{\lambda}_{\phantom{\lambda} ;\mu}=\frac{d^2x^{\lambda} }{
d\tau^2} +\Gamma^{\lambda}_{\mu \nu} 
\frac{dx^{\mu}}{ d\tau}\frac{dx^{\nu } }{ d\tau} =0
\label{25}
\end{equation}
(as obtained here with the proper time $d\tau$ appropriate to
massive particles and not the affine parameter $q$), we thus recover
the well known result that the center of a quantum-mechanical wave
packet follows the stationary classical trajectory. Further, since we
may also reexpress the stationary $\int p_{\mu}dx^{\mu}$ as $-mc\int d
\tau$, we see that we can also identify the quantum-mechanical eikonal
phase as $P(x)=-mc\int d\tau$, to thus enable us to make contact with
the $I_T$ action given in Eq. (\ref{6}).\footnote{We make contact with
the $I_T$ of Eq. (\ref{6}) since we start with the minimally coupled
Eq. (\ref{21}).} Though we have thus made contact with $I_T$, it is
important to realize that we were only able to arrive at Eq. (\ref{23})
after having started with the equation of motion of Eq. (\ref{21}),
an equation whose own validity requires that stationary variation of
the Klein-Gordon action from which it is derived had already been made,
with only the stationary classical action actually being a solution to
the Hamilton-Jacobi equation. The action $I_T=-mc\int d\tau$ as
evaluated along the stationary classical path is thus a part of the
solution to the wave equation, i.e. the output, rather than a part of
the input.\footnote{Thus we cannot appeal to
$I_T$ to put particles on geodesics, since we already had to put them
on the geodesics which follow from  Eq. (\ref{23}) in order to get to
$I_T$ in the first place. While the use of an action such as $I_T$ will
suffice to obtain geodesic motion, as we thus see, its use is not at all
necessary, with it being eikonalization of the quantum-mechanical wave
equation which actually puts massive particles on geodesics.} Thus in
the quantum-mechanical case we never need to assume as input the
existence of any point particle action such as $I_T$ at all. Rather, we
need only assume the existence of equations such as the standard
Klein-Gordon equation, with eikonalization then precisely putting
particles onto classical geodesics just as desired. To conclude then,
we see that not only does the equivalence principle hold for light
(even though it has no inertial mass or gravitational mass at all) and
hold for quantum-mechanical particles, we also see that the equivalence
principle need not be intimately tied to the classical test particle
action $I_T$ at all. Given this analysis, we turn now to consideration
of the curved space energy-momentum tensor.

\section{Structure of the energy-momentum tensor}

\subsection{Kinematic perfect fluids}

At the time of the development of general relativity, the prevailing
view of gravitational sources was that they were to be treated like
billiard balls, i.e. as purely mechanical kinematic particles which
carry energy and momentum; with the advent of general relativity then
requiring that such energy and momentum be treated covariantly, so that
the way the energy-momentum tensor was to be introduced in gravitational
theory was to simply covariantize the appropriate flat spacetime
expressions. Despite the subsequent realization that particles are very
far from being such kinematic objects (elementary particles are now
thought to get their masses dynamically via spontaneous symmetry
breaking), and despite the fact that the now standard $SU(3)\times
SU(2)\times U(1)$ theory of strong, electromagnetic and weak
interactions ascribes primacy to fields over particles, the
kinematically prescribed energy-momentum tensor (rather than an
$SU(3)\times SU(2)\times U(1)$ based one) is nonetheless still used in
treatments of gravity today. Apart from this already disturbing
shortcoming (one we shall remedy below by constructing the
energy-momentum tensor starting from fields rather than particles), an
additional deficiency of a purely kinematic prescription is that since
flat spacetime energy-momentum tensors had no need to know where the
zero of energy was (in flat spacetime only energy differences are
observable), their covariantizing left unidentified where the zero of
energy might actually be; and since gravity couples to energy itself
rather than to energy differences, this kinematic prescription is thus
powerless to address the cosmological constant problem, an issue to
which we shall return below. 

Historically there was of course good reason to treat particles
kinematically, since such a treatment did lead to geodesic motion for
the particles. Thus for the test particle action $I_T$ of Eq.
(\ref{6}), its functional variation with respect to the metric allows
one to define an energy-momentum tensor according to\footnote{With our
definition here and throughout of $T^{\mu\nu}$ as
$T^{\mu\nu}=2(\delta I_T/\delta g_{\mu\nu})/(-g)^{1/2}$, it is
$cT_{00}$ which then has the dimension of an energy density rather
than $T_{00}$ itself.}
\begin{equation}
\frac{2}{(-g)^{1/2}}\frac{\delta I_T}{ \delta g_{\mu
\nu}}=T^{\mu \nu}=\frac{mc}
{(-g)^{1/2}}                                                
\int d\tau \delta^4
(x-y(\tau))\frac{dy^{\mu}}                                      
{d\tau}\frac{dy^{\nu}}{d\tau}~~,
\label{26}
\end{equation}                                 
with its covariant conservation ($T^{\mu\nu}_{\phantom{\mu\nu};\nu}=0$) 
leading right back to the geodesic equation of Eq.
(\ref{5}).\footnote{While our ability to impose a conservation condition
on $T^{\mu\nu}$ would follow from Eq. (\ref{1}) since the Einstein
tensor $G^{\mu\nu}=R^{\mu\nu}-(1/2)g^{\mu\nu}
R^{\alpha}_{\phantom{\alpha}\alpha}$ obeys the Bianchi identity, the use of a conservation condition
in no way requires the validity of Eq. (\ref{1}). Specifically, in any
covariant theory of gravity in which the pure gravitational piece of the
action, $I_{GRAV}$, is a general coordinate scalar function of the
metric, the quantity $A^{\mu\nu}=(2/(-g)^{1/2})(\delta I_{GRAV}/\delta
g_{\mu\nu})$ will, because of covariance, automatically
obey $A^{\mu\nu}_{\phantom{\mu\nu};\nu}=0$, and through the
gravitational equation of motion $A^{\mu\nu}=T^{\mu\nu}$ then lead
to the covariant conservation of $T^{\mu\nu}$. The use of the Einstein
equations is only sufficient to yield
$T^{\mu\nu}_{\phantom{\mu\nu};\nu}=0$, but not at all necessary, with
the conservation ensuing in any general covariant pure metric theory
of gravity.} Moreover, an analogous situation is met when the
energy-momentum tensor is taken to be a perfect fluid of the form
\begin{equation}
T^{\mu \nu}=\frac{1}{c}\left[(\rho+p)U^{\mu}U^{\nu}+pg^{\mu\nu}\right]
\label{27}
\end{equation}                                 
with energy density $\rho$, pressure $p$ and a fluid
four-vector which is normalized to $U_{\mu}U^{\mu}=-1$. Specifically,
for such a fluid, covariant conservation leads to 
\begin{equation}
\left[ (\rho + p)U^{\mu}U^{\nu} +pg^{\mu \nu}\right]_{;\nu} =
[(\rho+p)U^{\nu}]_{;\nu}U^{\mu}                                               
+(\rho+p)U^{\mu}_{\phantom{\mu};\nu}U^{\nu} +p^{;\mu}=0~~,
\label{28}
\end{equation}                                 
and thus to
\begin{equation}
- [(\rho+p)U^{\nu}]_{;\nu}
+U_{\mu}p^{;\mu}=-                  
(\rho+p)U^{\mu}_{\phantom{\mu};\nu}U^{\nu}U_{\mu}=0~~,
\label{29}
\end{equation}                                 
with the last equality following since
$U_{\mu}U^{\mu}_{\phantom{\mu};\nu}=0$. With the insertion of Eq.
(\ref{29}) into (\ref{28}) then yielding
\begin{equation}
(\rho+p)U^{\mu}_{\phantom{\mu};\nu}U^{\nu}                                    
+p_{;\nu}[g^{\mu\nu} + U^{\mu}U^{\nu}]=0~~, 
\label{30}
\end{equation}                                 
viz. 
\begin{equation}
\frac{D^2x^{\mu}}{D\tau^2}=                                   
-[g^{\mu\nu} + U^{\mu}U^{\nu}]\frac{p_{;\nu}}{(\rho+p)}~~, 
\label{31}
\end{equation}                                 
we see that the geodesic equation then emerges whenever the right-hand
side of Eq. (\ref{31}) is negligible, a situation which is for instance
met when the fluid is composed of non-interacting pressureless dust.

Since Eqs. (\ref{26}) and (\ref{31}) do lead to geodesic motion, it is
generally thought that gravitational sources should thus be described
in this way. However, even within an a priori kinematic perfect fluid
framework, Eq. (\ref{31}) would in no way need to be altered if to the
perfect fluid of Eq. (\ref{27}) we were to add an additional
$T^{\mu\nu}_{EXTRA}$ which was itself independently covariantly
conserved. Since a geometric tensor such as the Einstein tensor
$G^{\mu\nu}$ is both covariantly conserved and non-existent in flat
spacetime, a curved space generalization of a flat space
energy-momentum tensor which would include a term of the form
$T^{\mu\nu}_{EXTRA}=G^{\mu\nu}$ would not affect Eq. (\ref{31}) at all.
Moreover, since the metric tensor $g^{\mu\nu}$ is also covariantly
conserved, the inclusion in $T^{\mu\nu}_{EXTRA}$ of a $\Lambda
g^{\mu\nu}$ term with $\Lambda$ constant would also leave Eq.
(\ref{31}) untouched, and while such a term would not vanish in the
flat spacetime limit, its presence in flat spacetime would only lead to
a non-observable overall shift in the zero of energy. Restricting to
the perfect fluid of Eq. (\ref{27}) is thus only sufficient to recover
Eq. (\ref{31}), and not at all necessary.

Within the kinematic perfect fluid framework, the use of such fluids as
gravitational sources is greatly facilitated if some equation of state
of the form $p=w\rho$ can be prescribed where $w$ would be a constant.
To see what choices are suggested for $w$ from flat space physics, we
consider a relativistic flat space ideal $N$ particle classical gas of
particles of mass $m$ in a volume $V$ at a temperature $T$. For this
system the Helmholtz free energy $A(V,T)$ is given as 
\begin{equation}
e^{-A(V,T)/NkT}=V\int
d^3pe^{-(p^2+m^2)^{1/2}/kT}~~, 
\label{32}
\end{equation}                                 
so that the pressure takes the simple form 
\begin{equation}
P=-\left(\frac{\partial A}{ \partial
V}\right)_T=\frac{NkT}{V}~~,
\label{33}
\end{equation}                                 
while the internal energy $U=\rho V$ evaluates in terms of Bessel
functions as  
\begin{equation}
U=A-T\left(\frac{\partial A}{ \partial
T}\right)_V=3NkT+Nm\frac{K_1(m/kT)}{K_2(m/kT)}~~.
\label{34}
\end{equation}                                 
In the high and low temperature limits (the radiation and matter eras)
we then find that the expression for $U$ simplifies to
\begin{eqnarray}
\frac{U}{V}\rightarrow
\frac{3NkT}{V}=3P~~&,&~~\frac{m}{kT}\rightarrow 0~~,
\nonumber \\
\frac{U}{V} \rightarrow
\frac{Nm}{V}+\frac{3NkT}{2V}=\frac{Nm}{V}+\frac{3P}{2} \approx
\frac{Nm}{V}~~&, &~~\frac{m}{kT}
\rightarrow \infty~~.
\label{35}
\end{eqnarray}                                 
Consequently, while $p$ and $\rho$ are nicely proportional to each other
in the high temperature radiation and the low temperature matter eras
(where $w(T\rightarrow\infty)=1/3$ and $w(T\rightarrow 0)=0$), we also
see that in transition region between the two eras their relationship is
altogether more complicated. Use of a $p=w\rho$ equation of state would
at best only be valid at temperatures which are very different from
those of order $m/K$, though for massless particles it would be of
course be valid to use $p=\rho/3$ at all temperatures, a point to which
we shall return below.

\subsection{Perfect Robertson-Walker fluids}

In trying to develop equations of state in curved space, one should
replace the partition function in Eq. (\ref{32}) by its curved space
generalization (i.e. one should covariantize it, just as is proposed
for $T^{\mu\nu}$ itself),\footnote{Typically one replaces
$(p^2+m^2)^{1/2}/kT$ by $(dx^{\mu}/d\tau)U^{\nu}g_{\mu\nu}/kT$ 
[$U^{\nu}$ is a four-vector] and replaces
$\int d^3p$ by a sum over a complete set of basis modes associated with
the propagation of a spinless massive particle in the chosen
$g_{\mu\nu}$ background.} and then follow the steps above to see what
generalization of Eq. (\ref{35}) might then ensue. However for curved
backgrounds of high symmetry, the use of the isometry structure of the
background can greatly simplify the discussion. Thus, for instance, for
the Robertson-Walker geometry of relevance to cosmology, viz.
\begin{equation}
ds^2=-c^2dt^2+R^2(t)\left( \frac{dr^2}{(1-kr^2)}+r^2d\Omega_2\right)~~,
\label{36}
\end{equation}                                 
the maximal 3-symmetry of the background entails that any rank two
tensor such as the energy-momentum tensor itself must have the generic
form
\begin{equation}
T^{\mu\nu}=[C(t)+D(t)]U^{\mu}U^{\nu}+D(t)g^{\mu\nu}~~,
\label{37}
\end{equation}                                 
to thus automatically be of a perfect fluid form with comoving
fluid 4-vector $U^{\mu}=(1,0,0,0)$ and general functions $C$ and $D$
which can only depend on the comoving time $t$. With the
energy-momentum tensor being covariantly conserved, $C$ and $D$ have to
obey
\begin{equation}
\frac{d}{dt}\left(R^3(C+D)\right)=R^3\frac{dD}{dt}~~,
\label{38}
\end{equation}                                 
with $D$ and $C$ thus being related according to
\begin{equation}
D=-\frac{d}{dR^3}\left(R^3C\right)~~.
\label{39}
\end{equation}                                 
While $D$ and $C$ must be also directly proportional to each other in
the Robertson-Walker case since each only depends on the single
parameter $t$, nonetheless, even though we therefore can set $D/C=w$ in
such cases, in general the quantity $w$ could still be a function of
$t$ and need not necessarily be a constant.\footnote{When $w$ is a
constant, we can then set $C=1/R^{3(w+1)}$.} In the above we have
purposefully not identified $C$ with a fluid $\rho$ and $D$ with a
fluid $p$, since should the energy-momentum tensor actually consist of
two types of fluid (this being the conventional dark matter plus
dark energy picture), even if both of them have their own independent
$w$ according to $p_1=w_1\rho_1$, $p_2=w_2\rho_2$, the sum of their
pressures would then obey $p_1+p_2=w_1\rho_1+w_2\rho_2$ and would not
in general be proportional to $\rho_1+\rho_2$, with the total
$\rho_1+\rho_2$ and
$p_1+p_2$ obeying
\begin{equation}
p_1+p_2=-\frac{d}{dR^3}\left(R^3(\rho_1+\rho_2)\right)~~,
\label{40}
\end{equation}                                 
and with neither $\rho_1+\rho_2$ or $p_1+p_2$ then scaling as a power of
$R$.\footnote{As well as needing to require that $w_1$ and $w_2$ both
be constant, to secure the conventional $\rho_1=1/R^{3(w_1+1)}$,
$\rho_2=1/R^{3(w_2+1)}$ in the two fluid case additionally requires
the separate covariant conservation of the energy-momentum tensor of
each fluid, so that the two fluids do not then exchange energy
and momentum with each other. A dynamics which is to secure this would
have to be identified in any cosmological model (such as a dark matter
plus quintessence fluid model) which uses two such fluids unless one of
the two fluids just happens to have $w=-1$, since this would correspond
to a cosmological constant whose energy momentum tensor
$T^{\mu\nu}=-\Lambda g^{\mu\nu}$ is in fact independently conserved.}
However, if the full energy-momentum tensor is to describe radiation
fluids alone, then no matter how many of them there might actually be
in total, in such a situation the full $T^{\mu\nu}$ must additionally
obey the tracelessness condition $T^{\mu}_{\phantom{\mu}\mu}=0$, to
then unambiguously fix $w=D/C$ to the unique value $w=1/3$.

\subsection{Schwarzschild fluid sources}

While the isometry of the Robertson-Walker geometry does automatically
lead us to the perfect fluid form given in Eq. (\ref{37}) (though not
necessarily to any particular equation of state unless
$T^{\mu}_{\phantom{\mu}\mu}=0$), the situation for lower symmetry
backgrounds is not as straightforward. Specifically, for a standard
static, spherically symmetric geometry of the form  
\begin{equation}
ds^2=-B(r)c^2dt^2+A(r)dr^2+r^2d\Omega_2~~,
\label{41}
\end{equation}                                 
the three Killing vector symmetry of the background entails that the
most general energy-momentum tensor must have the generic diagonal form
\begin{equation}
T_{00}=\rho(r) B(r)~~,~~T_{rr}=p(r) A(r)~~,~~T_{\theta\theta}=q(r)r^2~~,
~~T_{\phi\phi}=q(r)r^2{\rm sin}^2\theta 
\label{42}
\end{equation}                                 
with conservation condition
\begin{equation}
\frac{dp}{dr}+\frac{(\rho +p)}{2B}\frac{dB}{dr}+\frac{2}{r}(p-q)=0~~,
\label{43}
\end{equation}                                 
with the function $q(r)$ not at all being required to be equal to
$p(r)$.\footnote{Unlike the Robertson-Walker case, a static,
spherically symmetric geometry is only spherically symmetric about a
single point and not about all points in the spacetime.} Thus while a
flat space static, spherically symmetric perfect fluid would have $p=q$
and equation of state $p=w\rho$, it does not follow that a curved space
one would as well, with this being a dynamical and not a kinematic
issue whose resolution would require an evaluation of the covariant
partition function in the background of Eq. (\ref{41}).\footnote{Since
$q$ would be equal to $p$ in the flat space limit, one might
anticipate that for weak gravity the difference between $q$ and $p$
would still be small, though $q$ could differ radically from $p$ in the
strong gravity black hole limit.}

Despite these considerations, it turns out that a quite a bit of the
standard phenomenology associated with static, spherically symmetric
sources still holds even if $q$ and $p$ are quite different from each
other. Specifically, for the geometry of Eq. (\ref{41}) the components
of the Ricci tensor obey
\begin{equation}
\frac{R_{00}}{2B}+\frac{R_{rr}}{2A}+\frac{R_{\theta\theta}}{r^2}=
\frac{1}{r^2}\left[\frac{d}{dr}\left(\frac{r}{A}\right)-1\right]~~,~~
R_{\theta\theta}=-1+\frac{r}{2A}\left(\frac{1}{B}\frac{dB}{dr} -
\frac{1}{A}\frac{dA}{dr}\right)+\frac{1}{A}~~,
\label{44}
\end{equation}                                 
while the components of the energy-momentum tensor obey
\begin{eqnarray}
\left(T_{00}-
\frac{g_{00}T^{\alpha}_{\phantom{\alpha}\alpha}}{2}\right)&=&
\frac{B}{2}(\rho+p+2q)~~,
\nonumber \\
\left(T_{rr}-
\frac{g_{rr}T^{\alpha}_{\phantom{\alpha}\alpha}}{2}\right)&=&
\frac{A}{2}(\rho+p-2q)~~,
\nonumber \\
\left(T_{\theta\theta}-
\frac{g_{\theta\theta}T^{\alpha}_{\phantom{\alpha}\alpha}}{2}\right)&=&
\frac{r^2}{2}(\rho-p)~~,~~
\nonumber \\
\frac{1}{2B}\left(T_{00}-
\frac{1}{2}g_{00}T^{\alpha}_{\phantom{\alpha}\alpha}\right)
+\frac{1}{2A}\left(T_{rr}-
\frac{1}{2}g_{rr}T^{\alpha}_{\phantom{\alpha}\alpha}\right)
+\frac{1}{r^2}\left(T_{\theta\theta}-
\frac{1}{2}g_{\theta\theta}T^{\alpha}_{\phantom{\alpha}\alpha}\right)
&=&\rho~~,
\label{45}
\end{eqnarray}                                 
with the last expression conveniently being independent of $q$. If we
now do impose the Einstein equations of Eq. (\ref{1}), then in terms of
the quantity
\begin{equation}
\tilde{M}(r)=4\pi\int_0^rdr r^2\rho(r)
\label{46}
\end{equation}                                 
we immediately obtain
\begin{equation}
A^{-1}(r)=1-\frac{2 G}{r}\tilde{M}(r)~~,
\label{47}
\end{equation}                                 
to then find that $B(r)$ obeys
\begin{equation}
\frac{1}{B}\frac{dB}{dr}=\frac{2G}{r(r-2G\tilde{M})}
\left(\tilde{M}+4\pi r^3p\right)~~.
\label{48}
\end{equation}                                 
We recognize Eqs. (\ref{47}) and (\ref{48}) as being of precisely the
same form as the standard expressions which are obtained (see e.g.
\cite{Weinberg1972}) for $A$ and $(1/B)dB/dr$ when $q(r)$ is equal to
$p(r)$, though the substitution of these expressions into Eq. (\ref{43})
leads to 
\begin{equation}
\frac{dp}{dr}+\frac{(\rho +p)G}{r(r-2G\tilde{M})}
\left(\tilde{M}+4\pi r^3p\right)+\frac{2}{r}(p-q)=0~~,
\label{49}
\end{equation}                                 
an equation which differs from the standard expression by the presence
of the $2(p-q)/r$ term. If the matter density terminates at some finite
$r=R$, then outside of the fluid the geometry is a standard exterior
Schwarzschild geometry with metric
\begin{equation}
B(r>R)=A^{-1}(r>R)=1-\frac{2 MG}{r}~~.
\label{50}
\end{equation}                                 
Matching this exterior solution to the interior solution at $r=R$ then
yields for $M$ the standard
\begin{equation}
M=4\pi\int_0^Rdr r^2\rho(r)~~,
\label{51}
\end{equation}                                 
with the integration constant required for Eq. (\ref{48}) then being
fixed to yield for $B(r)$ the standard expression
\begin{equation}
B(r)={\rm exp}\left(-2G\int_r^{\infty}dr 
\frac{(\tilde{M}+4\pi r^3p}{r(r-2G\tilde{M})}\right)~~.
\label{52}
\end{equation}                                 
As we thus see, the functional forms of Eqs. (\ref{47}), (\ref{51}) and
(\ref{52}) are completely unaffected by whether or not $q$ is equal to
$p$, with any difference between $q$ and $p$ only showing up in Eq.
(\ref{49}). As far as the geometry outside of the fluid is concerned,
the structure of the exterior Schwarzschild metric is the standard one
with the standard form for the total mass $M$ as given in Eq.
(\ref{51}). However, inside the fluid the dynamics could be different
from the conventional treatment because of modifications to the
equation of state. However, for weak gravity where $p$ is already of
order $G$, should the term of order $G$ in $dp/dr$ be given exactly
by $G\rho\tilde{M}/r^2$, the quantity $p-q$ would then only begin in
order $G^2$ and the standard lowest order in $G$ hydrostatic treatment
of sources such as stars would not be affected. Nonetheless, even if
that is to be the case (something which is not immediately clear), for
strong gravity inside sources (where there are currently no data), any
difference between $p$ and $q$ could have substantial consequences, and
so we should not in general expect a static, spherically symmetric
source to possess an energy-momentum tensor of the form
$T^{\mu\nu}=[\rho(r)+p(r)]U^{\mu}U^{\nu}+p(r)g^{\mu\nu}$
just because it does so in flat space.\footnote{In passing we
additionally note that once $q$ is not equal $p$, for a traceless fluid
with $T^{\mu}_{\phantom{\mu}\mu}=-\rho+p+2q=0$, Eq. (\ref{43}) reduces
to $dp/dr+(\rho +p)/2B)(dB/dr)+(3p-\rho)/r=0$, and dependent on how $p$
depends on $\rho$, it may be possible for radiation to support a static,
stable source. (When $\rho=3p$ the only solution is $p\sim 1/B^2$ which
would require $B(r)$ to be singular at the point $r=R$ where $p$
vanishes.)}

\subsection{Scalar field fluid sources}

As we have thus seen, wisdom gained from experience with kinematic
particle sources in flat space serves as a quite limited guide to the
structure of gravitational sources in curved spacetime. However, that
is not their only shortcoming, with their connection to the structure
of the energy-momentum tensor which is suggested by fundamental theory
being quite remote. Thus we need to discuss what is to be expected of
the energy-momentum tensor in a theory in which the action is built out
of fields rather than particles, and in which the fields develop
masses by spontaneous symmetry breakdown. However, before going to the
issue of dynamical masses, we first need to see how we can connect our
above analysis of geodesic motion of eikonalized fields to the
structure of the energy-momentum tensor. 

To illustrate what is involved, it is convenient to consider a
massive complex flat spacetime scalar field, and to get its
energy-momentum tensor (and to subsequently enforce a tracelessness
condition for it when we restrict to massless fields), we take as
action the non-minimally coupled curved space action\footnote{To
construct the correct energy-momentum tensor in flat space, it is
necessary to first vary the curved space action with respect to the
metric and then take the flat limit.}
\begin{equation}
I_M=-\int
d^4x(-g)^{1/2}\left[\frac{1}{2}S^{;\mu}
S^{*}_{;\mu} +\frac{1}{2}m^2SS^{*}
-\frac{\xi}{12}SS^{*}R^{\mu}_{\phantom{\mu}\mu}\right]~~.
\label{53}
\end{equation}                                 
Its variation with respect to the scalar field yields the equation of
motion 
\begin{equation}
S^{;\mu}_{\phantom{\mu};\mu}+\frac{\xi}{6}SR^{\mu}_{\phantom{\mu}\mu}
-m^2S=0~~,
\label{54}
\end{equation}                                 
while its variation with respect the metric yields the energy-momentum 
tensor
\begin{eqnarray}
T_{\mu \nu} &=&
\left(\frac{1}{2}-\frac{\xi}{6}\right)\left(S_{;\mu}
S^{*}_{;\nu} +S_{;\nu}S^{*}_{;\mu}\right) 
-\frac{(3-2\xi)}{6}g_{\mu\nu}S^{;\alpha} S^{*}_{;\alpha}
-\frac{\xi}{6}\left(SS^{*}_{;\mu;\nu}+S^{*}S_{;\mu;\nu}\right)
\nonumber \\             
&&+\frac{\xi}{6}g_{\mu\nu}\left(
S^{*}S^{;\alpha}_{\phantom{;\alpha};\alpha}
+SS^{*;\alpha}_{\phantom{*;\alpha};\alpha}\right)
-\frac{1}{2}g_{\mu\nu}m^2SS^{*}                            
-\frac{\xi}{6}SS^{*}\left(R_{\mu\nu}
-\frac{1}{2}g_{\mu\nu}R^{\alpha}_{\phantom{\alpha}\alpha}\right)        
~~. 
\label{55}
\end{eqnarray}                                 
With the trace of this energy-momentum tensor evaluating to
\begin{equation}
T^{\mu}_{\phantom {\mu} \mu} =(\xi
-1)\left(S^{;\mu}S^{*}_{;\mu}
+\frac{1}{2}S^{*}S^{;\mu}_{\phantom{\mu};\mu}
+\frac{1}{2}SS^{*;\mu}_{\phantom{*\mu};\mu}\right) -m^2SS^{*}
\label{56}
\end{equation}                                 
in field configurations which obey Eq. (\ref{54}), we see that the
choice $\xi=1$, $m=0$ will enforce the tracelessness of $T^{\mu\nu}$. 
Bearing this in mind we thus set $\xi=1$,\footnote{When $\xi=1$, the
coupling of the scalar field to the geometry is conformal, with
the massless action $I_M=-\int d^4x(-g)^{1/2}\left[S^{;\mu}
S^{*}_{;\mu}/2 
- SS^{*}R^{\mu}_{\phantom{\mu}\mu}/12\right]$ being invariant under
the local conformal transformation $S(x)\rightarrow e^{-\alpha(x)}S(x)$,
$g_{\mu\nu}(x)
\rightarrow e^{2\alpha(x)}g_{\mu\nu}(x)$.} so that the flat space limit
of the $\xi=1$ theory then takes the form 
\begin{eqnarray}
T_{\mu \nu} &=&
\frac{1}{3}\left(\partial_{\mu}S
\partial_{\nu}S^{*} +\partial_{\nu}S\partial_{\mu}S^{*}\right) 
-\frac{1}{6}g_{\mu\nu}\partial_{\alpha}S\partial^{\alpha}S^{*}
-\frac{1}{6}\left(S\partial_{\mu}\partial_{\nu}S^{*}
+S^{*}\partial_{\mu}\partial_{\nu}S\right)
\nonumber \\             
&&+\frac{1}{6}g_{\mu\nu}\left(
S^{*}\partial_{\alpha}\partial^{\alpha}S
+S\partial_{\alpha}\partial^{\alpha}S^{*}\right)
-\frac{1}{2}g_{\mu\nu}m^2SS^{*}
\nonumber \\             
&=&\frac{1}{3}\left(\partial_{\mu}S
\partial_{\nu}S^{*} +\partial_{\nu}S\partial_{\mu}S^{*}\right) 
-\frac{1}{6}g_{\mu\nu}\partial_{\alpha}S\partial^{\alpha}S^{*}
-\frac{1}{6}\left(S\partial_{\mu}\partial_{\nu}S^{*}
+S^{*}\partial_{\mu}\partial_{\nu}S\right)
-\frac{1}{6}g_{\mu\nu}m^2SS^{*}~~.  
\label{57}
\end{eqnarray}                                 
In a plane wave solution to the $\partial_{\mu}\partial^{\mu}S=m^2S$
wave equation of the form $S(x)=e^{ik\cdot x}/V^{1/2}(E_k)^{1/2}$ where
$k^{\mu}k_{\mu}=-m^2$, $E_k=(k^2+m^2)^{1/2}$ and $V$ is the 3-volume,
$T_{\mu\nu}$ then readily evaluates to 
\begin{equation}
T_{\mu\nu}=\frac{k_{\mu}k_{\nu}}{VE_k}~~,
\label{58}
\end{equation}                                 
and even though the wavefront to the massive plane wave is geodesic,
this particular energy-momentum tensor does not look anything like a
perfect fluid form, since for a single massive plane wave
$k^{\mu}=(E_k,0,0,k)$ propagating geodetically in the $z$-direction
$T_{\mu\nu}$ evaluates to
\begin{equation}
T_{\mu\nu} = \frac{1}{V}\pmatrix{
E_k&0&0&-k\cr
0&0&0&0 \cr
0&0&0&0 \cr
-k&0&0&k^2/E_k \cr}~~.
\label{59}
\end{equation}                                 
To get a perfect fluid we need to incoherently add an entire family of
these plane waves.\footnote{This is equivalent to using (the zero
temperature limit of) the partition function discussed earlier.} Thus
if we take a set of six plane waves moving in the
$\pm x$, $\pm y$, $\pm z$ directions, all with the same $k=|\vec{k}|$
and $E_k$, viz.
$k^{\mu}=(E_k,k,0,0)$, $k^{\mu}=(E_k,-k,0,0)$, $k^{\mu}=(E_k,0,k,0)$,
$k^{\mu}=(E_k,0,-k,0)$, $k^{\mu}=(E_k,0,0,k)$, and
$k^{\mu}=(E_k,0,0,-k)$, and then incoherently add up their individual
contributions to $T_{\mu\nu}$, we obtain
\begin{equation}
T_{00}=\frac{6E_k}{V}~,~~T_{xx}=T_{yy}=T_{zz}=\frac{2k^2}{E_kV}~~,~~
T^{\mu}_{\phantom {\mu} \mu} =-\frac{6m^2}{E_kV}~~. 
\label{60}
\end{equation}                                 
We recognize Eq. (\ref{60}) as being of precisely the perfect fluid
form with $\rho=6E_k/V$, $p=2k^2/E_kV$. As such, we not only get a
perfect fluid form, we even see that in the event that the mass is
introduced kinematically as in the action of Eq. (\ref{53}), then for
small $k\ll m$ we obtain $p \ll \rho$, with the effective $p/\rho=w$
being zero, while for very large $k$ (or equivalently for zero $m$) we
obtain $p=\rho/3$. As such the above procedure shows how to obtain a
perfect fluid form starting from field theory, with its generalization
to curved space requiring an equivalent incoherent averaging over the
basis modes associated with the curved space wave equations appropriate
to the curved space backgrounds of interest, a procedure which we
indicated earlier is not guaranteed to automatically yield the
straightforward $T_{xx}=T_{yy}=T_{zz}$ condition found in flat space.

\subsection{Dynamical fluid sources}

While the above analysis shows how the perfect fluid form for massive
particles can be obtained from field theory, and shows why the test
particle action and energy-momentum tensor of Eqs. (\ref{6}) and
(\ref{26}) are not at all germane to the issue, this analysis still
fails to take into consideration the fact that elementary particle
masses are not kinematic, but rather that they are acquired dynamically
by spontaneous breakdown. To investigate what is to happen in the
dynamical mass case, it is convenient to consider a spin one-half matter
field fermion $\psi(x)$ which is to get its mass through a real
spin-zero Higgs scalar boson field $S(x)$. In order to illustrate the
difference between dynamic and kinematic masses in the sharpest way
possible, we take the action to possess no intrinsic mass scales and the
energy-momentum tensor to be traceless. We thus give neither the fermion
nor the scalar field any kinematic mass at all, and in order to secure
the tracelessness of the energy-momentum tensor couple the scalar field
conformally to gravity, to thus yield as curved space matter 
action\footnote{In Eq. (\ref{61}) the general-relativistic Dirac
matrices $\gamma^{\mu}(x)$ and the fermion spin connection
$\Gamma_{\mu}(x)$ are defined as
$\gamma^{\mu}(x)=V^{\mu}_a(x)\hat{\gamma}^a$ and
$\Gamma_{\mu}=[\gamma^{\nu}(x),\partial_{\mu}\gamma_{\nu}(x)]/8
-[\gamma^{\nu}(x),\gamma_{\sigma}(x)]\Gamma^{\sigma}_{\mu\nu}/8$ where 
$V^{\mu}_a(x)$ is a vierbein and $\hat{\gamma}^a$ is a
special-relativistic Dirac gamma matrix which, with the
$ds^2=-(dx^0)^2+(dx^1)^2+(dx^2)^2+(dx^3)^2=\eta_{ab}dx^adx^b$ metric,
obeys
$\hat{\gamma}^a\hat{\gamma}^b+\hat{\gamma}^b\hat{\gamma}^a=-2\eta^{ab}$,
while
$\bar{\psi}$ is given by
$\bar{\psi}=\psi^{\dagger}\hat{D}$ where $\hat{D}$ is a Hermitian
flat spacetime Dirac matrix which effects
$\hat{D}\hat{\gamma}^{a}\hat{D}^{-1}=\hat{\gamma}^{a\dagger}$. In our
notation, here and throughout Hermiticity in the fermion kinetic energy
sector is implicit, with
$i\bar{\psi}\gamma^{\mu}(x)[\partial_\mu+\Gamma_\mu(x)]          
\psi$ denoting
$(i/2)\bar{\psi}\gamma^{\mu}(x)[\partial_\mu+\Gamma_\mu(x)]          
\psi 
-(i/2)\bar{\psi}[\stackrel{\leftarrow}{\partial_\mu}+\Gamma_{\mu}(x)]
\gamma^{\mu}(x)\psi
=(i/2)\bar{\psi}\gamma^{\mu}(x)[\partial_\mu+\Gamma_\mu(x)]          
\psi +{\rm Hermitian~conjugate}$, and
$i
\bar{\psi} \gamma_{\mu}(x)[
\partial_{\nu}
+\Gamma_\nu(x)]\psi$ 
denoting $(i/4) \bar{\psi} \gamma_{\mu}(x)[
\partial_{\nu}+\Gamma_\nu(x)]\psi +(i/4)\bar{\psi} \gamma_{\nu}(x)[
\partial_{\mu}+\Gamma_\mu(x)]\psi
+{\rm Hermitian~conjugate}$ in the associated energy-momentum tensor
constructed below as the symmetric
$T_{\mu\nu}=2g_{\mu\alpha}g_{\nu\beta}(\delta I_M/\delta
g_{\alpha\beta})/(-g)^{1/2}=-(1/2)[(V_{\mu a}(x)/|V(x)|)(\delta
I_M/\delta V^{\nu}_{a}(x))+(V_{\nu a}(x)/|V(x)|)(\delta I_M/\delta
V^{\mu}_{a}(x))]$ where
$|V(x)|=(-g)^{1/2}$. In this construction we note that $\delta
g_{\mu\nu}=V_{\mu a}\delta V_{\nu}^a+\delta V_{\mu a}V_{\nu}^a$, 
$\delta|V(x)|=-|V(x)|V_{\mu}^{a}(x)\delta V^{\mu}_{a}(x)$, and $\delta
V_{\sigma c}=-V_{\mu c}V^{a}_{\sigma}\delta V^{\mu}_{a}$, to thereby
yield $(V_{\nu a}(x)/|V(x)|)(\delta/\delta V^{\mu}_{a}(x))[\int d^4x
(-g)^{1/2}\bar{\psi}\gamma^{\sigma}(x)[\partial_{\sigma}
+\Gamma_{\sigma}(x)]\psi] =
\bar{\psi}\gamma_{\nu}(x)[\partial_\mu+\Gamma_\mu(x)]\psi
+(1/8)\bar{\psi}\gamma^{\sigma}(x)
([\gamma_{\nu},\gamma_{\mu}])_{;\sigma}\psi
-g_{\nu\mu}\bar{\psi}\gamma^{\sigma}(x)[\partial_{\sigma}
+\Gamma_{\sigma}(x)]\psi$, from which Eq. (\ref{64}) then follows.
(The author wishes to thank Dr. R. K. Nesbet for helpful comments
on the material presented here.)}
\begin{equation}
I_M=-\int d^4x(-g)^{1/2}\left[\frac{1}{2}S^{;\mu}
S_{;\mu}-\frac{1}{12}S^2R^\mu_{\phantom         
{\mu}\mu}
+\lambda S^4
+i\bar{\psi}\gamma^{\mu}(x)[\partial_\mu+\Gamma_\mu(x)]             
\psi -hS\bar{\psi}\psi\right]
\label{61}
\end{equation}                                 
where $h$ and $\lambda$ are dimensionless coupling
constants.\footnote{ As such, the action of Eq. (\ref{61}) is the most
general curved space matter action for the $\psi(x)$ and $S(x)$ fields
which is invariant under the local conformal transformation
$S(x)\rightarrow e^{-\alpha(x)}S(x)$, $\psi(x)\rightarrow
e^{-3\alpha(x)/2}\psi(x)$,
$\bar{\psi}(x)\rightarrow e^{-3\alpha(x)/2}\bar{\psi}(x)$,
$g_{\mu\nu}(x)\rightarrow e^{2\alpha(x)}g_{\mu\nu}(x)$.} Variation of
this action with respect to 
$\psi(x)$ and
$S(x)$ yields the equations of motion
\begin{equation}
i \gamma^{\mu}(x)[\partial_{\mu} +\Gamma_\mu(x)]                              
\psi - h S \psi = 0~~,
\label{62}
\end{equation}                                 
and 
\begin{equation}
S^{;\mu}_{\phantom{\mu};\mu}+\frac{1}{6}SR^\mu_{\phantom{\mu}\mu}
-4\lambda S^3 +h\bar{\psi}\psi=0~~,
\label{63}
\end{equation}                                 
while variation with respect to the metric yields (without use of any
equation of motion) an energy-momentum tensor of the form
\begin{eqnarray}
T_{\mu \nu} &=& i \bar{\psi} \gamma_{\mu}(x)[
\partial_{\nu}                    
+\Gamma_\nu(x)]                                                                 
\psi+\frac{2}{3}S_{;\mu} S_{;\nu} 
-\frac{1}{6}g_{\mu\nu}S^{;\alpha} S_{;\alpha}
-\frac{1}{3}SS_{;\mu;\nu}
+\frac{1}{3}g_{\mu\nu}SS^{;\alpha}_{\phantom{;\alpha};\alpha}  
\nonumber \\             
&&                          
-\frac{1}{6}S^2\left(R_{\mu\nu}
-\frac{1}{2}g_{\mu\nu}R^\alpha_{\phantom{\alpha}\alpha}\right)         
-g_{\mu\nu}\left[\lambda S^4
+i\bar{\psi}\gamma^{\alpha}(x)[\partial_{\alpha}+\Gamma_{\alpha}(x)]             
\psi -hS\bar{\psi}\psi\right]~~, 
\label{64}
\end{eqnarray}                                 
with use of the matter field equations of motion then permitting us to
rewrite $T_{\mu\nu}$ as  
\begin{eqnarray}
T_{\mu \nu} &=& i \bar{\psi} \gamma_{\mu}(x)[
\partial_{\nu}                    
+\Gamma_\nu(x)]                                                                 
\psi+\frac{2}{3}S_{;\mu} S_{;\nu} 
-\frac{1}{6}g_{\mu\nu}S^{;\alpha} S_{;\alpha}
-\frac{1}{3}SS_{;\mu;\nu}
\nonumber \\             
&&+\frac{1}{12}g_{\mu\nu}SS^{;\alpha}_{\phantom{;\alpha};\alpha}                            
-\frac{1}{6}S^2\left(R_{\mu\nu}
-\frac{1}{4}g_{\mu\nu}R^\alpha_{\phantom{\alpha}\alpha}\right)         
-\frac{1}{4}g_{\mu\nu}h S\bar{\psi}\psi~~. 
\label{65}
\end{eqnarray}                                 
Additional use of the matter field equations of motion
then confirms that this energy-momentum tensor is indeed traceless.

In the presence of a spontaneously broken non-zero constant expectation
value $S_0$ for the scalar field, the energy-momentum tensor is then
found to simplify to  
\begin{equation}
T_{\mu \nu} = i \bar{\psi}
\gamma_{\mu}(x)[\partial_{\nu}+\Gamma_\nu(x)]\psi
-\frac{1}{4}g_{\mu\nu}hS_0\bar{\psi}\psi
-\frac{1}{6} S_0^2\left(R_{\mu\nu}-\frac{1}{4}g_{\mu\nu}
R^\alpha_{\phantom{\alpha}\alpha}\right)~~,
\label{66}
\end{equation}                                 
with flat space limit
\begin{equation}
T_{\mu \nu} = i \bar{\psi}
\gamma_{\mu}\partial_{\nu}\psi
-\frac{1}{4}\eta_{\mu\nu}hS_0\bar{\psi}\psi~~.
\label{67}
\end{equation}                                 
With the fermion now obeying 
\begin{equation}
i \gamma^{\mu}\partial_{\mu}\psi - h S_0 \psi = 0~~,
\label{68}
\end{equation}                                 
the tracelessness of the flat space energy-momentum tensor of Eq.
(\ref{67}) is manifest. The quantization of the flat space limit theory
is straightforward since Eq. (\ref{68}) just describes a free fermion
with mass $m=hS_0$, and yields one particle plane wave             
eigenstates $\vert k \rangle$ of four momentum $k^{\mu} =(E_k, \vec{k})$
where $E_k = (k^2 +m^2)^{1/2}$. For a positive energy, spin-up, Dirac
spinor propagating in the $z$-direction, in analog to Eq.
(\ref{59}) we obtain the matrix elements
\begin{equation}
T_{\mu\nu} = \frac{1}{V}\pmatrix{
E_k&0&0&-k\cr
0&0&0&0 \cr
0&0&0&0 \cr
-k&0&0&k^2/E_k \cr}+
\frac{1}{V}\pmatrix{
-m^2/4E_k&0&0&0\cr
0&m^2/4E_k&0&0 \cr
0&0&m^2/4E_k&0 \cr
0&0&0&m^2/4E_k \cr}~~.
\label{69}
\end{equation}                                 
with trace
\begin{equation}
T^{\mu}_{\phantom{\mu}\mu}=\frac{1}{V}\left[-E_k
+\frac{k^2}{E_k}+\frac{m^2}{E_k}\right]=0~~.
\label{70}
\end{equation}                                 
In Eq. (\ref{69}) we recognize a two-component structure to
the fermion energy-momentum tensor, a standard kinematic
piece in which $\langle k \vert T_{00} \vert k\rangle =
E_k/V$, $\langle k \vert T_{33}\vert k\rangle 
=k^2/E_kV$, and a dynamic part coming from the symmetry
breaking in which $\langle k \vert T_{00} \vert k\rangle =
-m^2/4E_k$, $\langle k \vert T_{11} \vert
k\rangle =\langle k \vert T_{22} \vert k\rangle =\langle k \vert T_{33}
\vert k\rangle = m^2/4E_k$. On incoherently averaging over the
directions of $\vec{k}$, the energy momentum tensor is then found to
take the form
\begin{equation}
T_{\mu\nu}=(\rho+p)U_{\mu}U_{\nu}+p\eta_{\mu\nu}+\Lambda
\eta_{\mu\nu}~~,
\label{71}
\end{equation}                                 
where
\begin{equation}
\rho=\frac{6E_k}{V}~~,~~p=\frac{2k^2}{VE_k}~~,
~~\Lambda=\frac{3m^2}{2VE_k}~~,
\label{72}
\end{equation}                                 
and where the tracelessness of $T^{\mu\nu}$ is enforced by the
relation
\begin{equation}
3p-\rho+4\Lambda=0~~.
\label{73}
\end{equation}                                 
As an energy-momentum tensor, Eq. (\ref{71})  provides an explicit
example of a phenomenon we referred to earlier, namely that it is
possible to add on to a kinematic energy-momentum tensor an additional
tensor which is itself covariantly conserved without affecting the
covariant conservation of the kinematic energy-momentum tensor itself.
As well as showing how a cosmological constant type term $\Lambda$ can
naturally arise in dynamical mass theories, the great virtue of using
Eq. (\ref{71}) is that the tracelessness condition of Eq. (\ref{73})
constrains the value of $\Lambda$ to be neither smaller nor larger than
$\rho -3p$, an issue which we shall revisit below when we discuss the
dark energy problem, with the tracelessness condition thus being
seen to give us control of the cosmological constant.

It is instructive to compare this dynamical theory of fermion masses
with a strictly kinematic fermion mass theory in which the action is
given by
\begin{equation}
I_M=-\int d^4x(-g)^{1/2}\left[
i\bar{\psi}\gamma^{\mu}(x)[\partial_\mu+\Gamma_\mu(x)]             
\psi -m\bar{\psi}\psi\right]~~,
\label{74}
\end{equation}                                 
with the flat space fermion wave equation being given by 
\begin{equation}
i \gamma^{\mu}\partial_{\mu}                              
\psi - m \psi = 0~~,
\label{75}
\end{equation}                                 
the flat space energy-momentum tensor being given by
\begin{equation}
T_{\mu \nu} = i \bar{\psi}
\gamma_{\mu}\partial_{\nu}\psi~~,
\label{76}
\end{equation}                                 
and the trace being given by the non-zero
\begin{equation}
T^{\mu}_{\phantom{\mu} \mu}=m\bar{\psi}\psi~~.
\label{77}
\end{equation}                                 
This time, an incoherent averaging over plane wave states yields 
\begin{equation}
T_{\mu\nu}=(\rho+p)U_{\mu}U_{\nu}+p\eta_{\mu\nu}~~,~~
\rho=\frac{6E_k}{V}~~, ~~p=\frac{2k^2}{VE_k}~~,
\label{78}
\end{equation}                                 
with trace $T^{\mu}_{\phantom{\mu} \mu}=3p-\rho \neq 0$. With both Eq.
(\ref{71}) and Eq. (\ref{78}) leading to the covariant conservation of 
the kinematic $(\rho+p)U_{\mu}U_{\nu}+p\eta_{\mu\nu}$, we see that both
energy-momentum tensors lead to eikonalized geodesic motion, with the
validity of the equivalence principle in no way requiring the use of Eq.
(\ref{78}).\footnote{Analogously, if we take the scalar $S(x)$ to be a
constant in Eq. (\ref{12}), Eq. (\ref{12}) then reduces to Eq.
(\ref{5}), with test particle motion being geodesic whether mass is
kinematic or dynamic.} The key distinction between Eqs. (\ref{71}) and
(\ref{78}) is that Eq. (\ref{71}) contains not just the energy on the
fermion field, but also that of the Higgs field that gave it its mass,
an energy which couples to gravity. In dynamical theories of mass
generation, gravity is thus sensitive to the Higgs field associated
with the fermion, while in kinematic case it of course is not. 

\subsection{Implications of elementary particle physics}

While it is conventional to use Eq. (\ref{78}) and its analogs
in current standard model cosmological studies, it is actually Eq.
(\ref{71}) and its analogs which is suggested by elementary particle
physics. Specifically, in the standard $SU(3)\times SU(2)\times U(1)$
theory of strong, electromagnetic and weak interactions, the only mass
scale present in the fundamental Lagrangian is the wrong sign, tachyonic
mass term in the Higgs potential $V(S)=\lambda S^4-\mu^2S^2/2$ with all
fermions and gauge bosons getting their masses dynamically through a
non-vanishing Higgs field expectation value, and all couplings between
fields being dimensionless. As regards the Higgs field, it could be a
fundamental field with a bona fide fundamental tachyonic mass, or it
could be a dynamical manifestation of an underlying symmetry breaking
through bilinear fermion condensates, with $S(x)$ then only being a
Ginzburg-Landau type long range order parameter (and $V(S)$ its
effective Ginzburg-Landau action) which arises when the fermion
bilinear takes a non-zero expectation value in a spontaneously broken
vacuum. In such a case the underlying theory of fermions and gauge
bosons would possess no intrinsic mass scales at all and would then
have a traceless energy-momentum tensor. Beyond these two possibilities
for the Higgs field, inspection of the action of Eq. (\ref{61}) reveals
yet another origin for the wrong sign mass term, namely that it could
even arise from curvature in a theory with no fundamental mass scale at
all, with $R^\mu_{\phantom{\mu}\mu}$ serving as $\mu^2$ in a non-flat
background. Thus either the energy-momentum tensor of $SU(3)\times
SU(2)\times U(1)$ is traceless (dynamical Higgs) or its trace is given
(fundamental Higgs) by the non-zero $T^{\mu}_{\phantom{\mu}
\mu}=\mu^2S_0^2$ (c.f. Eq. (\ref{56})) when $S$ is constant. At best
then the trace can only be related to the Higgs field itself, and would
thus not be given by the $T^{\mu}_{\phantom{\mu} \mu}=3p-\rho$ expected
of the kinematic fluid of Eq. (\ref{78}). The use of the perfect fluids
commonly employed in standard gravity and cosmology would thus appear
to be at variance with the standard $SU(3)\times SU(2)\times U(1)$
model of particle physics.\footnote{In the standard particle physics
model radiative corrections will lead to the generation of a trace
anomaly in theories whose starting Lagrangian has no fundamental mass
scales at all. However, such anomalies while then making the trace of
the renormalized energy-momentum tensor non-zero, do not themselves
introduce any new mass scale (the dimension four trace is
proportional to terms which are quadratic in the curvature tensor) and
would not in any way make the trace equal to the kinematic
$3p-\rho$ required of Eq. (\ref{78}). Moreover, it is possible to
actually cancel the trace anomaly altogether, by a judicious
choice of fields, a judicious choice of geometric background, or by a
renormalization group fixed point at which the coefficient of the trace
anomaly is then zero. That the trace anomaly is to be cancelled would
appear to be part of the standard cosmological wisdom anyway, since no
such anomaly term contribution is ever included alongside the kinematic
ordinary matter, kinematic dark matter and dark energy components which
make up the energy content of the standard cosmological model, and so
we shall not consider the trace anomaly any further here.}
Hence, even before we enter into the issue of the fact that the use of
standard gravity in astrophysics and cosmology leads to the dark matter
and dark energy problems, we see that the very use of such dark matter
and dark energy sources at all in the kinematic way in which they are
commonly used is already not in accord with the standard model of
particle physics. With this word of caution in mind we shall now
discuss the status of standard gravity with its conventional kinematic
perfect fluid sources, and in looking for departures from standard
gravity then allow for theories in which the exact energy-momentum
tensor is traceless. Since the use of a traceless energy-momentum
tensor in the Einstein equations of Eq. (\ref{1}) would lead to a Ricci
scalar which would have to vanish in every conceivable situation (to
rapidly then bring standard gravity into conflict with data), when we do
consider the source of gravity on the right-hand side of the
gravitational equations of motion to be traceless, we will have to
modify the left-hand side as well in a way which would make it
automatically traceless as well. As we shall see, such a departure from
the standard theory will readily allow us to resolve both the dark
matter and the dark energy problems. Having now explored constraints
on the right-hand side of Eq. (\ref{1}) in some detail, we turn next to
an exploration of its left-hand side, and begin first with the
Newtonian gravity which was its antecedent.

\section{Newtonian gravity}

\subsection{The Newtonian potential}

The Newtonian prescription for determining the non-relativistic
potential $\phi ({\vec r})$ at any point $\vec{r}$ due to a set of
static, mass sources, $m_i$, at points $\vec{r_i}$ is to sum over them
according to
\begin{equation}
\phi ({\vec r})= -\sum_i^N \frac{m_iG}{\vert {\vec r} -{\vec {r_i}}
\vert} 
\label{79}
\end{equation}                                 
where $G$ is Newton's constant, with the motions of material test
particles then being determined via
\begin{equation}
\frac{d^2\vec{r}}{dt^2}=-\vec{\nabla}\phi~~.
\label{80}
\end{equation}                                 
As such, Eq. (\ref{79}) contains the full content of Newton's law of
gravity, and for any given set of gravitational sources, any candidate
theory of gravity needs to recover the associated $\phi(\vec{r})$ in
any kinematic region in which Eq. (\ref{79}) has been confirmed, to the
level of precision required by available data. To determine what are
the appropriate kinematic regions requires first testing Newtonian
predictions in a candidate region using sources which are already known
and prescribed in advance of data, and after success has been achieved,
and only after, is one then free to use the law again in that same
region to infer the existence of other, previously unknown, sources. At
the present time one can say that in this way Newton's law of gravity
has been well-established on distance scales from the order of
millimeters (the smallest distance scale on which there has been
testing) out to distances of the order of $10^{15}$ cm or so (viz.
solar system distance scales). At much larger distances all tests which
use only the sources which were known in advance of observation
have been found to fail, with additional (dark matter) sources always
having to be invoked after the fact, with the validity of Newton's
law yet to be confirmed on those distances.\footnote{Testing the
validity of Newton's law on a distance scale such as galactic would
require knowing the dark matter distribution in a galaxy in advance of
measurement of the orbital velocities of the material in it.} The
establishing of a given law in one kinematic region cannot be regarded
as evidence for its validity in others.

As regards solar system tests of Newton's law of gravity, these tests
are spectacular, with the planets being found to move according
to Eq. (\ref{80}) to very high accuracy when the sun and the planets
themselves are taken to be the sources. To a very good approximation
the planets move around the sun with a centripetal acceleration of the
form
\begin{equation}
\frac{v^2}{r}=\frac{GM_{\odot}}{r^2}~~,
\label{81}
\end{equation}                                 
and have Keplerian fall-off of velocity with distance of the form
\begin{equation}
v=\frac{G^{1/2}M_{\odot}^{1/2}}{r^{1/2}}
\label{82}
\end{equation}                                 
where $M_{\odot}$ is the mass of the sun. In fact so reliable was this
law found to be in the solar system when only known visible sources
were used, that when the orbit of Uranus was found to depart slightly
from the Newtonian expectation, a perturbation due to a then
undetected nearby planet was proposed, with the subsequent detection
of Neptune giving dramatic confirmation of Newton's law of gravity
at that distance scale.

While Newton's law had always been thought to hold on all solar and
sub-solar distance scales, recently it has come in to question
on very small laboratory scales of order millimeters or so. While the
meter or so distance region had actually been quite throughly searched
(though to no avail) when it was thought that there might be a
so-called fifth force operative on those distance scales (fifth since
it would be in addition to the strong, electromagnetic, weak and
gravitational forces), recent studies of large extra dimension physics
had led to a reopening of the issue with a focus on the even smaller
millimeter region distance scale. While there has been a longstanding
theoretical interest in the possible existence of additional spacetime
dimensions beyond the four established ones, it had generally been
presupposed that such dimensions would be truly microscopic, possibly
being as small as the $10^{-33}$ cm Planck length. In an attempt to
resolve the longstanding hierarchy problem of understanding why there
was such a huge disparity between the $M_{EW}=10^{3}$ GeV electroweak
and the $M_{PL}=10^{19}$ GeV gravitational mass scales, Arkani-Hamed,
Dimopoulos and Dvali \cite{Arkani-Hamed1998} found a candidate extra
dimension based solution in which rather than be microscopic, the
additional dimensions beyond four would instead need to be of the order
of millimeters. In such a situation gravitational flux lines could then
spread out to such distances in the extra dimensions, and therefore
lead to modifications to standard gravity at the millimeter level.
Following on the work of Arkani-Hamed, Dimopoulos and Dvali, Randall
and Sundrum \cite{Randall1999a,Randall1999b} found an alternate way to
address the hierarchy problem which required the geometry in the extra
dimensions not to be the Minkowski one that had always been assumed in
higher dimensional theories, but rather to be anti-de Sitter. In such
a situation the extra dimensions could then not only be large but even
infinite in size, with the curvature of the anti-de Sitter space acting
as a sort of refractive medium which would sharply inhibit the
penetration of gravitational flux lines into the higher dimensions. In
such theories there could again be departures from Newton's law at the
millimeter level, which thus prompted a renewed search for possible
millimeter region departures. Since for the moment, none has yet
actually been detected \cite{Adelberger2003}, the issue should be
regarded as open, and so we shall assume here the validity of Eq.
(\ref{79}) on all solar and sub-solar distance scales including
millimeter ones and below, with the deriving of the Newtonian
phenomenology on such scales thus being set as a requisite for any
candidate gravitational theory.

\subsection{The second order and fourth order Poisson equations}

While Eq. (\ref{79}) contains the full content of Newton's law of
gravity, in order to actually perform the needed sum over sources, it
very convenient to recast Newton's law in the form of a second
order Poisson equation. Thus if we set 
\begin{equation}
\nabla^2\phi(\vec{r})=g(\vec{r})~~,
\label{83}
\end{equation}                                 
we can write the potential as 
\begin{equation}
\phi(\vec{r})= -\frac{1}{4\pi}\int
d^3\vec{r^{\prime}}\frac{g(\vec{r^{\prime}})}{\vert
\vec{ r} -\vec {r^{\prime}}\vert}~~,
\label{84}
\end{equation}                                 
with the potential exterior and interior to a spherically symmetric
static source of radius $R$ then being given by
\begin{equation}
\phi(r>R)= -\frac{1}{r}\int_0^R
dr^{\prime}r^{\prime 2}g(r^{\prime})~~,~~
\phi(r<R)= -\frac{1}{r}\int_0^r
dr^{\prime}r^{\prime 2}g(r^{\prime})-\int_r^R
dr^{\prime}r^{\prime }g(r^{\prime})~~.
\label{85}
\end{equation}                                 
However, Eq. (\ref{83}) is not the only Poisson type equation which will
lead to a $1/r$ potential. Thus if we for instance consider the fourth
order
\begin{equation}
\nabla^4\phi(\vec{r})=h(\vec{r})~~,
\label{86}
\end{equation}                                 
we can write the potential as 
\begin{equation}
\phi(\vec{r})= -\frac{1}{8\pi}\int
d^3\vec{r^{\prime}}h(\vec{r^{\prime}})\vert
\vec{ r} -\vec {r^{\prime}}\vert~~,
\label{87}
\end{equation}                                 
with the potential exterior and interior to a spherically symmetric
static source of radius $R$ then being given by
\begin{eqnarray}
\phi(r>R)= -\frac{r}{2}\int_0^R
dr^{\prime}r^{\prime 2}h(r^{\prime})
-\frac{1}{6r}\int_0^R
dr^{\prime}r^{\prime 4}h(r^{\prime})~~,
\nonumber \\
\phi(r<R)= -\frac{r}{2}\int_0^r
dr^{\prime}r^{\prime 2}h(r^{\prime})
-\frac{1}{6r}\int_0^r
dr^{\prime}r^{\prime 4}h(r^{\prime})
-\frac{1}{2}\int_r^R
dr^{\prime}r^{\prime 3}h(r^{\prime})
-\frac{r^2}{6}\int_r^R
dr^{\prime}r^{\prime }h(r^{\prime})~~.
\label{88}
\end{eqnarray}                                 
Moreover, recovering a $1/r$ potential is not even restricted to the
above choices of Poisson equation, since the sixth order
$\nabla^6\phi(\vec{r})=k(\vec{r})$ would yield an exterior potential of
the generic form $\phi \sim 1/r+r+r^3$, with the pattern repeating for
all higher even number of derivative Poisson equations. Characteristic
of all of these possible Poisson equations is the fact that we recover
a $1/r$ term for each and every one of them, and in all of them the
terms which depart from a pure $1/r$ form are all only important at
larger distances (viz. just the distances on which the dark matter
problem is encountered), with there being no modifications to the $1/r$
law at very small distances. Providing that the $r$, $r^3$ and so on
type terms are all negligible on solar system distance scales, all of
these Poisson equations will thus reproduce the standard Newtonian
solar system phenomenology. 

Now at first glance the expressions for the coefficients of the $1/r$
terms in Eqs. (\ref{85}) and (\ref{88}) appear to differ since they are
given as different moments of the source. However, this is not of
concern since measurements of $\phi(r)$ in the $r>R$ region where there
are data can never uncover any information regarding the behavior of
the integrand in Eq. (\ref{85}) in the $r<R$ region. What is required of
the $1/r$ term in the $r>R$ region is only that it conform with the
large $r$ behavior required by the original Newton law of Eq.
(\ref{79}) (and thus the all $r>R$ behavior since $1/r$ is the unique
behavior of the Newton term in the entire exterior region), viz. that
for a set of sources all with the same mass $m$ the potential behave as
\begin{equation}
\phi(r >R)= -\frac{NmG}{r}~~,
\label{89}
\end{equation}                                 
i.e. that the coefficient of the $1/r$ term be linear in the number of
sources $N$. To achieve such a linearity in the second order Poisson
equation case one can choose to set 
\begin{equation}
g(r<R)=mG\sum_{i=1}^N\frac{\delta(r-r_i)}{r^2}
\label{90}
\end{equation}                                 
in the $r<R$ region, though as we had just noted, this is not mandated
by Eq. (\ref{79}). (While use of this particular $g(r<R)$  is sufficient
to give Eq. (\ref{89}), it is not necessary.) What is necessary is only
that there be $N$ discrete sources, with Eq. (\ref{79}) only counting
the number of them in the $r>R$ region. To achieve exactly the same
result through the use of the fourth order Eq. (\ref{88}), it is
convenient, though not mandatory, to consider as source\footnote{We
choose the particular $[\nabla^2 -(r^2/12)\nabla^4](\delta(r-r_i)/r^2)$
source rather than the more straightforward
$\nabla^2(\delta(r-r_i)/r^2)$ one, since unlike the latter, the former
one happens to be positive definite. An additional virtue of this
particular choice of source is that it only couples to the fourth moment
integral and not to the second moment one; and with the
$\delta(r-r_i)/r^2$ source only coupling to the second moment
integral and not to the fourth moment one, the logical independence of
the two moments is thus established.}
\begin{equation}
h(r<R)=-\gamma c^2\sum_{i=1}^N\frac{\delta(r-r_i)}{r^2}
-\frac{3\beta c^2}{2}\sum_{i=1}^N\left[\nabla^2
-\frac{r^2}{12}\nabla^4\right]\left[\frac{\delta(r-r_i)}{r^2}\right]~~,
\label{91}
\end{equation}                                 
with its insertion in Eq. (\ref{88}) yielding 
\begin{equation}
\phi(r >R)= -\frac{N\beta c^2}{r}+\frac{N\gamma c^2r}{2}~~.
\label{92}
\end{equation}                                 
As we see, the coefficient of the $1/r$ term is again linear in the
number of sources, with Newton's law of gravity only counting the
number of discrete sources that are present in the $r<R$ region.
Comparing Eqs. (\ref{89}) and (\ref{92}), we see that we can once and
for all define $\beta =mG/c^2$ at the individual discrete source level
(a level which could be microscopic), without ever needing to decompose
$\beta c^2$ into separate $m$ and $G$ pieces, with only the
Schwarzschild radius of the source (viz. $2mG/c^2$ in the second order
case and
$2\beta$ in the fourth order) ever being measurable
gravitationally.\footnote{Even if the microscopic $\beta c^2$ were not
to be equal to the product of the conventionally defined $m$ and $G$
(not that we know what the gravitational coupling of a microscopic
source is), the use of the coefficient $N\beta c^2$ would then lead to a
estimate for the number of atoms in a macroscopic source such as the
sun which would be different from the conventional one, a point which
is not of concern since without appealing to the $mG$ coefficient, we
do not know how many atoms there anyway are in the sun, with the
standard estimate of the number being based on the a priori use of the
$mG$ coefficient.} 

It is important to stress that it is the presence of two independent
singularities in the source of Eq. (\ref{91}) which leads to the
logical independence of the $\beta$ and $\gamma$ coefficients, since if
we could approximate the source $h(r)$ by a constant source $h(r)=h$,
we would instead have to conclude that the coefficients of the two
potential terms would be related according to
\begin{equation}
\phi(r >R)= -\frac{NhR^5}{30r}-\frac{N hR^3r}{6}~~,
\label{93}
\end{equation}                                 
and thus have a ratio of order the radius squared of the source. This
is to be contrasted with the choice of a constant source $g(r)=g$ in the
second order case where the potential is then given by
\begin{equation}
\phi(r >R)= -\frac{NgR^3}{3r}~~.
\label{94}
\end{equation}                                 
Since a comparison of Eqs. (\ref{89}) and (\ref{94}) would entail that
one can make the identification $mG=gR^3/3$, a view of gravitational
sources has developed that macroscopic gravitational sources can be
treated as being continuous rather as being a collection of
independent discrete microscopic sources. However, apart from not at
all being mandated by the successful use of Eq. (\ref{89}), the issue
is in fact not even addressable if one only makes measurements in the
exterior $r>R$ region. However, in the fourth order Poisson case the
issue does become relevant, because here one does measure more than
just one moment of the source, with the higher derivative theory thus
being able to probe deeper into the source. Thus within the fourth order
theory, an experimental determination of the ratio of the coefficients
of the $1/r$ and $r$ terms which would indicate that the ratio of the
two coefficients is very far from being of order the square of the
radius of a given macroscopic source, would indicate that the
macroscopic source would have to be composed of microscopic components
which are discrete. Quite remarkably then, higher order derivative
theories open up the possibility of establishing discreteness at the
microscopic level from macroscopic measurements alone.\footnote{As a
historical aside, we note that since the use of higher order gravity
can thus provide for a macroscopic manifestation of the existence of
atoms, it thus meets an objective which was first sought by Boltzmann,
one which was part of his motivation for developing kinetic theory (and
which was subsequently achieved by a determination of a finite
value for Avogadro's number by macroscopic means).} Moreover, below we
shall actually apply the potential of Eq. (\ref{92}) to galactic
rotation curves, to find that the $\beta^{*}$ and
$\gamma^{*}$ coefficients of the fourth order theory stellar potential 
\begin{equation}
\phi^*(r >R)= -\frac{\beta^*c^2}{r}+\frac{\gamma^* c^2r}{2}
\label{95}
\end{equation}                                 
actually are very far from possessing a ratio $\beta^*/\gamma^*$ which
is anything like the square of a typical stellar radius (numerically we
find $(\beta^*/\gamma^*)^{1/2}\sim 10^{23}$ cm, with the
$\gamma^*c^2r/2$ term first becoming competitive with the
$\beta^*c^2/r$ term on galactic distance scales). Within the framework
of a gravity which is based on fourth order equations then we
can conclude that matter must be discrete at the microscopic level.

While the second order Poisson equation of Eq. (\ref{83}) does
yield the Newtonian potential of Eq. (\ref{89}) as solution, there is
actually a critical difference between the use of the second order
Newton potential and Poisson equation on the one hand and the use of
their fourth order counterparts given as the potential of Eq.
(\ref{92}) and the Poisson equation of Eq. (\ref{86}) on the other.
Specifically, while the fourth order potential of Eq. (\ref{92})
does nicely reduce to the second order potential of Eq. (\ref{89}) in
the small $r$ limit, the fourth order Poisson equation itself never
reduces to the second order one in any limit at all. It is thus
possible for solutions to two different equations to approximate each
other very closely in some kinematic region even while the equations to
which they are solutions never approximate each other at all. The
equations which describe any candidate alternate theory of gravity thus
do not at all need to reduce to those which describe the standard
theory. Rather, it is only the solutions to the alternate theory which
have to recover the solutions to the standard theory in any of the
kinematic regions in which the standard theory solutions have
themselves been tested.\footnote{If the alternate equations of motion
did reduce to the standard ones in some limit, then of course so would
their solutions. However, the converse is not true, since the alternate
theory solutions are still able to reduce to the standard theory
solutions even if there is no such parallel for the equations of motion
themselves.} Outside of such regions the alternate theory can then
provide for departures from the standard theory expectations, something
we will capitalize on below to explain galactic rotation curve
systematics without the need to invoke dark matter. The essential
difference between the second order Poisson equation and the fourth
order one, is that once the second order Poisson equation is specified
as being the fundamental equation of motion, there is no possibility to
ever get any departure from Newton's law on any distance scale at all,
something which is allowed for in a higher order theory. Since the
Einstein equations of Eq. (\ref{1}) were explicitly constructed so that
their non-relativistic weak gravity limit would be the second order
Poisson equation, the Einstein equations thus lock non-relativistic
weak gravity into being Newtonian on all distance scales, and to thus
be so even on distances well beyond the solar system ones where
Newton's law of gravity was originally established. In this sense then
the dark matter problem could originate in the lack of reliability of
the extrapolation of standard solar system wisdom to altogether larger
distance scales.

\subsection{Universal acceleration scale and the MOND theory}

While the above discussion has focused on the standard Newtonian picture
and those allowed departures from it which are associated with a
critical distance scale, an alternate form of allowed departure has been
suggested by Milgrom \cite{Milgrom1983a,Milgrom1983b,Milgrom1983c}, one
which has proven to be remarkably instructive and fruitful.
Specifically, on noting that the centripetal accelerations of particles
in orbits around galaxies were typically much smaller than those
associated with the orbits of planets around the sun or satellites
around the earth, Milgrom suggested that the determining factor was not
in fact a distance scale at all but rather an acceleration one, with
departures from the standard wisdom occurring whenever accelerations of
particles dropped below some critical universal acceleration scale
$a_0$. Quite remarkably, Milgrom found that it is actually possible to
choose such an $a_0$ so that its effects would be of significance for
galaxies and yet leave solar system phenomenology
untouched.\footnote{Phenomenologically $a_0$ was found to be of order
$10^{-8}$ cm/sec$^2$, to be compared with an acceleration at the outer
limits of the solar system of order $10^{-3}$ cm/sec$^2$ due to the
pull of the sun.} In contrast to proposals which seek to modify the
right-hand side (viz. the gravitational side) of Eq. (\ref{80}),
Milgrom instead proposed to modify the left-hand side (viz. the
inertial side) with the proposal thus being known as MOND (modified
Newtonian dynamics). Specifically Milgrom replaced Eq. (\ref{80}) by 
\begin{equation}
 \mu\left(\frac{a}{a_0}\right)\vec{a}=\vec{f}
\label{96}
\end{equation}                                 
where $\vec{a}=d^2\vec{r}/dt^2$ is the ordinary acceleration,
$\vec{f}=-\vec{\nabla}\phi$ is the gravitational force and $\mu(a/a_0)$
is the modification. In order for this modification to reduce to the
standard Newton law when $a \gg a_0$, the function $\mu(x)$ has to
behave as $\mu(x \gg 1)\rightarrow 1$. Guided by the fact that many
prominent spiral galaxies possess rotational velocity curves which are
flat (i.e. the rotational velocity $v$ is independent of the
distance $R$ from the center of the galaxy), and that the velocities of
these particular galaxies obey the empirical Tully-Fisher law which
relates $v$ to the total luminosity $L$ of the galaxy as $v^4 \sim L$,
Milgrom further suggested that at very small $x$ the function $\mu(x)$
should behave as $\mu(x\ll 1)\rightarrow x$, since the centripetal
accelerations 
$a=v^2/R \ll a_0$ of orbits in a galaxy of mass $M$ would then obey
\begin{equation}
\frac{a^2}{a_0}=\frac{v^4}{a_0R^2}=\frac{MG}{R^2}~~,
\label{97}
\end{equation}                                 
and thus yield the $R$ independent $v^4=a_0MG$. Then, if the mass
involved was just the luminous mass of the galaxy, it would be related
to the luminosity of the galaxy by $M=(M/L)L$ where $M/L$ is the mass
to light ratio of the galaxy, with both the flatness of galactic
rotation curves and the Tully-Fisher relation thereby being obtained
from luminous matter alone. 

As a proposal, the relation of Eq. (\ref{96}) is perhaps a little bit
too strong a departure from the standard picture, since as such it would
modify the inertial side of Newton's second law of motion not just for
gravitational processes but for non-gravitational ones as well,
processes for which there are no known problems. Additionally, since
the proposal is a modification of the inertial side of Newton's second
law of motion, in an accelerating coordinate system Eq. (\ref{96})
would not readily recover Eqs. (\ref{4}) and (\ref{5}). To get a
sense of what might instead happen, one possible way to effect a
covariant generalization of Eq. (\ref{96}) is to replace the
acceleration $\vec{a}$ by the contravariant $D^2x^{\lambda}/D\tau^2$ of
Eqs. (\ref{4}) and (\ref{5}), to introduce a four-vector $V^{\lambda}$
which reduces to $V^{\lambda}=(1,0,0,0)$ in a non-accelerating
coordinate system, and to take $\mu(x)$ to be a general coordinate
scalar function of
\begin{equation}
x=\frac{V_{\lambda}}{a_0}\frac{D^2x^{\lambda}}{D\tau^2}
\label{98}
\end{equation}                                 
where the parameter $a_0$ is then a general coordinate scalar with the
dimensions of acceleration. While the presence of such a $V_{\lambda}$
would lead to frame-dependent effects, such departures from standard
wisdom need not necessarily be in conflict with observation if $\mu(x)$
is sensibly close to one in the kinematic regime where tests of the
equivalence principle and preferred frame effects have so far been made.

Nonetheless, since it would be more economical not to have to make
so radical a conceptual departure from standard wisdom, an
alternative procedure for introducing a universal acceleration
$a_0$ would be to have it instead appear on the gravitational
side of Eq. (\ref{80}), with Eq. (\ref{80}) then being replaced by  
\begin{equation}
\vec{a}=\nu\left(\frac{f}{a_0}\right)\vec{f}
\label{99}
\end{equation}                                 
instead. To get a sense of what the function $\nu(y)$ might look like,
we consider a particularly simple choice for $\mu(x)$ which meets
its required large and small $x$ limits, viz.
\begin{equation}
\mu(x)= \frac{x}{(1+x^2)^{1/2}}~~,
\label{100}
\end{equation}                                 
to then find that $\nu(y)$ and $\vec{a}$ are given by
\begin{equation}
\nu(y)= \left(\frac{1}{2}+\frac{(y^2+4)^{1/2}}{2y}\right)^{1/2}
\label{101}
\end{equation}                                 
\begin{equation}
\vec{a}=\left(\frac{1}{2}+\frac{(f^2+4a_0^2)^{1/2}}{2f}
\right)^{1/2}\vec{f}~~.
\label{102}
\end{equation}                                 
With the two relevant limits of the function $\nu(y)$ being 
$\nu(y \gg 1) \rightarrow 1$, $\nu(y \ll 1) \rightarrow 1/y^{1/2}$, we
thus recover Eq. (\ref{80}) when $f \gg a_0$, and come right back to Eq.
(\ref{97}) when $f \ll a_0$, with Eqs. (\ref{96}) and (\ref{99})
leading to the same gravitational phenomenology. Having now presented
an analysis of Newtonian gravity and some possible allowed departures
from it, we turn next to a discussion of Einstein gravity.

\section{Einstein gravity}

\subsection{Einstein gravity for static sources and the Newtonian limit}

In his construction of a covariant theory of gravity, even though
Einstein used a fundamental principle, viz. the equivalence principle,
to identify the metric $g_{\mu\nu}$ as the gravitational field, his
choice for the dynamical equation of motion which was to then fix
the metric, viz. Eq. (\ref{1}), appears to have been based on a
phenomenological consideration, namely that the theory recover the
standard second order Poisson equation in the non-relativistic, weak
gravity limit. As such, the second order Einstein equations can be
viewed as being a covariant generalization of the second order Poisson
equation of Eq. (\ref{83}), replacing it by ten equations (reducible to
six by gauge invariance) of which only the $(00)$ component equation
would be of order one in the standard non-relativistic, weak gravity
limit in which the components of $T_{\mu\nu}$ other than $T_{00}$ are
taken to be negligible compared to $T_{00}$ itself. The great virtue of
the Einstein equations is that in prescribing a covariant
generalization of the second order Poisson equation, it not only
incorporated Newtonian gravity, but also  provided it with general
relativistic corrections which could then be tested, leading to the
three classic tests of general relativity, the gravitational red shift,
gravitational bending of light, and the precession of planetary orbits.
With the Einstein equations not only satisfying these three tests, but
doing so in truly spectacular fashion, by and large the consensus in
the community is that the issue is therefore considered settled, with
the Einstein equations being the true equations of gravity which no
longer need to be open to question.

However, it is important to note that while Einstein's original
construction was motivated by the second order Poisson equation, the
application of Eq. (\ref{1}) to the three classic solar system tests is
not actually sensitive to this particular aspect of the Einstein
equations at all. Specifically, what is needed for the three tests is
only knowledge of the metric exterior to a static, spherically
symmetric source, viz. the Schwarzschild metric as given earlier as
Eqs. (\ref{41}) and (\ref{50}), a metric which obeys $R_{\mu\nu}=0$ in
the source-free region. As such, the exterior Schwarzschild solution is
parameterized by a parameter $MG$ associated with a given source, and
while this parameter can be determined from a matching of the exterior
solution of Eq. (\ref{50}) to the interior solution of Eq. (\ref{47})
to yield Eq. (\ref{51}), no use of Eq. (\ref{51}) is actually made in the
analysis of the classic tests. While the $R_{\mu\nu}=0$ condition does
indeed follow from Eq. (\ref{1}) in the source-free
$T_{\mu\nu}=0$ region, Eq. (\ref{1}) does not follow from this fact, i.e.
knowing that $R_{\mu\nu}$ vanishes in the source-free region is not
sufficient to determine what $R_{\mu\nu}$ is to be equal to in regions
where it does not vanish. To emphasize the point, we note that while
functional variation of the second order Einstein-Hilbert action 
\begin{equation}
I_{EH}=-\frac{c^3}{16\pi G}\int
d^4x(-g)^{1/2}R^{\alpha}_{\phantom{\alpha}\alpha}
\label{103}
\end{equation}                                 
with respect to the metric yields the Einstein tensor according to
\begin{equation}
\frac{16\pi G}{c^3(-g)^{1/2}}\frac{\delta I_{EH}}{ \delta g_{\mu
\nu}}=G^{\mu \nu}=R^{\mu\nu}-
\frac{1}{2}g^{\mu\nu}R^{\alpha}_{\phantom{\alpha}\alpha}~~,
\label{104}
\end{equation}                                 
variation of the fourth order general coordinate scalar actions
\begin{equation}
I_{W_1}=\int d^4x(-g)^{1/2}[R^{\alpha}_{\phantom{\alpha}\alpha}]^2
\label{105}
\end{equation}                                 
and
\begin{equation}
I_{W_2}=\int d^4x(-g)^{1/2}R_{\alpha\beta}R^{\alpha\beta}
\label{106}
\end{equation}                                 
respectively yield \cite{DeWitt1964}
\begin{equation}
\frac{1}{(-g)^{1/2}}\frac{\delta I_{W_1}}{ \delta g_{\mu
\nu}}=W^{\mu \nu}_{(1)}= 
2g^{\mu\nu}(R^{\alpha}_{\phantom{\alpha}\alpha})          
^{;\beta}_{\phantom{;\beta};\beta}                                              
-2(R^{\alpha}_{\phantom{\alpha}\alpha})^{;\mu;\nu}                           
-2 R^{\alpha}_{\phantom{\alpha}\alpha}
R^{\mu\nu}                              
+\frac{1}{2}g^{\mu\nu}(R^{\alpha}_{\phantom{\alpha}\alpha})^2~~,
\label{107}
\end{equation}                                 
and 
\begin{equation}
\frac{1}{(-g)^{1/2}}\frac{\delta I_{W_2}}{ \delta g_{\mu
\nu}}=W^{\mu \nu}_{(2)}=
\frac{1}{2}g^{\mu\nu}(R^{\alpha}_{\phantom{\alpha}\alpha})   
^{;\beta}_{\phantom{;\beta};\beta}+
R^{\mu\nu;\beta}_{\phantom{\mu\nu;\beta};\beta}                     
 -R^{\mu\beta;\nu}_{\phantom{\mu\beta;\nu};\beta}                        
-R^{\nu \beta;\mu}_{\phantom{\nu \beta;\mu};\beta}                          
 - 2R^{\mu\beta}R^{\nu}_{\phantom{\nu}\beta}                                    
+\frac{1}{2}g^{\mu\nu}R_{\alpha\beta}R^{\alpha\beta}~~.
\label{108}
\end{equation}                                 
Then, since the Schwarzschild solution is one in which the Ricci
tensor vanishes, in the Schwarzschild solution it follows that covariant
derivatives of the Ricci tensor vanish as well, with both $W^{\mu
\nu}_{(1)}$ and $W^{\mu \nu}_{(2)}$ thus vanishing when $R_{\mu\nu}=0$.
Moreover, this same argument immediately generalizes to general
coordinate scalar actions based on even higher powers of the Ricci
scalar or Ricci tensor. The exterior Schwarzschild solution is thus an
exterior solution to any pure metric theory of gravity which uses any 
Ricci tensor based higher order action in place of the Einstein-Hilbert
action,\footnote{This would not necessarily be true for any higher order
Riemann tensor based action since the Schwarzschild metric is not
Riemann flat, only Ricci flat.} with all of them thus satisfying the
three classic tests.\footnote{While the vanishing of the Ricci tensor
implies the vanishing of its derivatives as well, the vanishing of the
derivatives can be achieved without the Ricci tensor itself needing to
vanish, with the vanishing of tensors such as $W^{\mu\nu}_{(1)}$ or
$W^{\mu\nu}_{(2)}$ thus being achievable by additional,
non-Schwarzschild type solutions as well, a point we shall return to
below.} Thus, in complete parallel to our earlier discussion of the
difference between the second and higher order Poisson equations, we
see that even though there is no limit in which any linear combination
of $W^{\mu\nu}_{(1)}$ and $W^{\mu \nu}_{(2)}$ reduces to the Einstein
tensor of Eq. (\ref{104}), nonetheless the use of any such combination
as a gravitational tensor which could replace $G_{\mu\nu}$ in Eq.
(\ref{1}) would still lead to the Schwarzschild solution. The
Schwarzschild solution thus bears the same relation to the
Einstein-Hilbert action and its higher order generalizations as the
Newton potential does to the Poisson equation and its higher order
generalizations, with use of the Einstein equations only being
sufficient to secure the three classic tests but not at all necessary. 
Hence, as had first been noted by Eddington as long ago as only shortly
after he himself led an expedition in 1919 to confirm that light did
indeed bend as it passed by the sun, the success of the three classic
tests does not secure the validity of the Einstein equations, with
other options being possible. Indeed, at that time Eddington issued a
challenge to the community to tell him which gravitational action then
was the correct one to use, a challenge that the community has never
answered even though it goes directly to the heart of gravitational
theory.

Beyond the three classic tests of general relativity, the only other
test on solar system sized distance scales is the decay of the orbit of
a binary pulsar, an effect which again only involves an exterior metric,
this one due to the presence of two stellar sources rather than just the
one involved in the solar system tests. The decay of the orbit is due to
gravitational radiation reaction, with each star in the binary responding
to retarded gravitational signals emitted by the other and not to
instantaneous ones. Such retardation effects will exist in any
covariant metric theory of gravity since in all of them gravitational
information cannot be communicated faster than the speed of light, and
in all of them there will thus be some decay of the binary pulsar orbit.
While there thus will be orbit decay not only in Einstein-Hilbert based
gravity but also in its higher order alternatives, nonetheless all such
alternatives face the (calculationally daunting) challenge of actually
obtaining the specific amount of decay which has actually been observed,
with calculations having so far only been carried through (and with 
stunning success) is the second order Einstein theory itself.

\subsection{The Einstein static universe and the cosmological constant}

The one situation in which the full content of the Einstein equations
is felt is in cosmology, since in that case both sides of Eq. (\ref{1})
are non-zero, with cosmological observations being made inside the
universe and thus in the region where $T_{\mu\nu}$ is explicitly
non-zero. In order to prepare for the discussion of the dark
energy problem and the accelerating universe to be given below, it is
useful to recall Einstein's effort to produce a static universe
solution to the Einstein equations through the introduction of a
cosmological constant. In order to see the structure of the Einstein
model it is instructive to start not with a static model but rather
with the general time dependent Robertson-Walker geometry
of Eq. (\ref{36}) and the generic source of Eq. (\ref{37}) with its two
separate time-dependent functions $C(t)$ and $D(t)$. For such a set-up
the Einstein equations of Eq. (\ref{1}) yield
\begin{equation}
3c\ddot{R}=-4\pi G(C+3D)R
\label{109}
\end{equation} 
and 
\begin{equation}
\dot{R}^2+kc^2=\frac{8 \pi G}{3c}CR^2~~.
\label{110}
\end{equation} 
A solution in which $C$, $D$ and $R$ are all constant
is then obtained when 
\begin{equation}
\frac{D}{C}=-\frac{1}{3}~~,~~k=\frac{8\pi G}{ 3c^3}CR^2~~. 
\label{111}
\end{equation} 
On taking $C$ to be positive, we see
that $k$ would need to be positive too (viz. topologically closed
3-space), while $D$ would have to be negative, with the Einstein static
universe thus being supportable by a source with $D/C=-1/3$. Fluids in
which $D/C$ is negative are known as quintessence fluids and have
recently come into prominence \cite{Caldwell1998} through the discovery
of the accelerating universe, with a quintessence fluid with $C+3D
<0$ then leading to a net cosmic acceleration ($\ddot{R}>0$) in Eq.
(\ref{109}), though its role in the Einstein model is to put the
acceleration precisely at the $\ddot{R}=0$ borderline where it it is
neither positive nor negative. In his actual construction, Einstein
himself did not use the language of quintessence fluids. Rather,
Einstein used an energy-momentum tensor containing just an ordinary
non-relativistic matter fluid with a density $\rho_m$ and zero pressure
$p_m$, and actually modified Eq. (\ref{1}) by adding a cosmological
constant term to its left-hand side according to 
\begin{equation}
-\frac{c^3}{8\pi G}\left(R^{\mu\nu}-
\frac{1}{2}g^{\mu\nu}R^{\alpha}_{\phantom{\alpha}\alpha}
\right)+\Lambda g^{\mu\nu}=T^{\mu\nu}~~.
\label{112}
\end{equation}
With the cosmological constant term itself having the form of a
perfect Robertson-Walker geometry fluid with energy density
$\rho_{\Lambda}=c\Lambda>0$ and pressure $p_{\Lambda}=-c\Lambda<0$ (so
that the cosmological constant term itself can be thought of as being a
quintessence fluid with
$p_{\Lambda}=-\rho_{\Lambda}$ and
$\rho_{\Lambda}+3p_{\Lambda}<0$), we can relate $\rho_m$ and $\Lambda$
to $C$ and
$D$ according to
$C=\rho_m/c+\Lambda$, $D=-\Lambda$.
The requirement that $C$ and $D$ obey $C+3D=0$ then requires
that $\rho_m$ and $\Lambda$ be related according to the very specific
fine-tuned relation 
\begin{equation}
2c\Lambda=\rho_m~~.
\label{113}
\end{equation} 
Since on its own the ordinary $\rho_m$ fluid would lead to $\ddot{R}<0$
in Eq. (\ref{109}) while on its own a positive cosmological constant would
lead to $\ddot{R}>0$, as we thus see, to get the Einstein model to be
static we need to fine-tune the attraction generated by the energy
density $\rho_m$ and the repulsion generated by the cosmological
constant $\Lambda$ so that they precisely cancel each other.
Interestingly, we shall encounter a reflection of such a fine-tuning
below in our discussion of the dark energy problem.

With Slipher and Hubble's discovery of the expansion of the universe
Einstein realized that there was no further need for a static
cosmology,\footnote{Even in his  model, any departure from the
$2c\Lambda=\rho_m$ fine-tuning condition would lead to a non-static
cosmology.} and so he abandoned the cosmological constant. However,
with the recent discovery of the acceleration of the universe (i.e. not
just an expanding $\dot{R}>0$, but also an accelerating $\ddot{R}>0$),
the cosmological constant has come back into prominence, not just as a
fundamental term of the form inserted on the left-hand side of Eq.
(\ref{112}) but also as a dynamically generated spontaneous symmetry
breakdown contribution to the energy-momentum tensor on the right-hand
side of Eq. (\ref{112}), with current cosmology requiring these two
types of cosmological constant to cancel each other to
unbelievable precision.

From a theoretical perspective, the very fact that one is able to
introduce the $\Lambda g_{\mu\nu}$ term on the left-hand side of Eq.
(\ref{112}) at all is a reflection of the fact that Einstein's very
choice of the original Eq. (\ref{1}) did not
derive from a fundamental principle but only from the phenomenological
need to recover Newton's law.\footnote{Phenomenological that is unless
there is some as yet unknown principle which would actually require the
validity of Newton's law on all distance scales, since for the moment
Newton's law itself is only phenomenological.} Indeed, it was
phenomenology which lead him first to introduce and then subsequently
to remove the cosmological constant term. It is this lack of a
principle which renders the choice of gravitational action non-unique,
with there then being no principle with which to control the
contribution of the cosmological constant. In addition, the use of Eq.
(\ref{1}) dictates that the low energy limit of the theory is the
second order Poisson equation not just on solar system distances but on
all distances. And so we turn now to a study of the motions of
particles on larger distance scales to see just how well the second
order Poisson equation and its Newtonian potential solution then fare.

\section{The dark matter problem}

\subsection{The galactic rotation curve problem}

It is really quite remarkable how early in the study of galaxies and
then clusters of galaxies that mass discrepancies started to appear.
Specifically, not all that long after Shapley's work on the
shape and size of the Milky Way galaxy and Hubble's demonstration that
the spiral nebulae were distant galaxies in their own right, Oort's
measurements \cite{Oort1932} of the velocities of stars in our local
solar neighborhood and Zwicky's \cite{Zwicky1933} and then Smith's
\cite{Smith1936} measurements of the velocity dispersions of galaxies
in clusters of galaxies all pointed to discrepancies. Since the
discrepancy that Zwicky detected was based on the use of the Newtonian
gravity virial theorem, Zwicky was able to conclude that unless the
virial theorem itself did not hold for clusters (either because Newton's
law did not hold on those distances or because the cluster had not yet
virialized), there then had to be more mass in the cluster than he
could detect. As such, it was thought at the time that any such missing
mass would be in the form of astrophysical objects which were too faint
to be detectable, with the notion that the missing matter might be of an
entirely different form that could never be detected optically at all
coming only altogether later when it was realized that modern elementary
particle physics could actually provide such so called Wimp (weakly
interacting massive particle) objects. 

While there continued to be further indications of mass discrepancies
over time (for a history see e.g. \cite{Trimble1995}), the situation
started to become critical following the HI 21 cm radio studies of the
spiral galaxies NGC 300 and M33 by Freeman \cite{Freeman1970} and of M31
by Roberts and Whitehurst  \cite{Roberts1975} which  found no sign of the
expected Keplerian fall-off of rotational velocity $v$ with distance $R$
from the center of the galaxy. With hydrogen gas being distributed in
spiral galaxies out to much further distances than the stars themselves,
the HI studies were thus the ideal way to probe the outer reaches of
spiral galaxies, and thus the ideal probe to test for the Keplerian
fall-off expected of the rotation velocities of particles which were
located far beyond the visible stars (viz. the visible mass) of the
galaxies. Moreover, the measurement of such rotation velocities would be
unhindered by projection effects if studies were made of spirals that
were close to being edge on along our line of sight. 

The continuing persistence of the cluster problem and concern in
general about mass discrepancies prompted detailed exploration of
outer region galactic rotation curves, with early systematic surveys
being made by Bosma \cite{Bosma1981} and Begeman \cite{Begeman1989} as
part of the University of Groningen Westerbork Synthesis Radio Telescope
survey. What was found in these and in many related studies was that
the outer region rotation velocities were systematically larger than
the Keplerian expectation, with none giving any indication of a fall-off
at all. In Fig. 1 we display the rotational velocity curves of a
sample of eleven spiral galaxies which have been identified by Begeman,
Broeils and Sanders \cite{Begeman1991} as being particularly reliable and
characteristic of the general pattern of behavior of galactic rotation
curves which has been found, with their paper giving complete data
references. With the stellar surface matter density of each spiral galaxy
in the sample behaving as
\begin{eqnarray}
\Sigma(R)&=&\Sigma_0e^{-R/R_0}~~,
\nonumber \\
~~N^{*}&=&2\pi\int_0^{\infty} dR
R\Sigma(R)=2\pi\Sigma_0R_0^2~~,
\label{114}
\end{eqnarray} 
where $R_0$ is the scale length and $N^{*}$ is the total number of
stars,\footnote{We assume that light traces mass, so that the surface
matter density is proportional to the measured luminous surface brightness
with a mass to light ratio which is independent of $R$.} in Fig. 1 we
have plotted the rotation velocities (in units of km/sec) in each given
galaxy as a function of $R$ as measured in units of that particular
galaxy's own scale length, with the rotation curves being displayed in the
figure in order of increasing luminosity from the lowest (DDO 154) to
the highest (NGC 2841) in the sample. In Table 1 we provide some
characteristics of the galaxies, and note that the luminosities of the
galaxies in the selected sample range over a factor of order a
thousand.

\begin{figure}
\epsfig{file=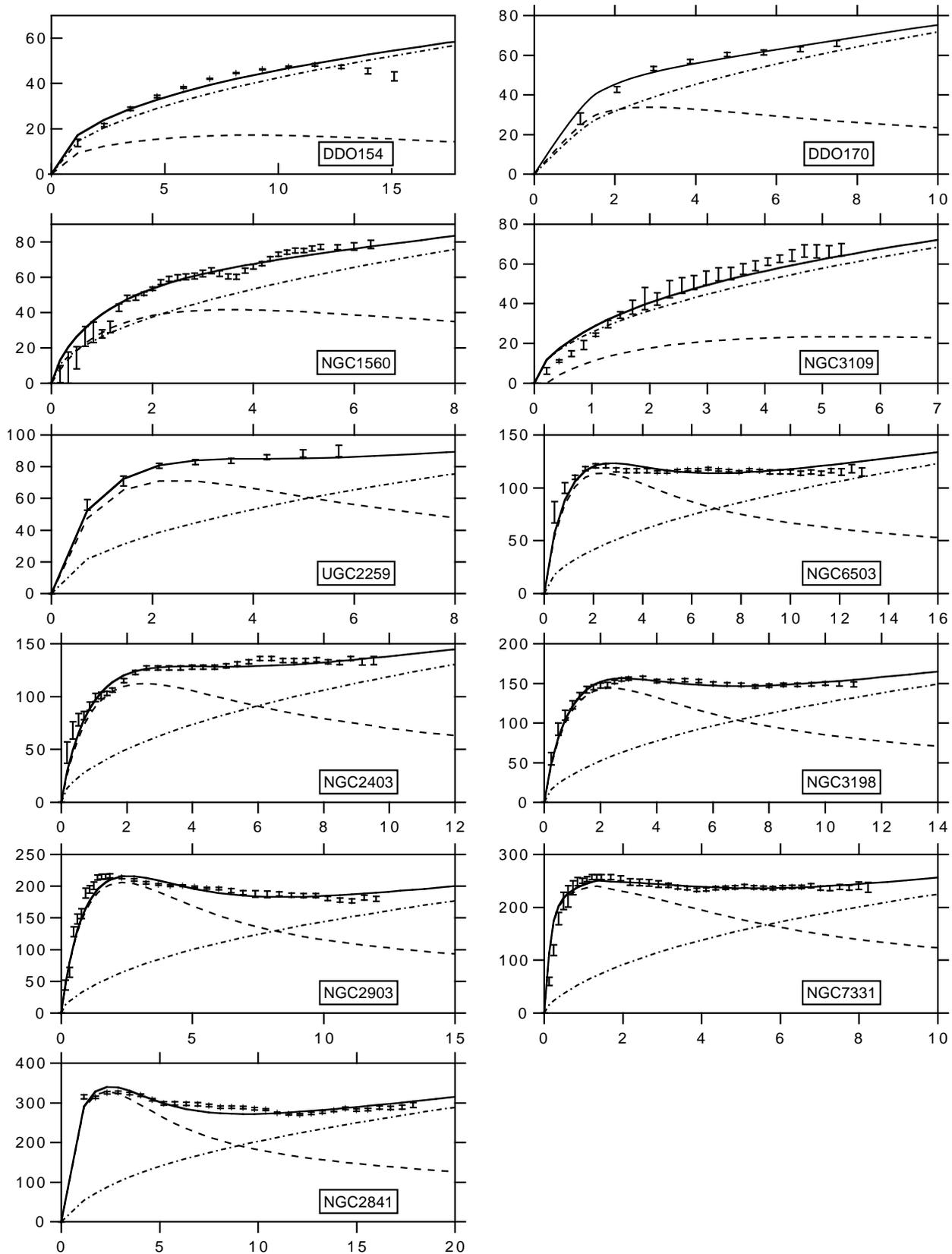,scale=0.85}
\caption{Some typical measured galactic rotation velocities (in km/sec)
with their quoted errors as plotted as a function of $R/R_0$ where
$R_0$ is each galaxy's own optical disk scale length. The dashed
(falling) curve shows the velocities that would be produced by the
luminous Newtonian contribution alone. The dash-dotted (rising) curve
shows the velocities that would be produced by the linear potentials
of the alternate conformal gravity theory which is discussed
below, and the full curve shows the overall predictions of that
theory, with no dark matter being assumed.}
\label{Fig. 1}
\end{figure}

\noindent
\begin{table}[t]
\centerline{Table 1: Characteristics of the eleven galaxy sample}
\medskip
$$
\begin{array}{ccccccccc}

{\rm Galaxy}~~&~~{\rm Distance}~~&~~ {\rm Luminosity}~~&~~~R_0~~~&
~~(v^2/c^2R)_{{\rm last}} ~~&~ (M/L) ~~& \\
{}& ({\rm Mpc}) & (10^9L_{B_\odot}) & ({\rm kpc}) &(10^{-30}{\rm
cm}^{-1})  & 
(M_{\odot}L_{B_\odot}^{-1})&  \\

{}&{}&{}&{}&{}&{} \\ 

 {\rm DDO}\phantom{1}~154  &\phantom{0}3.80& \phantom{0}0.05 & 0.48&
1.51&   0.71   \\

 {\rm DDO}\phantom{1}~170  &          12.01& \phantom{0}0.16 & 1.28&1.63&
 5.36   \\

 {\rm NGC}~1560&            \phantom{0}3.00& \phantom{0}0.35 & 1.30&2.70&
 2.01 &   \\

 {\rm NGC}~3109&            \phantom{0}1.70& \phantom{0}0.81 & 1.55&1.98&
  0.01  &  \\

 {\rm UGC}~2259&            \phantom{0}9.80& \phantom{0}1.02 & 1.33&3.85&
3.62  &   \\

 {\rm NGC}~6503&            \phantom{0}5.94& \phantom{0}4.80 & 1.73&2.14&
 3.00  &   \\

 {\rm NGC}~2403&            \phantom{0}3.25& \phantom{0}7.90 & 2.05&3.31&
 1.76  &   \\

 {\rm NGC}~3198&            \phantom{0}9.36& \phantom{0}9.00 & 2.72&2.67& 
 4.78  &   \\

 {\rm NGC}~2903&            \phantom{0}6.40&           15.30 & 2.02&4.86&
 3.15  &   \\

 {\rm NGC}~7331&                      14.90&           54.00 & 4.48&5.51& 
 3.03  &  \\

 {\rm NGC}~2841&            \phantom{0}9.50&           20.50 &2.39&7.25
  & 8.26  & 

\end{array}
$$
\end{table}
\bigskip
\bigskip

To get a quick sense of what the luminous Newtonian expectation for
galaxies might look like, we note that for an infinitesimally thin
disk of stars with distribution $\Sigma(R)=\Sigma_0e^{-R/R_0}$ and 
potential $V^*(r)=-\beta^*c^2/r=-M_{\odot}G/r$, the total centripetal
acceleration is then given by \cite{Freeman1970} 
\begin{equation}
\frac{v_{{\rm lum}}^2}{R}=g_{\beta}^{{\rm lum}}=
\frac{N^*\beta^*c^2 R}{2R_0^3}\left[I_0\left(\frac{R}{2R_0}
\right)K_0\left(\frac{R}{2R_0}\right)-
I_1\left(\frac{R}{2R_0}\right)
K_1\left(\frac{R}{2R_0}\right)\right]~~.
\label{115}
\end{equation} 
(In Appendix A we provide a straightforward derivation of this relation.)
From the behavior of the modified  Bessel functions, the $v(R)$ velocity
profile associated with Eq. (\ref{115}) is found to consist of an initial
rise from $R=0$ up to a peak at $R=2.2R_0$ which is then followed by a
steady fall-off. With $\Sigma(R)$ typically reaching to about four scale
lengths before becoming negligible, the visible stellar material in the
galaxy is thus contained almost entirely within this region (the optical
disk), with Eq. (\ref{115}) asymptoting to the Keplerian
\begin{equation}
\frac{v_{{\rm lum}}^2}{R} \rightarrow \frac{N^*\beta^*c^2}{R^2}
\label{116}
\end{equation} 
far outside the optical disk. To treat the luminous Newtonian contribution
more accurately we need to allow for the fact that the optical disk
actually has a thickness (with its own scale height $z_0$), that the HI
gas comes with its own surface brightness distribution and can have a
total mass as high as 15 per cent or so of the luminous mass, and that
the two largest galaxies in the sample also have a central bulge. Using
the calculational tools presented in Appendices A, B and C the
contributions of all of these various components can readily be determined
for each galaxy in the sample, with the exact luminous Newtonian
expectation for $v(R)$ reported in the typical calculation of
\cite{Mannheim1997} being exhibited as the dashed curves in Fig. 1.
In generating this total luminous Newtonian contribution, the total
number of stars $N^*$ in each galaxy was adjusted so that the luminous
Newtonian contribution would account for the entire rotational velocity
at the $R=2.2R_0$ peak (the maximum disk prescription), with the obtained
values for $N^*$ being presented in Table 1 as galactic mass to light
ratios $M/L$ which thus express $N^*$ in terms of the measured total
luminosity of each galaxy. As we see, normalizing to this maximum disk
prescription then completely accounts for the entire initial rise in the
rotation curve, to therefore indicate that the inner region is well
described by the luminous Newtonian contribution alone. However, as
we also see from Fig. 1, in the outer regions of the galaxies the
luminous Newtonian contributions totally underaccount for the measured
velocities.\footnote{The maximal disk prescription takes the luminous
Newtonian contribution to be as large as it possibly can be without
overshooting the inner region data. Since this simultaneously fixes the
normalization of the  Newtonian contribution to the outer region, it thus
represents the best that the luminous Newtonian contribution can possibly
do in the outer region. And even if we were to change the overall
normalization of $N^*$, this would anyway not change the shape of the
luminous Newtonian contribution, with its behavior at large $R$ still
falling off even then.} This then is the mass discrepancy problem in
galaxies, and as such it is actually far more severe than the cluster
problem, since it is based not on any statistical averaging over the
velocities of many galaxies in a cluster or on any assumption regarding
their virialization, but only on looking at individual orbits in galaxies
one by one just as was done for planetary orbits in the solar system. Not
only do we find such discrepancies in galaxies, we wryly note if the
galactic data were the only data that were available to us and we had no
solar system data at all, then from the measured galactic rotational
velocities alone it would not be possible to infer an inverse square
gravitational law at all.\footnote{It is amusing to ponder how gravity
theory might have evolved if the earth had been the only object in orbit
around the sun.}

\subsection{The dark matter solution for galaxies}

Given the prior experience with first the prediction and then the
discovery of the planet Neptune, it was most natural to try to explain
the galactic mass discrepancies by positing the presence of non-luminous
dark matter in  galaxies. In fact, there was already some theoretical
basis for doing so since Ostriker and Peebles \cite{Ostriker1973} had
suggested that there should be non-luminous spherical halos (non-luminous
since there was no sign of any luminous such halos of the size they
envisaged) surrounding unbarred spiral galaxies, due to the fact that in
Newtonian gravity such disk shaped structures would not be gravitationally
stable, though spherical ones would. However, any such stabilizing halo
would only need to envelop the optical disk in order to stabilize it and
would thus have no immediate need to extend beyond four disk scale
lengths or so. Moreover, since both elliptical galaxies and clusters of
galaxies are also found to have mass discrepancies, it would appear that
whatever is causing such spherically shaped systems to possess these
discrepancies, it could not be stability since such spherical
distributions of luminous Newtonian matter are perfectly capable of being
stable on their own. Nonetheless, the difficulty with not taking galactic
dark matter to be spherically distributed is that if it were also to lie
in a disk then it would not be stable either. With the measured
rotational velocities far exceeding the luminous Newtonian contributions
in the outer galactic regions, within Newtonian gravity any proposed dark
matter which would be capable of affecting the shape of the rotation
curves in those outer regions would explicitly have to be located in
those selfsame regions itself. Moreover, with the measured velocities in
the outer galactic regions being at least twice as big as the luminous
Newtonian contributions, the dark matter contribution to the $v^2/R$
centripetal accelerations would thus have to be three to four times as
big as the luminous one. Thus not only would dark matter have to be
located in precisely the region where there are very few stars, the total
amount of dark matter in galaxies would have to greatly exceed the total
amount of visible matter, matter which itself is predominantly located in
the inner region optical disk. In this sense then the galactic dark matter
solution is not entirely equivalent to the Neptune solution to the solar
system Uranus problem, since the Neptune modification to solar system
orbits was a very minor modification to the Newtonian gravity produced by
the sun and not an effect which was of the same order as it. From the
point of view of gravity theory, a modification to the galactic luminous
Newtonian expectation which is comparable with it in magnitude is a quite
drastic modification, while from the point of view of observational
astronomy, the notion that such a huge amount of matter is not only not
optically observable but is not even predominantly located in the region
occupied by the visible stars is equally drastic, to thus underscore how
very serious the dark matter problem is.

Beyond the actual assumption of the existence of galactic dark matter at
all, its explicit use in galaxies still leaves much to be desired.
Specifically, in order to actually fit the galactic rotation curve data
explicitly using dark matter, it is conventional to take the spherical
dark matter density
$\sigma(r)$  to be the distribution associated with an isothermal
Newtonian sphere in hydrostatic equilibrium, viz.
\begin{equation}
\sigma(r)=\frac{\sigma_0}{(r^2+r_0^2)}~~,
\label{117}
\end{equation} 
as modified by the introduction of a core radius $r_0$ to prevent the
distribution from diverging at $r=0$. Given such a distribution, the
associated centripetal accelerations in the galactic plane are given by
(see e.g. Eq. (\ref{A29}))
\begin{equation}
\frac{v_{{\rm dark}}^2}{R}=g_{\beta}^{{\rm
dark}}=\frac{4\pi\beta^{*}c^2\sigma_0}{R}
\left[1-\frac{r_0}{R}{\rm arc tan}\left(\frac{R}{r_0}\right)\right]~~.
\label{118}
\end{equation} 
While Eq. (\ref{118}) leads to rotation curves which are asymptotically
flat, an actual fitting to an observed flat rotation curve (such as the
high luminosity ones displayed in Fig. 1) is not automatically achievable
simply by adding $g_{\beta}^{{\rm dark}}$ to $g_{\beta}^{{\rm lum}}$ of
Eq. (\ref{115}). Rather, the two free parameters in $g_{\beta}^{{\rm
dark}}$ have to be adjusted galaxy by galaxy. As we see from a typical
galaxy such as NGC 3198, the observed rotation curve is pretty flat
from the inner region peak near $R=2.2R_0$ where (c.f. Eq. (\ref{115}))
\begin{equation}
v_{{\rm lum}}^2 \sim \frac{0.4N^*\beta^{*}c^2}{R_0}~~,~~v_{{\rm lum}}^4
\sim \left[0.32\pi\Sigma_0\beta^{* 2}c^4\right]N^{*}~~,
\label{119}
\end{equation} 
all the way out to the last detected point at $R\sim 10R_0$. While the
asymptotic limit of Eq. (\ref{118}) leads to
\begin{equation}
v_{{\rm dark}}^2 \rightarrow 4\pi\beta^{*}c^2\sigma_0~~,
\label{120}
\end{equation} 
there is no apparent reason why the dark matter $\sigma_0$ should be
related to the luminous parameters $N^*$ and $R_0$ via
\begin{equation}
\sigma_0=\frac{N^*}{10\pi R_0}~~,
\label{121}
\end{equation} 
and yet without such a fine-tuning between the dark halo and optical disk
parameters, the asymptotic halo contribution would not match the inner
region peak velocity at all, being either larger or smaller than it.
Moreover, even if this particular fine-tuning is invoked, at best that
would only match the velocity at $R=2.2R_0$ to the velocity at $r=10R_0$
and not require the rotation curve to be flat at all points in between.
To achieve such intermediate region flatness, in addition it is necessary
to also fine-tune the halo core radius parameter $r_0$ as well (so as to
match at $R=6R_0$ or so), and only after this is also done is the flatness
of the rotation curve over the entire $2.2R_0\leq R \leq 10R_0$ region
then secured. Unsatisfactory as this use of two fine-tuned parameters
is, the situation is even worse than that since such a fine-tuning has to
be made individually for each and every galaxy; and while a two parameter
per galaxy prescription is basically found to work well
phenomenologically for the rotation curves that have so far been
observed, at two halo parameters per galaxy this would already lead to
22 free halo parameters for just the eleven galaxy sample of Fig. 1
alone. 

Beyond the isothermal sphere shape for $\sigma_0$, numerical $N$-body cold
dark matter (CDM) simulations of hierarchical clustering have also yielded
halos, of which the typical one found by Navarro, Frenk and White
\cite{Navarro1996,Navarro1997} is quite close in form to the profile
\begin{equation}
\sigma(r)=\frac{\sigma_0}{r(r+r_0)^2}~~.
\label{122}
\end{equation} 
Unlike the halos of Eq. (\ref{117}), with the  halos of Eq. (\ref{122})
being cuspy ones which diverge at $r=0$, they make a quite substantial
(though still finite) contribution to rotational velocities in the inner
region, and have been criticized on this score \cite{Sellwood2001} since
they are found to have some difficulty in fitting galaxies such as dwarfs
and low surface brightness galaxies\footnote{Viz. galaxies with a small
$N^{*}$ or a small central surface brightness $\Sigma_0$.} where the inner
region luminous Newtonian contribution is small, a region where the true
nature of the halo should then be manifest. Moreover, even though these
cuspy dark matter halos do provide good fitting to high surface brightness
galaxies, since these particular halos would contribute in the inner
region, getting good fitting in the high surface brightness case entails
giving up the nice accounting of the inner region which the luminous
Newtonian contribution provides, and instead requiring an interplay
between dark and luminous matter even in that region. Very recently
attempts have been made to remedy the cuspiness problem
\cite{Navarro2004,Hayashi2004} leading to halos close in form to
\begin{equation}
\sigma(r)=\sigma_0{\rm
exp}\left[-\frac{2}{\alpha}\left(\frac{r^n}{r_s^n}-1\right)\right]~~,
\label{123}
\end{equation} 
and the issue should be regarded as being open at the present time.
However, apart from issues such as this, there is actually a far more
serious concern for essentially all dark halo models, namely even if one
can generate dark matter halos dynamically, it is not clear what fixes
which halo is to go with which given luminous matter distribution, i.e.
if the only information available is the measured surface brightness of
a given galaxy, from that how do we then determine which values
of $\sigma_0$ and $r_0$ to use with each such given galaxy so as to be
able to predict the structure of the galactic rotation curve in advance
of its measurement, and thereby render dark matter theory falsifiable.
Forms such as those of Eqs. (\ref{122}) and (\ref{123}) only give the
generic shapes of halos, and do not provide any a priori determination of
values for the $\sigma_0$ and $r_0$ parameters or their relation to the
luminous $\Sigma(R)=\Sigma_0e^{-R/R_0}$ profile, leaving $\sigma_0$ and
$r_0$ to be free parameters to be fit to galactic data galaxy by galaxy,
to then quickly generate large numbers of free parameters as more and more
galactic rotation curves are considered.

\subsection{Clues from the data}

In fact perhaps the most serious challenge to the whole dark matter idea
is something which is intrinsic to the very motivation which gave rise
to it in the first place. Specifically, since galactic rotation curves
could not be fitted by luminous Newtonian matter alone, any additional
matter which was to be invoked would then have to be non-luminous, and
thus be essentially decoupled from the luminous matter. However, as
we see from relations such as the $v^4 \sim L$ Tully-Fisher relation which
holds for the prominent high luminosity spiral galaxies, the velocity
which is fixed by the total gravitational potential is in fact correlated
with the luminosity, and thus whatever is producing this total
gravitational potential has to know about the luminous matter even as it
is to totally dominate over it in the potential. Moreover, the
Tully-Fisher relation is not the only regularity in the data, and in
order to get some guidance from the data as to what is needed of dark
matter theory (or of any alternate theory for that matter), we now look
at some particularly instructive regularities in the galactic data
themselves.

As regards the surface brightness, as first noted by Freeman, as well as
obey the Tully-Fisher relation, high surface brightness spiral galaxies
are found to all have the same central surface brightness $\Sigma_0^F$,
with all other types of spiral galaxy typically being found to have
central surface brightnesses which are lower than this $\Sigma_0^F$, with
the Freeman value for $\Sigma_0^F$ serving as an upper limit on the
general $\Sigma_0$. Moreover, from Eq. (\ref{119}) we see that Freeman
limit galaxies have an inner region peak velocity which obeys
\begin{equation}
v_{{\rm lum}}^4=0.32\pi\Sigma_0^F\beta^{* 2}c^4N^{*}~~,
\label{124}
\end{equation} 
with all galaxies with this common $\Sigma_0^F$ then immediately obeying
the Tully-Fisher relation, and doing so entirely because of the
behavior of the luminous matter distribution alone, to thus not
immediately point in the direction of dark matter.

While most attention in galactic rotation curves has focussed on the flat
ones since they are so very striking, nonetheless, as we see from Fig. 1
not all rotation curves are in fact flat. Rather, as first recognized by
Casertano and van Gorkom \cite{Casertano1991}, the data fall into three
broad classes. Specifically, the rotation curves of the low luminosity
dwarf galaxies are typically found to be rising at the last detected data
points, the rotation curves of the intermediate to high luminosity
galaxies are found to be flat, while the rotation curves of the very
highest luminosity galaxies are found to be mildly falling. While there
are many flat rotation curves, there are also many which are seen to
be rising, and if these curves are to eventually flatten off, this has yet
to be observed. Of these various classes, it is only the high luminosity,
flat rotation curve galaxies which are Freeman limit, Tully-Fisher
galaxies, with the lower luminosity, non-flat galaxies not enjoying this
particular form of universality at all. 

To explore the rotation curves of these various classes of galaxies in
more detail, and to try find a universality which both Freeman limit
and sub-Freeman limit galaxies can enjoy, it is very instructive to
focus not on the total rotational velocities, but rather on the
difference between the total velocity and the luminous Newtonian
expectation. With the luminous Newtonian contributions all falling
far outside the optical disk, for galaxies for which the total
velocities are flat, the difference between the total and the luminous
Newtonian expectation can not itself flat, but must instead actually
be rising in the detected region. Inspection of the flat rotation curves
in the sample of Fig. 1 shows no galaxy for which the difference between
the total and the luminous Newtonian expectation is itself flat, so if the
effect of galactic halos is to produce contributions to rotational
velocities which are to be asymptotically flat (as per Eq. (\ref{120})),
no sign of any such asymptotic behavior is yet manifest in the data. 
A similar shortfall pattern is also found to obtain for the dwarf
galaxies, galaxies whose rotation curves are not flat, with the shortfall
between the measured rotational velocities and the luminous Newtonian
expectation again being seen to be rising. In fact the situation for
dwarfs is actually more severe than for bright spirals, since while the
ratio of the total velocity to the luminous Newtonian expectation at the
last detected data point is a factor of two for the bright spirals, for
dwarfs the ratio is a factor of three (and thus a factor of nine in the
centripetal acceleration). Mass discrepancies in dwarfs are thus even
more pronounced than in the bright spirals.

As we see, the characteristic feature of all the eleven galaxies in the
sample is that the shortfall between the measured rotational velocities
and the luminous Newtonian expectation is a shortfall which explicitly
rises with distance from the center of each galaxy. It is thus of
interest to see whether this rise might have any universal structure which
is common to all the galaxies in the sample. To this end it is very
instructive to evaluate the centripetal acceleration at the last detected
data point in each galaxy, with the obtained values being exhibited in
Table 1. As we see from the table, the quantity $(v^2/c^2R)_{{\rm last}}$
is found to basically increase with luminosity. With the asymptotic
luminous Newtonian contribution to the centripetal acceleration being
given by Eq. (\ref{116}) where $N^{*}$ has been fixed once and for all
by the inner region maximum disk procedure, its subtraction from
$(v^2/c^2R)_{{\rm last}}$ is found to yield a function which satisfies a
two parameter formula containing one term which is linear in $N^{*}$ and a
second term which, quite extraordinarily, is independent of it altogether,
with the full $(v^2/c^2R)_{{\rm last}}$ and $v^2_{{\rm last}}$ being
given by
\cite{Mannheim1997}
\begin{eqnarray}
\left(\frac{v^2}{c^2R}\right)_{{\rm last}}&=&
\frac{\beta^*N^*}{R^2}+\frac{\gamma^*N^*}{2}+\frac{\gamma_0}{2}~~,~~
\nonumber \\
v^2_{{\rm last}}&=&
\frac{\beta^*c^2N^*}{R}+\frac{\gamma^*c^2N^*R}{2}+\frac{\gamma_0c^2R}{2}
\label{125}
\end{eqnarray} 
where $\gamma^*$ and $\gamma_0$ are two new universal parameters with
values
\begin{equation}
\gamma^*=5.42\times 10^{-41} {\rm cm}^{-1}~~,~~\gamma_0=3.06\times
10^{-30} {\rm cm}^{-1}~~.
\label{126}
\end{equation} 
To get a sense of the relative importance of these various terms, we note
that with the solar $\beta^*$ being given by $\beta^*=1.48\times 10^5$ cm,
the quantity $(\beta^*/\gamma^*)^{1/2}$ evaluates to
$(\beta^*/\gamma^*)^{1/2}=.52\times 10^{23}$ cm, with the $\beta^*N^*/R^2$
and $\gamma^*N^*/2$ terms thus being competitive with each other on
none other than galactic distance scales. With the large galaxies
typically having of order $10^{11}$ stars, for them the $\gamma^*N^*/2$
and $\gamma_0/2$ are of the same order of magnitude, with all three
contributions in Eq. (\ref{125}) being of importance, and with the falling
luminous Newtonian contribution having its most significant effect in the
most luminous (largest $N^*$) of the galaxies. However, for small $N^*$
the $\gamma_0c^2R/2$ term completely dominates in $v^2_{{\rm last}}$ with
the last data points then rising with $R$, just as exhibited in Fig. 1 for
the dwarf galaxies, and with Eq. (\ref{125}) precisely capturing the
essence of the pattern identified in \cite{Casertano1991}.

In order to assess the significance of Eq. (\ref{125}), we note that
there is nothing at all special about the last data point in each galaxy.
Such data points are fixed purely by the observational limit of the radio
detectors used to survey the galaxies, a limit which is not fixed by the
galaxies themselves but by their distances from us. And yet we are
obtaining a universal formula. It appears to us that the only way to
achieve such an outcome is to have shortfalls which are rising not just
at the last currently detected $R$ but at all possible $R$ beyond the
optical disk, so that no matter where an individual detector just happens
to cut off, it will be doing so on a curve that is universally rising, to
thus always yield the same Eq. (\ref{125}). Since the eleven galaxy
sample is large enough to be characteristic of galactic rotation curve
systematics and since it was by no means chosen by the authors of
\cite{Begeman1991} in order to obtain a formula such as that of Eq.
(\ref{125}), we thus believe we can identify Eq. (\ref{125}) as a
universal characteristic of galactic rotation curve systematics, one
which is of interest in its own right and which dark matter theory is
therefore required to reproduce. With Eq. (\ref{125}) reducing to the
$N^*$-independent relation $v^2_{{\rm last}}=\gamma_0c^2R/2$
in the small $N^*$ limit, the small $N^*$ limit exposes the departure from
the luminous Newtonian expectation in the starkest terms. And it could be
that $\Lambda$CDM (CDM with a cosmological constant $\Lambda$) generated
dark matter halo models are having difficulty fitting dwarfs because 
relations such as $v^2_{{\rm last}}=\gamma_0c^2R/2$ might not be natural
to them.

In regard to Eq. (\ref{125}), we additionally note that the quantity
which measures the departure of the centripetal acceleration
$(v^2/R)_{{\rm last}}$ from the luminous Newtonian expectation is a
universal acceleration $\gamma_0c^2/2$ ($\gamma_0+\gamma^*N^*$ is never
smaller than $\gamma_0$) whose magnitude is given by
$\gamma_0c^2/2 =1.4\times 10^{-9}$ cm/sec/sec, a value which is 
close to that found for the universal acceleration parameter $a_0$ of the
MOND theory discussed above. Further support for a role for a
universal acceleration was noted by McGaugh \cite{McGaugh1999} who studied
the behavior of the quantity $M_{{\rm dyn}}(R)/M_{{\rm lum}}(R)$ as a
function of the measured centripetal acceleration $v^2(R)/R$ at points
$R$ within galaxies. Here $M_{{\rm dyn}}(R)=Rv^2(R)/G$ is the total
amount of matter interior to $R$ as would be required by Newtonian gravity
to produce the measured $v^2(R)/R$, while $M_{{\rm lum}}(R)$ is the
total amount of luminous matter which is detected within the same region.
In his study of a quite extensive sample of galaxies McGaugh found that
mass discrepancies [viz. $M_{{\rm lum}}(R)$ less than the needed $M_{{\rm
dyn}}(R)$] systematically occurred in galaxies whenever the measured
$v^2(R)/R$ fell below a universal value of order $10^{-8}$ cm/sec/sec or
so, a value which is in accord with that of both the 
acceleration $a_0$ associated with MOND and the $\gamma_0c^2/2$
acceleration associated with Eq. (\ref{125}). Thus from both
McGaugh's phenomenological study and from the phenomenological Eq.
(\ref{125}), we see that it is indeed an  acceleration scale (or
equivalently an inverse distance scale times the square of the speed of
light) which determines exactly when luminous Newtonian matter fails to
account for observed data. As such, the acceleration criterion which
Milgrom originally identified has to be regarded as being a bona fide
property of nature. As formulated, the MOND theory actually possesses
not one but two logically independent components. It comes not just
with this acceleration criterion which indicates when the luminous
Newtonian prediction is to fail, but also it provides a model (Eq.
(\ref{99})) for how Newtonian gravity is to then be modified in the
region (the MOND region) where the luminous Newtonian expectation does
in fact fail. However, regardless of how the theory is to behave in the
MOND region, i.e. regardless of what particular form the function
$\nu(f/a_0)$ might take,\footnote{If in Eq. (\ref{125}) we ignore the
$\gamma^*N^*/2$ term and set
$a_0=\gamma_0c^2/2$, we can then write Eq. (\ref{125}) in the generic
form of Eq. (\ref{99}) with the function $\nu(f/a_0)$ being given by
$\nu(f/a_0)=1+a_0/f$, an issue we will elaborate on further below.}  and
regardless of whether one likes any particular form (such as the one of
Eq. (\ref{101}) with its square root structure) that might be used to fit
data, it remains true that Milgrom's acceleration criterion is the
criterion which determines when departures from the luminous Newtonian
expectation are to first occur, a criterion which fundamental theory is
therefore obliged to explain. As we thus see, from Milgrom's work, from
McGaugh's work and from Eq. (\ref{125}), there is a lot of universality
in the rotation curve data, universality which does not immediately
appear to be natural to dark matter and which therefore challenges it, 
a universality which we will use below to guide us when we consider
alternatives to dark matter.

\section{The dark energy problem}

\subsection{The cosmological dark matter and flatness problems}

As we go to scales even larger than galaxies, mass discrepancies are not
only found to persist, two new types of mass discrepancy begin to appear,
one for clusters of galaxies and the other for the entire universe
itself. While non-relativistic luminous Newtonian expectation shortfalls
are found not just for galaxies but for clusters of galaxies as well, for
such clusters general-relativistic shortfalls are also found, with the
use of the Schwarzschild metric for a cluster leading to far less
gravitational lensing by the cluster than is observed if only the
luminous inventory of the cluster is used as the Schwarzschild source.
Thus for clusters one can say that either the dark matter idea is
internally consistent since both non-relativistic and
general-relativistic shortfalls are found in one and the same system,
or one can say that absent dark matter, with clusters it is not just
Newtonian gravity which is failing but Einstein gravity too, with the
cluster geometry probed by light from a distant quasar not being
the Schwarzschild metric associated with the visible sources. In making a
statement such is this it is important to note that the detection of
gravitational lensing is generally regarded as being a great triumph for
Einstein gravity. However the fact of lensing as opposed to the amount
of lensing only requires that gravity be a metric theory to which light
couples. It is the amount of lensing which is sensitive to the specific
make-up of the lens and the specific geometry which it produces, and
unless there is dark matter in clusters, then the standard theory would
simply not be predicting enough. As with our discussion of $\Lambda$CDM
generated galactic halos, what is needed to make cluster dark matter into
a falsifiable theory is a prediction of the amount of lensing given only
a knowledge of the luminous content of the cluster. Currently the amount
of dark matter in clusters is inferred only after the lensing
measurements have been made. While making lensing predictions based only
on the known luminous content of a cluster thus serves as a goal for dark
matter, at the present time it also remains an objective for the alternate
gravitational theories we shall discuss below, including those for
which galactic rotation curve predictions can be made from luminous
information alone. Progress on cluster gravitational lensing should
thus be definitive for both dark matter and its potential alternatives.

The other large distance scale on which mass discrepancies are found
is the largest distance scale there is, viz. that of the universe itself, 
with the standard theory not only again encountering the issue of dark
matter, this time the problem is compounded by a need for dark energy as
well. To heuristically see why such problems might be anticipated, it is
instructive to consider the implications for a simple Newtonian cosmology
of the existence of a universal acceleration. Specifically, we recall
that in a critical density Newtonian cosmology of size the Hubble radius
$L=c/H$ (viz. a cosmology where $\rho_C=3Mc^2/4\pi L^3=3H^2c^2/8\pi G$),
the Hubble radius happens to be equal to the Schwarzschild radius of the
matter within it (viz. $2MG/c^2 =L$). For such a cosmology the
centripetal acceleration of a test particle located at the Hubble radius
is given by $v^2/L=MG/L^2=cH/2$. Using a value of $72\pm 8$ km/sec/Mpc
for $H$ \cite{Freedman2001}, we find that $cH/2$ evaluates to
$cH/2=3.5\times 10^{-8}$cm/sec/sec, and with this value being very close
to the values found for both the MOND
$a_0$ and the acceleration $\gamma_0c^2/2$ associated with Eq.
(\ref{125}), we see that the centripetal accelerations of Newtonian
cosmology occur precisely in the acceleration regime where luminous
Newtonian matter falls short for galaxies. We this anticipate, and
shall in fact shortly see, that a pure luminous matter based standard
Einstein gravity cosmology will not adequately fit available cosmic data
either. Moreover, in addition this time we will find that the situation
can not even be rectified by invoking dark matter, with it being this
very failure which has led to the introduction of yet another ad hoc
dark object, viz. dark energy.

In standard Einstein gravity a full general-relativistic treatment of
cosmology is based on the Friedmann equations of Eqs.
(\ref{109}), (\ref{110}) and (\ref{39}) 
\begin{equation}
\dot{R}^2+kc^2=\frac{8 \pi
G}{3c}CR^2~~,~~3\ddot{R}=-\frac{4\pi G}{c}(C+3D)R~~,
~~D=-\frac{d}{dR^3}\left(R^3C\right)~~,
\label{127}
\end{equation}
as written for the moment in terms of the generic source
$T^{\mu\nu}=[C(t)+D(t)]U^{\mu}U^{\nu}+D(t)g^{\mu\nu}$ given in Eq.
(\ref{37}). Given the Robertson-Walker metric of Eq. (\ref{36}), the
objective of cosmology is to then determine the parameters $k$ and $R(t)$
which appear in the metric and determine the matter content of the
universe as expressed through $C(t)$ and $D(t)$. As a result
of many years of observational effort, values for the Hubble parameter
$\dot{R}/R=H$ have now converged on an agreed value for $H$ as given
above. With $H$ and $G$ being known, it is convenient to introduce the
dimensionless quantities
\begin{equation}
\Omega_C(t)=\frac{8\pi
GC(t)}{3cH^2}~~,~~\Omega_k(t)=-\frac{kc^2}{\dot{R}^2(t)}~~,
\label{128}
\end{equation}
and the deceleration parameter
\begin{equation}
q(t)=-\frac{\ddot{R}R}{\dot{R}^2}~~,
\label{129}
\end{equation}
since in terms of them the Friedmann equations of Eq. (\ref{127}) can 
be written very compactly as
\begin{equation}
\dot{R}^2+kc^2=\dot{R}^2\Omega_C(t)~~,~~\Omega_C(t)+\Omega_k(t)=1~~,
~~q(t)=\frac{1}{2}\left(1+\frac{3D}{C}\right)\Omega_C(t)~~.
\label{130}
\end{equation}

Historically, the cosmological source was taken to be ordinary matter with
$C=\rho_m/c$, $D=p_m/c$, with equation of state $p_m=\rho_m/3$ for
radiation and high energy ($kT \gg mc^2$) matter and equation of
state $p_m=0$ for low energy ($kT \ll mc^2$) matter, with an inventory of
the visible universe using galaxy counts leading to a current era value
for 
\begin{equation}
\Omega_M(t)=\frac{8\pi G\rho_m(t)}{3c^2H^2}
\label{131}
\end{equation}
of order $0.01$ at the current time $t_0$.\footnote{Since such
inventorying does not undercount, $0.01$ thus serves as a lower bound 
on $\Omega_M(t_0)$.} Such a value for $\Omega_M(t_0)$ is quite close to
the critical value $\Omega_M=1$, the value at which the Friedmann
equation would require the parameter $k$ to be zero (flat
universe).\footnote{It is called the critical value since the sign of
the 3-curvature $k$ of the universe has the same sign as 
$\Omega_M(t_0)-1$.} As first noted by Dicke, this
closeness of $\Omega_M(t_0)$ to one creates a severe problem for the
Friedmann equation, the so-called flatness problem. Specifically,
regardless of the value of $k$, the use of the above normal matter
equations of state in the Friedmann equation leads to the occurrence of a
big bang singularity in the early universe, with
$\dot{R}(t=0)$ diverging at $t=0$. Near to such a singularity the same
Friedmann equation then requires that
$\Omega_M(t=0)$ be incredibly close to one, again regardless of the
choice of $k$. Then, with $\Omega_M(t=0)$ being close to one in the early
universe, and with the universe then expanding for an entire Hubble time
($1/H(t_0)\sim 10^{10}$ years), it is very hard to understand why the
current era $\Omega_M(t_0)$ would be anywhere near unity today instead of
having redshifted to close to zero by now. In fact for the Friedmann
equation to lead to a cosmic evolution in which the current era value of
$\Omega_M(t_0)$ would be close to one today requires fine-tuning the
initial value of $\Omega_M(t=0)$ at a typical early universe time to be
one to within one part in $10^{60}$ or so. As an evolution equation, the
Friedmann equation is highly unstable, with it not being able to
easily reconcile its initial singularity with the current era value of
$\Omega_M(t_0)$. More naturally, it would have $\Omega_M(t)$ fall
to zero in the early universe itself.

\subsection{Fine-tuning and the inflationary universe}

A beautiful solution to this conflict was proposed by
Guth \cite{Guth1981}, the so-called inflationary universe, in which a de
Sitter phase was to precede the current Robertson-Walker phase, with the
rapid exponential expansion of the de Sitter phase precisely providing a
dynamics which would lead to the needed initial condition for the
subsequent Robertson-Walker phase, no matter what the value of
$k$. An additional virtue of having such an early universe de Sitter phase
was that it also provided a solution to the horizon problem associated
with the observed uniformity of the cosmic microwave background across
the sky, with the de Sitter phase phase being able to causally connect
regions which a luminous ($\Omega_M(t_0)=0.01$) or even a
luminous plus dark ($\Omega_M(t_0)>0.01$) matter Friedmann universe
could not.\footnote{That a de Sitter phase would solve the horizon
problem had also been noted in
\cite{Brout1978} and \cite{Kazanas1980}.} 

Geometrically it is possible to describe a de Sitter phase using the
comoving Robertson-Walker coordinate system, and for it we can take
the Friedmann equation to be of the form associated with a
cosmological constant source, viz. 
\begin{equation} 
\dot{R}^2(t) +kc^2=\alpha c^2 R^2(t)~~, 
\label{132}
\end{equation}
where $\alpha=8\pi G\Lambda/3c^3$. For the allowable choices of sign for
$\alpha$ and $k$ solutions to Eq. (\ref{132}) can readily be found, 
and they are given as
\begin{eqnarray}
R(t,\alpha<0,k<0)&=&\left(\frac{k}{\alpha}\right)^{1/2}{\rm
sin}[(-\alpha)^{1/2}ct]~~,
\nonumber \\
R(t,\alpha=0,k<0)&=&(-k)^{1/2}ct~~,
\nonumber \\
R(t,\alpha>0,k<0)&=&\left(-\frac{k}{\alpha}\right)^{1/2}{\rm
sinh}(\alpha^{1/2}ct)~~,
\nonumber \\
R(t,\alpha>0,k=0)&=&R(t=0){\rm exp}(\alpha^{1/2}ct)~~,
\nonumber \\
R(t,\alpha>0,k>0)&=&\left(\frac{k}{\alpha}\right)^{1/2}{\rm
cosh}(\alpha^{1/2}ct)~~.
\label{133}
\end{eqnarray}
On defining 
\begin{equation} 
\Omega_{\Lambda}(t)
=\frac{\alpha c^2}{H^2}=\frac{8\pi
G\Lambda}{3cH^2}~~,
\label{134}
\end{equation}
we find that in the de Sitter phase the quantities $q(t)$ and
$\Omega_{\Lambda}(t)$ are related by 
$q(t)=-\Omega_{\Lambda}(t)$, with $q(t)$ and
$\Omega_{\Lambda}(t)$ then being found to be given by
\begin{eqnarray}  
\Omega_{\Lambda}(t,\alpha<0,k<0))&=&-q(t,\alpha<0,k<0)
~=~-{\rm tan}^2[(-\alpha)^{1/2}ct]~~,
\nonumber \\ 
\Omega_{\Lambda}(t,\alpha=0,k<0))&=&-q(t,\alpha=0,k<0)~=~0~~, 
\nonumber \\
\Omega_{\Lambda}(t,\alpha>0,k<0))&=&-q(t,\alpha>0,k<0)~=~{\rm
tanh}^2(\alpha^{1/2}ct)~~,
\nonumber \\
\Omega_{\Lambda}(t,\alpha>0,k=0))&=&-q(t,\alpha>0,k=0)~=~1~~,
\nonumber \\ 
\Omega_{\Lambda}(t,\alpha>0,k>0))&=&-q(t,\alpha>0,k>0)~=~{\rm
coth}^2(\alpha^{1/2}ct)
\label{135}
\end{eqnarray}
in the various cases. As we see from the solutions, when
$\alpha$ is taken to be positive, then no matter what the value of
$k$, at late times the expansion radius will grow as ${\rm
exp}(\alpha^{1/2}ct)$ with $\Omega_{\Lambda}(t)$ asymptoting to
one and $\Omega_{k}(t)=1-\Omega_{\Lambda}(t)$ to zero. What a period of
de Sitter inflation thus achieves is that it quenches the contribution 
of curvature to cosmic expansion, with the universe inflating so
much that curvature is no longer able to contribute to the
expansion. If at the end of the inflationary era all of the energy
density contained in $\Omega_{\Lambda}(t)$  can be
converted into a matter $\Omega_{M}(t)$, $\Omega_{\Lambda}(t)$ will
then play no further role and $\Omega_{M}(t)$ will then enter the
Robertson-Walker era with a value that will be extremely close to
one in a universe in which curvature is no longer capable of
influencing cosmic expansion. With $\Omega_k(t)$ thus remaining
negligibly small throughout the Robertson-Walker era, and with
$\Omega_{\Lambda}(t)$ have been disposed of, $\Omega_{M}(t)$ will
then continue to remain close to one, and will thus still be
extremely close to one today. The inflationary solution to the
flatness problem is thus to make an $\Omega_{M}(t)$ which is of
order one actually be equal to one (in each and every
Robertson-Walker epoch), with the shortfall between a luminous
contribution of $0.01$ and one itself then requiring a $0.99$
contribution through matter that will contribute to $\Omega_{M}(t)$
with exactly the same equation of state as ordinary matter but
whose visual absence would require it to be non-luminous. Inflation
thus leads to the need for cosmological dark matter in an amount
which would make the current era $\Omega_{M}(t_0)$ be equal to one, viz.
to precisely the critical density value. With such a value for
$\Omega_{M}(t_0)$, and with $D/C$ being zero for the non-relativistic
fluids (dark or luminous) of relevance at the three degree current era
temperature, inflation would thus entail that the current era $q(t_0)$
would be equal to $1/2$ (c.f. Eq. (\ref{130})), so that even while
inflation itself leads to a negative $q(t)$ (c.f. Eq. (\ref{135})) in
the inflationary era, following the transfer of energy density from
$\Omega_{\Lambda}(t)$ to
$\Omega_{M}(t)$ at its end, we transit into a
Robertson-Walker phase in which $q(t)$ is positive.\footnote{That a
positive cosmological constant $\alpha$ leads to a negative $q(t)$ is
something we shall return to below.} While inflation thus provides
a nice solution to both the flatness and the horizon problems, the
very fact that it produces an era of exponential expansion no
matter what the value of $k$ means that inflation inflates away any
dependence on $k$, with inflation thus not being sensitive to the
sign or the magnitude of $k$. The fact that $\Omega_k(t_0)$ is to
be negligible today does not mean that $k$ itself is zero. Rather
inflation enables the current era $\Omega_k(t_0)$ to be negligible
even if $k$ is non-zero, though this very insensitivity to $k$ means
that the physics of what is to actually fix $k$ is thus not 
addressed by inflation. Since the universe does have a global topology
(flat, open or closed), something does have to fix it, with physics
either beyond inflation (or perhaps instead of it) still being required,
a point we return to below.

\subsection{The accelerating universe and the cosmological dark energy
solution}

While the development of the inflationary universe model immediately
prompted a vigorous search for dark matter candidates, a search which so
far has only yielded massive neutrinos (though with masses which are too
small to be of significance for the needed $\Omega_{M}(t_0)$), curiously,
at no time did any astrophysical observations themselves ever indicate
that $\Omega_{M}(t_0)$ was in fact equal to one or that $q(t_0)$ was
equal to $1/2$. In fact the best estimates for $\Omega_{M}(t_0)$ as
obtained from using dark matter to fit the luminous shortfalls found
for clusters of galaxies, lead to an $\Omega_{M}(t_0)$ of no more than
$0.3$ \cite{Bahcall2000}. As regards determining $q(t_0)$, we note that
since the deceleration parameter involves a second derivative with
respect to time, its measurement requires a measurement of the change
in the first derivative (the Hubble parameter) over cosmologically
separated distances, and thus requires an extension of the Hubble plot
out to very high redshift. A sufficient such extension out to $z=1$ has
recently been made using type Ia supernovae as standard candles, and
has led
\cite{Riess1998,Perlmutter1998} to an epochal discovery, namely that far
from being equal to $1/2$, the current era deceleration parameter is not
even positive at all, with the current universe actually being in an
accelerating rather than a decelerating phase. However, before addressing
this remarkable finding of the supernovae data, we note first that a
fit to the 54 most reliable of the original 60 type Ia 
supernovae sample of
\cite{Perlmutter1998} using the luminosity function 
\begin{equation} 
d_L=\frac{2c}{H(t_0)}(1+z)\left(1-\frac{1}{(1+z)^{1/2}}\right)
\label{136}
\end{equation}
associated with a pressureless $\Omega_{M}(t_0)=1$, $\Omega_{k}(t_0)=0$
universe is found to yield a $\chi^2$ value of $92.9$,\footnote{The
various $\chi^2$ values reported here and below are taken from
\cite{Mannheim2001,Mannheim2003a}.} a value which thereby actually
renders an $\Omega_{M}(t_0)=1$, $\Omega_{k}(t_0)=0$ universe unviable.
Consequently, recent supernovae data rule out the $\Omega_{M}(t_0)=1$
inflationary universe solution to the flatness problem. 

To get a sense of what is required to fit the supernovae data, we note
that as the matter content is reduced in this same pressureless
$\Omega_{M}(t_0)$, $\Omega_{k}(t_0)$ universe, a reduction which requires
a concomitant increase in $\Omega_{k}(t_0)$, the associated $\chi^2$ is
found to decrease, reaching a value of $\chi^2=61.5$ when
$\Omega_{M}(t_0)$ reaches its minimum value of zero, a situation in
which $\Omega_{k}(t_0)$ has then risen to one (viz. $k$ now negative),
the deceleration parameter has dropped to zero, and the luminosity
function can be written in the closed form 
\begin{equation} 
d_L=\frac{c}{H(t_0)}\left(z+\frac{z^2}{2}\right)~~.
\label{137}
\end{equation}
Since this particular model is very close to a universe with just
regular luminous matter ($\Omega_{M}(t_0)=0.01$) and no other
components, we see that while this model still gives a $\chi^2$
which is somewhat larger than the number of independent degrees of
freedom, the model does not fare all that badly, suggesting that it
may have captured some of the essence of the Hubble plot
data. In fact paradoxically, a pure luminous model with
$\Omega_{M}(t_0)=0.01$, $\Omega_{k}(t_0)=0.99$ actually does better with
the supernovae data than a $k<0$ luminous plus dark matter model with
$\Omega_{M}(t_0)=0.3$, $\Omega_{k}(t_0)=0.7$.\footnote{Of course both of
these models will still have to be excluded since neither of them can
account for the uniformity of the cosmic microwave background, and both
of them suffer from the flatness problem.} Within
a pressureless $\Omega_{M}(t_0)$, $\Omega_{k}(t_0)$ model, the only
way to reduce the $\chi^2$ even lower than its $\Omega_{M}(t_0)=0$,
$\Omega_{k}(t_0)=1$ value is to make either $G$ or $\rho_m$ negative --
not that one of course can do so in the model, though it does show the
needed trend, with the data favoring an even more reduced deceleration
parameter, viz. one which would actually be negative, a trend which
having large amounts of dark matter actually works against. As it stands
then, we see that the supernovae data exclude the only value of
$\Omega_{M}(t_0)$ for which the Friedmann evolution equation has so far
been shown to not have a fine-tuning problem (viz. $\Omega_{M}(t_0)=1$),
and leaves us with a universe which has even more cosmic repulsion
than that achievable by the smallest permissible $\Omega_{M}(t_0)$. We
stress this point now to show that even before invoking the cosmological
constant (to get some additional cosmic repulsion akin to that which it
provided in the original Einstein static universe discussed earlier),
once $\Omega_{M}(t_0)$ is not equal to one we are right back in the
flatness problem, with the Friedmann equation again requiring
fine-tuning. Thus even while the introduction of the cosmological
constant will also entail fine-tuning, a fine-tuning of its value down
from that anticipated from fundamental physics, this particular
fine-tuning is quite distinct from that associated with the Friedmann
cosmic evolution equation, with the standard cosmological evolution
equation already having fine-tuning problems even without a
cosmological constant, fine-tuning problems which do not disappear
following its introduction.

As regards the possible addition of a cosmological constant to the
model, an addition in which  Eq. (\ref{130}) is then replaced by 
\begin{eqnarray}
\dot{R}^2+kc^2&=&\dot{R}^2[\Omega_M(t)+\Omega_{\Lambda}(t)]~~,~~
\Omega_M(t)+\Omega_{\Lambda}(t)+\Omega_k(t)=1~~,
\nonumber \\
~~q(t)&=&\frac{1}{2}\left(1+\frac{3p_m}{\rho_m}\right)\Omega_M(t)
-\Omega_{\Lambda}(t)~~,
\label{138}
\end{eqnarray}
we now see that a positive $\Omega_{\Lambda}(t)$ will nicely serve
to reduce $q(t)$ just as we would want. To get a sense of how big
we would need the $\Omega_{\Lambda}(t)$ contribution to be, we note that
for a pure $\Omega_{\Lambda}(t)=1$, $\Omega_{M}(t)=0$,
$\Omega_{k}(t)=0$ universe the luminosity function can be
written in the closed form
\begin{equation} 
d_L=\frac{c}{H(t_0)}(z+z^2)~~,
\label{139}
\end{equation}
with it leading to $\chi^2=75.8$ for the same 54 supernovae data points.
While this value is still too large to be acceptable, it is closer
to the number of degrees of freedom than the $\chi^2$ associated
with a pure $(\Omega_{M}(t),~\Omega_{\Lambda}(t),~ 
\Omega_{k}(t))=(1,0,0)$ universe. Hence within the family of universes
with $\Omega_{k}(t_0)=0$, we can anticipate a best fit which is
weighted more to $\Omega_{\Lambda}(t)$ than to $\Omega_{M}(t)$, with
the values $\Omega_{M}(t_0)=0.3$, $\Omega_{\Lambda}(t_0)=0.7$ 
being found to yield $\chi^2=57.7$ and the explicit data fit displayed
as the lower curve in Fig. 2. With these values of
$\Omega_{M}(t_0)$ and $\Omega_{\Lambda}(t_0)$ the deceleration
parameter given by Eq. (\ref{138}) evaluates to $q(t_0)=-0.55$, and is
thus seen to be quite negative.

\subsection{The freedom inherent in fitting the current supernovae
data}

To get a sense of the fitting freedom inherent in the supernovae data,
it is instructive to look not just at the curvature-free
$\Omega_{k}(t)=0$ family of solutions, but also at a particular class
of matter-free $\Omega_{M}(t)=0$ solutions, viz. the $\alpha>0$,
$k<0$ ones as given in Eqs. (\ref{133}) and (\ref{135}) with
$R(t,\alpha>0,k<0)=(-k/\alpha)^{1/2}{\rm sinh}(\alpha^{1/2}ct)$ and
$\Omega_{\Lambda}(t,\alpha>0,k<0)=-q(t,\alpha>0,k<0)={\rm
tanh}^2(\alpha^{1/2}ct)$. This particular family of solutions is one
which is accelerating in all epochs [$q(t,\alpha>0,k<0)$ is negative at
all $t$ for any positive $\alpha$], and on parameterizing
$\alpha$ in terms of the current era deceleration parameter
$q(t_0)=q_0$, its luminosity function is given
\cite{Mannheim2001,Mannheim2003a} as the one parameter 
\begin{equation} 
d_L=-\frac{c}{H(t_0)}\frac{(1+z)^2}{q_0}\left(1-\left[1+q_0-
\frac{q_0}{(1+z)^2}\right]^{1/2}\right)~~.
\label{140}
\end{equation}
Using this luminosity function a best fit of $\chi^2=58.62$ is obtained
with $q_0=-0.37$, a deceleration parameter which again is negative, and
is exhibited as the upper curve in Fig. 2.\footnote{In Fig. 2 we have
labelled this fit as the conformal gravity fit, as it will be
associated with the conformal gravity alternative to standard gravity
which will be discussed below. However, for the moment it need only
be regarded as just one of the possible phenomenological fits
obtainable by treating $\Omega_{M}(t_0)$ and $\Omega_{\Lambda}(t_0)$ as
free parameters in Eq. (\ref{138}).} With the
$(\Omega_{M}(t_0),\Omega_{\Lambda}(t_0),\Omega_{k}(t_0)) =(0.3,0.7,0.0)$
and the $(\Omega_{M}(t_0),\Omega_{\Lambda}(t_0),
\Omega_{k}(t_0))=(0.0,0.37,0.63)$ fits being of equal and
indistinguishable quality, the best fits in the
$(\Omega_{M}(t_0),\Omega_{\Lambda}(t_0))$ plane will to good
approximation thus lie along the line 
\begin{equation} 
\Omega_{\Lambda}(t_0)=1.1\Omega_{M}(t_0)+0.37~~,~~
\Omega_{k}(t_0)=0.63 -2.1\Omega_{M}(t_0)
\label{141}
\end{equation}
which runs through these particular two fits. Along this line we obtain
\begin{equation} 
q(t_0)=-0.6\Omega_{M}(t_0)-0.37~~,
\label{142}
\end{equation}
with the current era deceleration parameter thus being negative
everywhere along the best fit line. With $\Omega_{\Lambda}(t_0)$ being
zero on the best fit line when $\Omega_{M}(t_0)$ is given by
$\Omega_{M}(t_0)=-0.34$, we recover our earlier observation that in pure
matter universes with no cosmological constant, the best fits are
actually obtained with $\Omega_{M}(t_0)$ negative.

\begin{figure} 
\epsfig{file=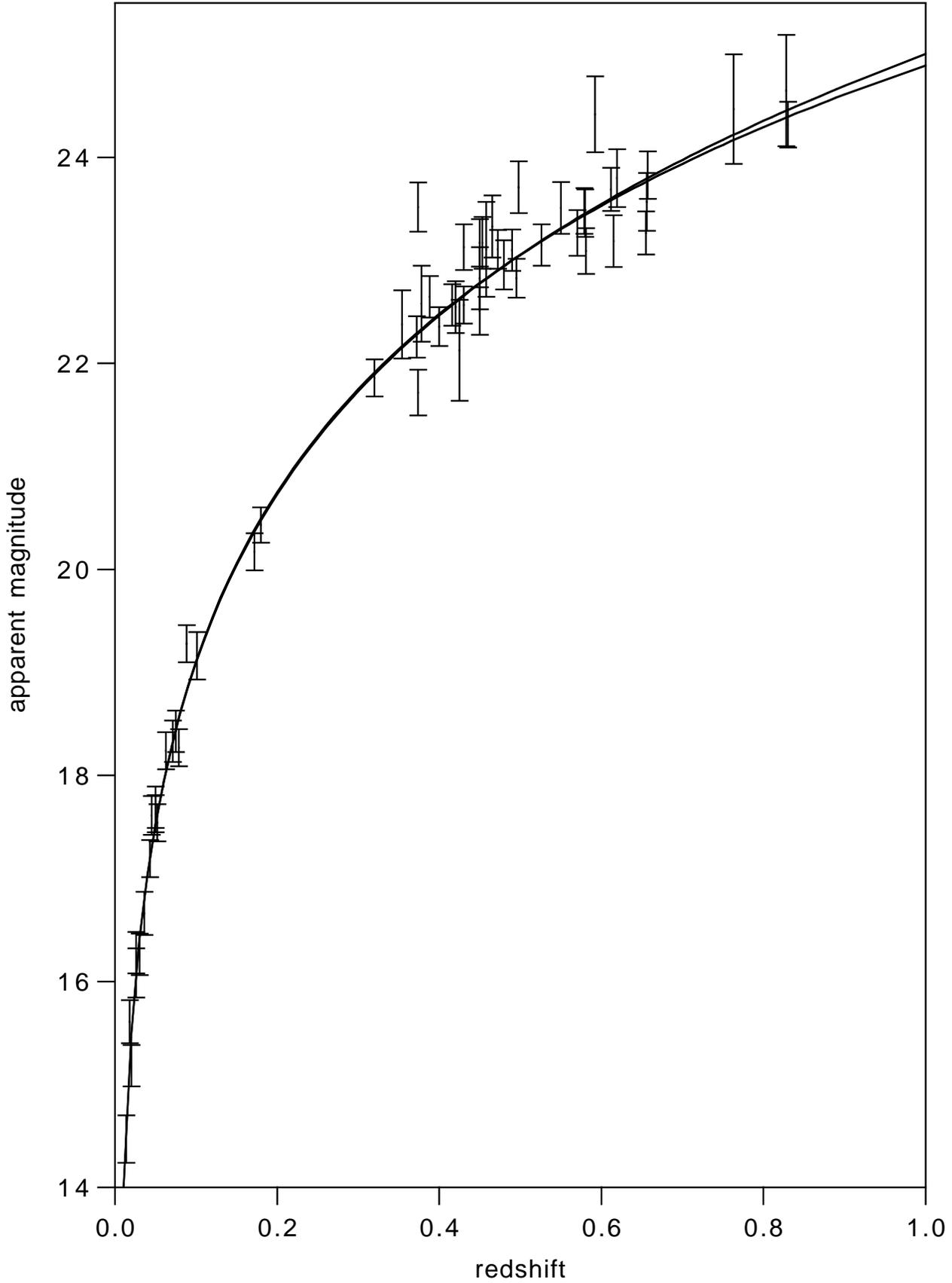,scale=0.9}
\caption{The $q_0=-0.37$ conformal gravity fit (upper curve) and the  
$\Omega_{M}(t_0)=0.3$, $\Omega_{\Lambda}(t_0)=0.7$ standard model fit 
(lower curve) to the $z<1$ supernovae Hubble plot data.}
\label{Fig. (2)}
\end{figure}

\subsection{The cosmic coincidence problem}

From the point of view of the standard model, there will be a problem no
matter where on the best fit line it finds itself, since along the best
fit line $\Omega_{M}(t_0)$ and $\Omega_{\Lambda}(t_0)$ are of the same
order of magnitude. In the presence of a cosmological constant, the
occurrence of an early universe initial singularity in the Friedmann
equation of Eq. (\ref{138}) this time translates into the requirement
that at the initial time the quantities $\Omega_{M}(t_0)$ and
$\Omega_{\Lambda}(t_0)$ must add up to one, viz.
\begin{equation} 
\Omega_{M}(t=0)+\Omega_{\Lambda}(t=0)=1~~.
\label{143}
\end{equation}
However, since $\rho_m$ redshifts while $\Lambda$ does not, in order
for these two quantities to be of the same order today, in the initial
universe $\Omega_{M}(t=0)$ must have been incredibly close to one
(typically to within one part in $10^{60}$ or so) and
$\Omega_{\Lambda}(t=0)$ must have been of order $10^{-60}$.
There is thus again a fine-tuning problem, with the cosmic evolution
Friedmann equation again not being able to naturally reconcile its
initial singularity with the current era values of $\Omega_M(t_0)$ and
$\Omega_{\Lambda}(t_0)$. With the most natural expectation for a
Friedmann universe with both matter and a cosmological constant being
that entirely in the early universe itself $\Omega_{M}(t)$
should go from one to zero and $\Omega_{\Lambda}(t)$ should go from
zero to one, the closeness of
their values some ten billion years later is often referred to as the
cosmic coincidence problem.

While inflation does not solve this cosmic coincidence problem, an early
universe inflationary era can nonetheless still occur (and indeed its
occurrence would still solve the horizon problem), with $\Omega_{k}(t)$
then still being negligible throughout the Robertson-Walker era. Within
the class of best fit solutions to the supernovae data this would then
single out the $\Omega_{M}(t_0)=0.3$, $\Omega_{\Lambda}(t_0)=0.7$ one,
and thus bring inflation into agreement with the $\Omega_{M}(t_0)=0.3$
dark matter cluster estimate. Further support for this
$\Omega_{M}(t_0)=0.3$, $\Omega_{\Lambda}(t_0)=0.7$ realization of
inflation has recently been obtained from study of the temperature
anisotropy of the cosmic microwave background, since such anisotropies
can actually be generated in a period of rapid early universe inflation
in which initially microscopic fluctuations could be amplified to
macroscopic size. The size to which such inflationary fluctuations will
grow at the time of recombination will thus act as a theoretical
standard yardstick which will be imprinted on the cosmic microwave
background, with the apparent size with which it will then appear to us
then reflecting the geometry through which photons have had to travel
to us since recombination. Through use of this technique the cosmic
microwave background anisotropy measurements are then found to have a
best fit in the $(\Omega_{M}(t_0),\Omega_{\Lambda}(t_0))$ plane which
scatters around the line \cite{deBernardis2000,Tegmark2004}
\begin{equation} 
\Omega_{\Lambda}(t_0)=1-\Omega_{M}(t_0)~~,~~
\Omega_{k}(t_0)=0
\label{144}
\end{equation}
with the compatibility of Eq. (\ref{141}) with Eq. (\ref{144})
remarkably bringing us right back to $\Omega_{M}(t_0)=0.3$,
$\Omega_{\Lambda}(t_0)=0.7$. Consequently, these values are now
regarded as being the ones both required of and established by the
standard cosmological theory, with the standard theory thus having
achieved its primary objective of determining the matter content of the
universe. 

However, even while the standard theory has now ostensibly done its
job, and while one should not understate how remarkable it is that there
actually is a specific single set of $\Omega_{M}(t_0)$,
$\Omega_{\Lambda}(t_0)$ values which fits currently available
cosmological data at all,\footnote{While it might be thought that the
lines of Eqs. (\ref{141}) and (\ref{144}) have to cross somewhere,
this is not quite the case since the lines are not actually of
infinite extent, with both of them only describing their respective data
in limited regions, regions which remarkably do in fact overlap.}  the
universe which emerges is still a highly perplexing one which not only
possesses huge amounts of still poorly understood dark matter, it needs
to augment this dark matter with an even more poorly understood
$\Omega_{\Lambda}(t_0)$. However, before looking at the implications of
such a value for $\Omega_{\Lambda}(t_0)$, it is important to note that
there could be an alternative to it within standard gravity, namely the
so-called quintessence model of \cite{Caldwell1998}. The quintessence
model was actually developed prior to the discovery of the accelerating
universe, with its objective at the time being to try to reconcile an
$\Omega_{M}(t_0)$ which would be less than one with inflation in a way
that would not require any early universe fine-tuning of the Friedmann
equation. To this end a quintessence source was added to the dark
matter source, one which would then have an $\Omega_{Q}(t_0)$ which
would be equal to $1-\Omega_{M}(t_0)$ today, and thus be of the same
order as $\Omega_{M}(t_0)$ today, to initially appear to replace the
$\Omega_{M}(t_0)$, $\Omega_{\Lambda}(t_0)$ cosmic coincidence by an
$\Omega_{M}(t_0)$, $\Omega_{Q}(t_0)$ cosmic coincidence instead.
However, it was found possible to construct a tracking potential for
the quintessence fluid which would have the effect of driving
$\Omega_{Q}(t)$ and $\Omega_{M}(t)$ to comparable current era values
today regardless of what early universe initial conditions were
presupposed, to thus render the cosmic evolution equation free of early
universe fine-tuning problems. In order for the quintessence fluid to
be able to achieve this objective, it would have to have an equation of
state in which $p/\rho=w$ is negative, and in order for it to have
escaped visual detection it would need to be non-luminous. Non-luminous
negative pressure fluids such as these are known as dark
energy,\footnote{Since it is their negative pressure which is abnormal
and not their energy density (which remains positive), a possibly
better designation for quintessence fluids would be dark pressure.}
with a cosmological constant being a special case of such fluids, one
with $w=-1$. With quintessence fluids having negative $w$, they can
thus lead to cosmic repulsion, and if $w$ is negative enough can
provide for fitting to the supernovae data of the same quality as the
$\Omega_{M}(t_0)=0.3$, $\Omega_{\Lambda}(t_0)=0.7$ cosmology fit shown
in Fig. 1. While quintessence is thus offered as a candidate solution
to the cosmic coincidence problem, it leaves open the question of what
is to happen to the cosmological constant, with quintessence only being
able to succeed if some mechanism can be found which would naturally
set the cosmological constant to zero. For quintessence models to be 
valid then still requires a solution to the cosmological constant
problem. 

\subsection{The cosmological constant problem}

Indeed the real problem with the cosmological constant is that rather
than provide an $\Omega_{\Lambda}(t_0)$ as small as $0.7$ or even one
equal to zero, the a priori expectation for $\Omega_{\Lambda}(t_0)$ is
that it will be huge, being as large as $10^{120}$ if fixed by the
quantum gravity Planck scale, and being as large as $10^{60}$ or so if
fixed by the electroweak phase transition scale, with use of either of
these values leading to absolute phenomenological disaster for the
standard theory. Moreover, even if $\Omega_{\Lambda}(t)$ is quenched to
zero at the end of the inflationary era, as the universe then cools
down it will go through phase transitions such as the electroweak one
at around $T_V=10^{15}$ degrees, at which point a brand new cosmological
constant will be generated by the change in free energy due to the phase
transition, with the difference between the free energy just above the
phase transition and its value at current temperatures way below it
being of order $\sigma T_V^4$ where $\sigma$ is Stefan's
constant. Not only is such a $\sigma T_V^4$ huge, since the free energy
is lowered in a phase transition, one would even expect it to produce a
negative contribution to $\Lambda$, with $\Lambda$ then differing both
in sign and magnitude from the required scale of the order of few
degrees which is associated with the positive
$\Omega_{\Lambda}(t_0)=0.7$. Apart from the electroweak scale, there
might even be a supersymmetry breaking scale as well, and while
supersymmetry could provide some possible dark matter wimp candidates,
the very same mechanism which would put their masses into at least the
TeV region (the current lower mass bound associated with their lack of
detection to date) would at the same time generate an even bigger
contribution to the cosmological constant than that associated with
electroweak symmetry breaking itself. Thus mechanisms which will
generate the cold dark matter particles needed for
$\Omega_{M}(t_0)=0.3$ will also generate huge contributions to
$\Omega_{\Lambda}(t_0)$ in a dynamics which would leave us nowhere near
the cosmic coincidence values at all. This then is the cosmological
constant problem.

At the present time within the standard model no convincing resolution
of the cosmological constant problem has been identified, and while
there are some suggestions, none of them are anywhere near being at the
stage where they could be falsified. Enormous efforts have been made to
find a dynamics which would naturally quench $\Omega_{\Lambda}(t_0)$
down from either $10^{60}$ or $10^{120}$ to a current value of $0.7$, so
far to no avail. Additionally, one can consider the possibility that in
addition to the overall dynamical cosmological constant which is
generated on the right-hand side of the Einstein equations as the
matter density goes through its various phase transitions, there
could also exist an a priori fundamental cosmological constant on the
left-hand side, one which will then almost, but not quite cancel the
dynamical one so that the current era observer then sees a net
$\Omega_{\Lambda}(t_0)=0.7$. Such a proposal requires that 
a fundamental cosmological constant exist as part and parcel of the
fundamental gravitational equations, that it be included in
gravitational theory in all epochs, and that it be preset with just the
value needed to lead to a net $\Omega_{\Lambda}(t_0)=0.7$ today. It is
not clear how one might test such an idea, with such a cancellation
having to be regarded as being a cosmic coincidence itself.

Another proposed mechanism is the so-called anthropic principle which
takes the view that there exists an infinite number of universes in
which $\Lambda$ takes on all possible values, with observers such as
ourselves only being able to exist and observe the universe in those
particular universes in which the value of $\Lambda$ permits the
emergence of intelligent beings. In a sense such a viewpoint is
actually a denial of physics, saying that when we can explain something
we do, and when we cannot, we appeal to the anthropic
principle.\footnote{A possibly more accurate name might be the
anthropic lack of principle, with it not being clear which phenomena
are to be attributed to the anthropic principle and which not.} From a
more technical viewpoint it may be the case that even with an infinite
number of universes, the cosmological constant could be Planck density
in all of them, and that while the Planck length itself might vary from
universe to universe, the relation of $\Lambda$ to it might not.
Additionally, it would be an extraordinary coincidence if the only
possible allowed range of $\Lambda$ needed for intelligent beings would
precisely be within the range needed to solve the early universe
Friedmann equation fine-tuning problem. Rather, one might expect a
somewhat broader range for $\Lambda$, and with the Friedmann fine-tuning
required range (viz. to one part in $10^{60}$) then being a very small
fraction of it, it is far more likely that the anthropic principle
would put us outside the Friedmann required range rather than in it.
Finally, if the anthropic principle is to be a part of physics, then
there has to be some falsifiable prediction that it makes which would
enable us to test it.

\subsection{Testing dark energy beyond a redshift of one}

However, regardless of any of these theoretical concerns, there is
actually a direct, model-independent test that one can make to see
whether it really is true that the real world $c\Lambda$ is indeed equal
to $0.7\rho_c\sim 0.6\times 10^{-8}$ergs/c.c. (viz.
$\alpha=\Omega_{\Lambda}H^2/c^2=8\pi G\Lambda/3c^3\sim 0.4\times
10^{-56}$/cm$^2$), this being the value which would make
$\Omega_{\Lambda}(t_0)$ be equal to $0.7$ today. Specifically, since
$\rho_m$ redshifts while $\Lambda$ does not,
$\rho_m$ would then dominate over $\Lambda$ at earlier times, with the
universe then having to be in a decelerating phase rather than an
accelerating one. For the given values of $\Omega_{M}(t_0)$ and
$\Omega_{\Lambda}(t_0)$ such a switch between acceleration and
deceleration would occur before $z=1$, and so a monitoring of the
Hubble plot above $z=1$ would then test for such a switch over. As such,
the very existence of a switch over near $z=1$ is a reflection of the
cosmic coincidence, with initial conditions having to be fine-tuned in
the early universe so that the switch from $\rho_m$-dominated
deceleration to $\Lambda$-dominated acceleration does not take place in
the early universe itself but is delayed for no less than ten billion
years. In Fig. 3 we have therefore plotted the higher $z$ Hubble plot
required of an $\Omega_{M}(t_0)=0.3$, $\Omega_{\Lambda}(t_0)=0.7$
universe, and if the standard model numbers are right, the luminosity
function must be found to follow the indicated curve.

To get a sense of the level of precision which is  required to make this
test, in Fig. 3 we have also plotted the expectations of the
$(\Omega_{M}(t_0),~\Omega_{\Lambda}(t_0))=(0,0)$ universe with
$q_0=0$ and the pure
$\Lambda$ $(\Omega_{M}(t_0),~\Omega_{\Lambda}(t_0))=(0,0.37)$ universe
with $q_0=-0.37$ which we discussed earlier, since the first of
these two universes represents the coasting borderline case between
acceleration and deceleration, while the latter represents the best fit
permanently accelerating pure $\Lambda$ universe.\footnote{For the
moment our interest in these particular cosmologies is only in having
cosmologies which are not decelerating above $z=1$ so as to provide a
contrast with the standard cosmology in that region. However,
shortly we shall see that such non-decelerating cosmologies are
of interest in their own right, as they are relevant to the
alternate conformal gravity solution to the cosmological constant
problem which we present below.} Discriminating between the three cases
(accelerating/decelerating, coasting, and permanently accelerating)
should eventually be achievable since at
$z=2$ these three typical universes respectively yield apparent
magnitudes
$m=26.75$, $27.04$ and $27.17$, while at $z=5$ they yield
$m=29.14$, $30.25$ and $30.40$, and should thus enable us to determine
what sort of phase the universe might be in at redshifts above one. 

\begin{figure}
\epsfig{file=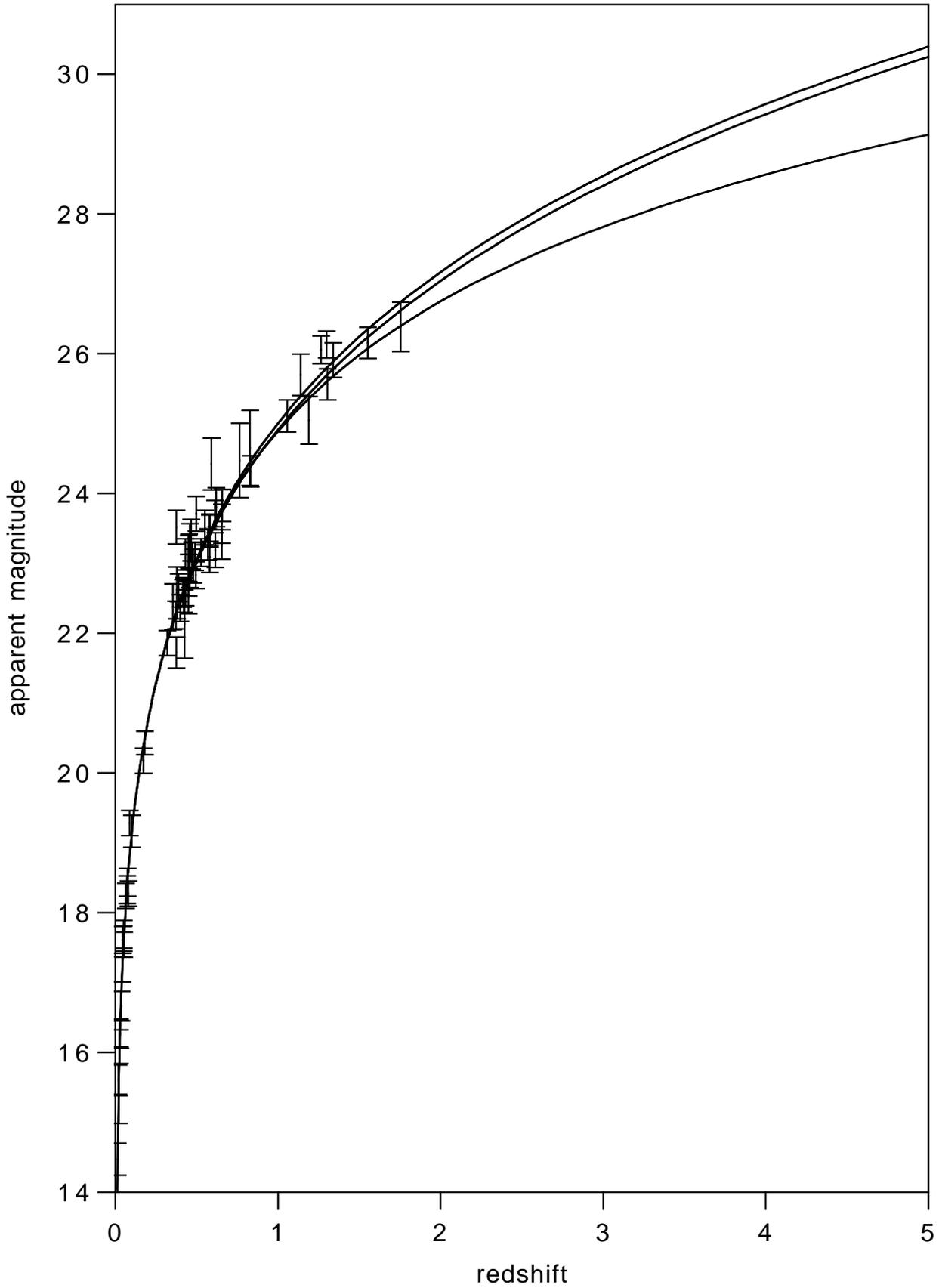,scale=0.9}
\caption{Hubble plot expectations for $q_0=-0.37$ (highest curve) and
$q_0=0$ (middle curve) conformal gravity and for
$\Omega_{M}(t_0)=0.3$,
$\Omega_{\Lambda}(t_0)=0.7$ standard gravity (lowest curve).}
\label{Fig. (3)}
\end{figure}

Now while full discrimination between these various options would
require obtaining as many data points above (ideally well
above) $z=1$ as there are below $z=1$, it is of interest to note
that a few $z>1$ data points have recently been obtained
\cite{Riess2001,Riess2004} including one (SN 1997ff) as high as
$z=1.75$. Of these new data points nine have been identified as being
particularly reliable and so we have extended our fits to
include them. In order to perform the fits, we note that in
\cite{Perlmutter1998} the 54 $z<1$ data points are presented as
apparent magnitudes $m$, while in \cite{Riess2004} the nine new $z>1$
data points are presented as distance moduli $m-M$. With these two
methods of characterizing the data being related via
$m=M+25+5{\rm log}_{10}d_L$ where $d_L$ is the luminosity distance as
measured in megaparsecs, in the fitting of the 54 $z<1$ data points of
\cite{Perlmutter1998} $M$ is treated as a free parameter, with the
notion that the supernovae are indeed standard candles then being
confirmed by the fact that in the very good fits of Fig. 2 a common
value for $M$ actually is obtained. On now fitting the combined 63
(=54+9) data points in fits which allow $M$ to again vary for the 54
points in the sample, the standard
$(\Omega_{M}(t_0),~\Omega_{\Lambda}(t_0))=(0.3,0.7)$ universe is found
to yield $\chi^2=74.5$ and $M=-19.15$, with the best
$\Omega_k(t_0)=0$ universe fit being found to be the slightly lower
mass (and thus more cosmically repulsive)
$(\Omega_{M}(t_0),~\Omega_{\Lambda}(t_0))=(0.21,0.79)$ universe with
$\chi^2=69.9$ and $M=-19.19$. For comparison, the pure
$\Lambda$ $(\Omega_{M}(t_0),~\Omega_{\Lambda}(t_0))=(0,0.37)$
universe with $q_0=-0.37$ yields $\chi^2=70.7$ and
$M=-19.14$, while the pure curvature
$(\Omega_{M}(t_0),~\Omega_{\Lambda}(t_0))=(0,0)$ universe with
$q_0=0$ yields $\chi^2=78.4$ and $M=-19.06$. As we see then, while 
a standard $\Omega_{M}(t_0)>0$,
$\Omega_{\Lambda}(t_0)=1-\Omega_{M}(t_0)$ cosmology with $q(z>1)>0$ and
an 
$\Omega_{M}(t_0)=0$, $\Omega_{\Lambda}(t_0)=0.37$  cosmology with
$q(z>1)<0$ both outperform the
$\Omega_{M}(t_0)=0$, $\Omega_{\Lambda}(t_0)=0$ cosmology with
$q(z>1)=0$ in the
$z>1$ region, statistically the $q(z>1)>0$ and $q(z>1)<0$
cosmologies are  equivalent to each other, with the current $z>1$ region
data being just as well fitted by a model which is accelerating above
$z=1$ as by one which is decelerating above it. Consequently, for the
moment it is not possible to determine whether the universe is in an
accelerating or a decelerating phase above
$z=1$, with both options being compatible with currently available data.
To be able to display the nine $z>1$ data points of \cite{Riess2004} on
the apparent magnitude plot given in  Fig. 3, we have taken the
fitted value of $M$ ($=-19.14$) and used it to extract values for $m$
for the nine points from their reported $m-M$ distance moduli, with
it being these inferred $m$ values which are then displayed in the
figure. While the
$\chi^2$ fitting itself is not at all sensitive to the value used for
$H(t_0)$ which is required for the luminosity distance $d_L$, the
extraction of an actual value for $M$ itself from the fitting is. For
instance with
$H(t_0)=72$ km/sec/Mpc, in megaparsec units the quantity $5{\rm
log}_{10}(c/H(t_0))$ evaluates to 18.099, while for $H(t_0)=65$
km/sec/Mpc it
evaluates to 18.321, a quite substantial difference of $0.222$. With an
increasing value for the Hubble parameter leading to a decreasing
$5{\rm log}_{10}(c/H(t_0))$, an increasing value for the Hubble
parameter thus also leads to an increasing derived $M$. In its turn an
increased $M$ leads to an increased (dimmer) value for $m$ as extracted
from a given distance modulus. On making this extraction, as we see from
Fig. 3, the extracted $z>1$ $m$ values then precisely straddle the three
indicated theoretical curves, to thereby prevent current data from
being able to differentiate between acceleration and deceleration in
the $z>1$ region.\footnote{The $z>1$ data are however able to exclude
the possibility that dust or cosmic evolution may be causing a dimming
of the $z<1$ supernovae, since both of those effects would dim the
$z>1$ supernovae even more, an effect which as noted in \cite{Riess2004}
the $z>1$ data now exclude, with the behavior of the
Hubble plot thus indeed being of cosmological origin.} With the SN
1997ff supernova at
$z=1.75$ being known to be lensed by two foreground galaxies which lie
at $z=0.5$ \cite{Benitez2002}, this particular supernova may actually be
dimmer (larger $m$) than it appears. Filling in the
$z>1$ Hubble plot with a large enough sample of data points and
allowing for systematic effects such as lensing should thus be
very informative.

Apart from the use of type Ia supernovae to explore the $z>1$ region, a
few other techniques have also been developed, with Daly and coworkers
having used the very powerful extended FRII radio sources as standard
yardsticks \cite{Guerra1998,Guerra2000} and Schaefer having explored the
use of gamma ray bursters as standard candles \cite{Schaefer2003}. For
the FRII approach some 20 radio sources which go out to $z=2$ have
already been fully analyzed (11 being above $z=1$), with there being
some more which go out to $z=3$ or so, while the gamma ray burster
technique will be able to go out to $z=4.5$. Global fits to the 54
$z<1$ supernovae studied above and the 20 FRII radio galaxies have
been made, with the
$(\Omega_{M}(t_0),~\Omega_{\Lambda}(t_0))=(0.3,0.7)$ standard cosmology
giving $\chi^2=74.4$ for the 74 total data points and an
$(\Omega_{M}(t_0),~\Omega_{\Lambda}(t_0))=(0,0.38)$ cosmology giving
$\chi^2=74.1$
\cite{Mannheim2003b}. The available radio source data are thus seen to
be just as compatible with both decelerating and accelerating universes
in the $z>1$ region as the supernovae data are. Further exploration of
the
$z>1$ Hubble plot using all of the available techniques will thus be
necessary in order to determine how the universe behaves in the $z>1$
region and ascertain whether in fact $\Omega_{\Lambda}(t_0)$  really is
$0.7$ in the real world. 

Having now described and diagnosed the dark matter and dark energy
problems, we turn next to an analysis of some alternatives to the
standard model in which these problems are resolved. And since the
cosmological dark energy problem is an intrinsically
general-relativistic one, to seek alternatives to dark matter and dark
energy will oblige us to have to consider alternatives to Einstein
gravity itself.

\section{Alternatives to Einstein gravity}

\subsection{Pure metric theories with additional curvature dependent
terms}

As we had noted earlier, while it is sufficient to use standard
Einstein gravity to obtain the standard Schwarzschild metric
phenomenology for the solar system, is not in fact necessary. Thus in
seeking alternatives to Einstein gravity, as well as look for
generalizations of it which contain the Einstein equations in an
appropriate limit, one can also look for generalizations which are able
to recover the Schwarzschild phenomenology while never reducing to the
Einstein theory in any limit at all, with the more general rule thus
being to generalize the Schwarzschild solution rather than the Einstein
equations themselves. We shall thus look at both kinds of
generalization (either reducing to the Einstein equations or bypassing
them altogether) and after detailing possible options, shall then look
for a fundamental principle which would enable us to unambiguously
select just one option from amongst them. 

Within the framework of general covariant pure metric theories of
gravity, the most straightforward generalization of the standard
Einstein theory is to augment the Einstein-Hilbert action of Eq.
(\ref{103}) with additional general coordinate invariant pure
metric terms, additional terms the smallness of whose coefficients or
the specific structure of which (vanishing in Ricci flat geometries)
would cause them to make negligible contributions in the solar system.
The choice of possible additional terms is actually unlimited, since it
not only encompasses appropriate contractions of arbitrarily high
powers of the Riemann tensor (such as the fourth order actions of Eqs.
(\ref{105}) and (\ref{106})),\footnote{In a quantization of the
standard Einstein theory an infinite set of such higher order
terms is generated via radiative correction counter terms -- though
these terms then come with Planck scale coefficients which are indeed
negligible compared to the $\hbar=0$ Einstein term itself.} provided
candidate additions are able to avoid violating locality, in principle
such additions need not even be expressible as a power series
expansion of Riemann tensor terms at all.\footnote{The addition to the
Einstein-Hilbert action of terms which behave as
$1/R^{\alpha}_{\phantom{\alpha}\alpha}$ has been considered by
\cite{Carroll2005} in an attempt to explain the acceleration of the
universe as a gravitational effect rather than one which is generated
by dark energy.} Moreover, within the class of additions to the
Einstein-Hilbert action, one must also include the addition that
Einstein himself made, viz. a cosmological constant term, as well.
Without a fundamental guiding principle almost nothing can be said
either in favor or against any candidate such addition, with this
freedom being due to the fact that it was not an appeal to a
fundamental principle which led to the specific choice of the
Einstein-Hilbert action in the first place. While it is sometimes
stated that the choice of the Einstein-Hilbert action can be justified
on grounds of simplicity, simplicity itself is not a law of nature, and
the simplicity of the Einstein approach to gravity is not so much in
the simplicity of its equations (or the ease with which they might be
solved) but in the simplicity of its concepts. For Einstein, the
simplicity of the concepts and the intrinsic elegance and beauty of the
theory were paramount, and one could argue that in adding in dark
matter and dark energy in the ad hoc way that is currently being done,
one gives up much of the inner beauty of the theory.

\subsection{Additional fields}  

Within the framework of general covariant theories of gravity, the next
most straightforward generalization of Einstein gravity is to introduce
additional macroscopic gravitational fields to accompany the metric
tensor itself, with the most common choice being additional scalar
fields. In and of itself, the addition of covariantly coupled scalar
fields does not imperil the validity of the equivalence principle,
since instead of coupling through the $I_T$ action of Eq. (\ref{6}), a
test particle will now couple though the $I_T$ action as augmented by
the $I_T^{(4)}$ action of Eq. (\ref{11}), to lead to a particle
equation of motion 
\begin{equation}
\left(mc+\hat{\kappa}S\right)
\frac{D^2x^{\lambda} }{ D\tau^2}=
\left(mc+\hat{\kappa}S\right)\left(
\frac{d^2x^{\lambda} }{ d\tau^2} +\Gamma^{\lambda}_{\mu \nu} 
\frac{dx^{\mu}}{d\tau}\frac{dx^{\nu } }{ d\tau} \right) 
= -\hat{\kappa}S_{;\beta} \left( g^{\lambda
\beta}+
\frac{dx^{\lambda}}{d\tau}                                                      
\frac{dx^{\beta}}{d\tau}\right)
\label{145}
\end{equation}
which still involves the covariant 4-acceleration
$D^2x^{\lambda}/D\tau^2$. While Eq. (\ref{145}) and its Eq.
(\ref{12}) antecedent can lead to departures from geodesic motion (an
issue we return to below when we consider covariant generalizations of
MOND), it still respects the equivalence principle requirement that the
Christoffel symbol term and the ordinary acceleration term
$d^2x^{\lambda}/d\tau^2$ always appear in exactly the combination
indicated. Within the class of scalar tensor theories the most
venerable is the Brans-Dicke theory with scalar tensor sector action
\cite{Brans1961}
\begin{equation}
I_{BD}=-\int d^4x(-g)^{1/2}\left(S
R^{\alpha}_{\phantom{\alpha}\alpha}
-w\frac{S_{;\mu}S^{;\mu}}{S}\right)
\label{146}
\end{equation}
where $S$ is the scalar field and $w$ is a
constant.\footnote{With its $1/S$ dependence 
the scalar field kinetic energy is not expressible as a power series
expansion in $S$.} In this theory there is no fundamental
$G$, with $G$ instead being induced as an effective gravitational
constant in solutions to the theory in which $S$ is non-zero. With $S$
being able to take different values in different epochs, the
Brans-Dicke theory therefore allows for a varying effective $G$, and
with it then being this epoch-dependent effective $G$ which will then
normalize $\Omega_M$ in each epoch, while not solving the Friedmann
equation fine-tuning problem the Brans-Dicke theory does at least allow
one to reformulate it as a need to find a dynamics in which $S$ could
naturally change by 60 orders of magnitude from the early universe
until today. While such a mechanism is a general one which actually
holds for any theory in which an effective $G$ is induced by a scalar
field, it turns out that solar system constraints on the Brans Dicke
theory itself are so tight, that there is no room to choose its
parameter $w$ in a way which could lead to any potential departure from
the standard theory. While this is thus a shortcoming of the
Brans-Dicke theory, it is not actually a generic shortcoming of all
scalar tensor theories of gravity, only a shortcoming of the scalar
tensor theory which is based on the particular action of Eq.
(\ref{146}), with every individual candidate scalar tensor theory of
gravity needing to be evaluated on a case by case basis.

\subsection{Additional spacetime dimensions}

The third kind of generalization of the standard Einstein theory is
one which has now become extremely popular and widespread, namely
increasing the number of spacetime dimensions. Such ideas originated
with Kaluza and Klein in the nineteen twenties who considered the
possible existence of a fifth dimension, with the concept of extra
dimensions remerging much later following the development of 
superstring theories. Characteristic of these higher dimensional
approaches is not just the existence of extra dimensions, but also of
extra fields. The Kaluza-Klein approach is based on a 5-dimensional
metric, a 15-component quantity whose 4-dimensional decomposition
consists of a 10-component rank two tensor graviton, a 4-component
vector field, and a 1-component scalar field, while the string theory
approach typically involves scalar dilaton fields and not just rank two
tensor gravitons.\footnote{String theory typically also generates
Planck scale higher order Riemann tensor action terms.} To account for
the apparent absence of any dimensions beyond four, it had generally
been thought that all but four of the higher dimensions would be
compactified down to Planck length size. More recently though it was
recognized by Randall and Sundrum \cite{Randall1999a,Randall1999b} that
if we lived on a brane (viz. membrane) which was embedded in a higher
dimensional anti-de Sitter bulk, the anti de Sitter bulk could (for at
least some choices of brane matter fields) inhibit the propagation of
gravitational signals into it no matter how large in size it might be. 

While these higher dimensional theories all contain a massless graviton,
it does not follow that a 4-dimensional observer confined to the brane
will then see standard 4-dimensional Einstein gravity on it, with the
gravity on the brane being modified by the very embedding itself, a
modification which actually opens up a possible new way to address the
dark matter and dark energy problems. In order to see how such
modifications come about, we note first that before imposing any
dynamics whatsoever, for the Gaussian normal metric
\begin{equation}
ds^2=g_{MN}dx^Mdx^N=dw^2+q_{\mu\nu}dx^{\mu}dx^{\nu}
\label{147}
\end{equation}
considered in the typical 5-dimensional bulk case,\footnote{Here we take
$w$ to denote the fifth coordinate, $M$ and $N$ to denote 5-dimensional
coordinates and $\mu$ and $\nu$ to denote the usual four coordinates on
the brane, and have chosen the geometry so that the brane is located at
$w=0$ and the normal to the brane is given by $n_M=(0,0,0,0,1)$. In this
coordinate system even though the components of the induced metric
$q_{\mu\nu}$ only range over the 4-space indices, $q_{\mu\nu}$ is
nonetheless a function of all five coordinates.}  use only
of the purely geometric Gauss embedding formula entails that the induced
Einstein tensor $^{(4)}G_{\mu\nu}$ on the brane (viz. the one calculated
with the induced metric $q_{\mu\nu}$ seen by an observer at $w=0$) is
given by\footnote{See e.g. \cite{Mannheim2005a} where a more detailed
treatment of the embedding issues and formulae discussed here may be
found.}
\begin{eqnarray}
^{(4)}G_{\mu \nu}&=&{2\over 3}
G_{\alpha\beta}q^{\alpha}_{\phantom{\alpha}\mu}
q^{\beta}_{\phantom{\beta}\nu}+ {2\over 3}
G_{55}q_{\mu\nu}-{1 \over 6}G^{A}_{\phantom{A}A}q_{\mu\nu}-E_{\mu\nu}
\nonumber \\
&&-K^{\alpha}_{\phantom{\alpha}\alpha}K_{\mu
\nu}+K^{\alpha}_{\phantom{\alpha}\mu}K_{\alpha\nu}+ {1 \over
2}(K^{\alpha}_{\phantom{\alpha}\alpha})^2q_{\mu\nu}
-{1 \over
2}K_{\alpha\beta}K^{\alpha\beta}q_{\mu\nu}
\label{148}
\end{eqnarray}   
where $G_{MN}$ is the 5-dimensional Einstein tensor (viz. the one 
calculated with the full 5-dimensional metric $g_{MN}$), $E_{\mu\nu}$
is the particular projection 
\begin{equation}
E_{\mu\nu}=q^{\beta}_{\phantom{\beta}\mu}
q^{\delta}_{\phantom{\delta}\nu}C^{5}_{\phantom{5}\beta
5\delta}
\label{149}
\end{equation}   
of the 5-dimensional Weyl tensor $C_{ABCD}$, and $K_{\mu\nu}$ (the
$(\mu\nu)$ component of the extrinsic curvature
$K_{MN}=(g^A_{\phantom{A}M}-n^An_M)(g^B_{\phantom{B}N}-n^Bn_N)n_{B;A}$
of the brane) is given by
\begin{equation}
K_{\mu\nu}=q^{\alpha}_{\phantom{\alpha}\mu}
q^{\beta}_{\phantom{\beta}\nu}n_{\beta;\alpha}~~.
\label{150}
\end{equation}   
Similarly, in terms of the acceleration vector $a_N=n^An_{N;A}$ which
lies in the brane ($n^Na_N=0$ when $n^Nn_N=1$), one can derive a second
purely geometric relation  
\begin{equation}
n^CK_{\mu\nu;C}-R_{BD}q^{B}_{\phantom{B}\mu}q^{D}_{\phantom{D}\nu}
=-^{(4)}R_{\mu \nu}
+{1 \over 2}(a_{\mu;\nu}+a_{\nu;\mu})
-K^{A}_{\phantom{A}A}K_{\mu\nu}
-a_{\mu}a_{\nu}~~.
\label{151}
\end{equation} 
Consequently, no matter what form we may assume for the brane world
energy-momentum tensor $T_{MN}$, we immediately see that even if we 
do take $G_{MN}$ to obey the Einstein equations in the higher
dimensional bulk space, viz.
\begin{equation}
G_{MN}=-\kappa_5^2T_{MN}~~,
\label{152}
\end{equation} 
it does not at all follow that $^{(4)}G_{\mu \nu}$ has to then obey
the standard Einstein equations in the 4-space. 

To see what equation $^{(4)}G_{\mu \nu}$ does in fact obey, we note that
if we take the bulk to possess a 5-dimensional cosmological constant
$\Lambda_5$ and the brane to possess some energy-momentum tensor
$\tau_{\mu\nu}$ which is confined to it, viz.
\begin{equation}
T_{MN}=-\Lambda_5g_{MN}
+\tau_{\mu\nu}\delta_M^{\mu}\delta_N^{\nu}\delta(w)~~,
\label{153}
\end{equation} 
then since every term on the right-hand side of Eq. (\ref{151}) is
continuous at the brane\footnote{Both $^{(4)}R_{\mu \nu}$ and $a_{\mu}$
lie in the plane of the brane, and even though the extrinsic curvature
has a step function discontinuity at the brane, terms
quadratic in it do not.} it follows that the discontinuities on the
left-hand side must cancel,\footnote{Since $K_{\mu\nu}$ is a step
function, its covariant derivative is a delta function.} so that on
crossing the brane we obtain
\begin{equation}
\int_{0^-}^{0^+}dw \left[\partial_{w}K_{\mu\nu}
+\kappa_5^2\left(\tau_{\mu\nu}
-\frac{1}{3}g_{\mu\nu}
\tau^{\alpha}_{\phantom{\alpha}\alpha}\right)\delta(w)\right] =0~~,
\label{154}
\end{equation} 
with the fully covariant Israel junction conditions \cite{Israel1966} 
\begin{equation}
K_{\mu\nu}(w=0^+)-K_{\mu\nu}(w=0^-)=-\kappa_5^2\left[\tau_{\mu\nu}-{1
\over 3}q_{\mu\nu}(w=0)\tau^{\alpha}_{\phantom{\alpha}\alpha}\right]
\label{155}
\end{equation} 
then emerging.\footnote{In a $D$ dimensional bulk and a $D$-space brane
the factor $1/3$ is replaced by $1/(D-2)$.} For the brane world set-up a
$Z_2$ symmetry about the brane is imposed so that the extrinsic
curvatures on the two sides of the brane are related by
$K_{\mu\nu}(w=0^+)=-K_{\mu\nu}(w=0^-)$, with the extrinsic curvature at
the brane then being given by
\begin{equation}
K_{\mu\nu}(w=0^+)=-K_{\mu\nu}(w=0^-)=-{\kappa_5^2 \over
2}\left[\tau_{\mu\nu}-{1 \over
3}q_{\mu\nu}(w=0)\tau^{\alpha}_{\phantom{\alpha}\alpha}\right]~~.
\label{156}
\end{equation} 
Given this discontinuity and the original 5-dimensional Einstein
equations of Eq. (\ref{152}) which fix the value of $G_{MN}$ at the
brane, the extraction of the continuous piece of Eq. (\ref{148}) is then
direct and yields
\cite{Shiromizu2000}
\begin{equation}
^{(4)}G_{\mu \nu}={1\over 2}\kappa_5^2\Lambda_5q_{\mu\nu}(w=0)
-\kappa_5^4 \Pi_{\mu\nu}
-\bar{E}_{\mu\nu}(w=0)
\label{157}
\end{equation}   
where 
\begin{equation}
\bar{E}_{\mu\nu}(w=0)={1 \over
2}[E_{\mu\nu}(w=0^+)+E_{\mu\nu}(w=0^-)]~~
\label{158}
\end{equation}   
and
\begin{equation}
\Pi_{\mu \nu}=-{1\over 4}\tau_{\mu\alpha}\tau_{\nu}^{\phantom{\nu}\alpha}
 +{1\over 12}\tau^{\alpha}_{\phantom{\alpha}\alpha}\tau_{\mu \nu}
+{1\over 8}q_{\mu\nu}(w=0)\tau_{\alpha\beta}\tau^{\alpha\beta}
-{1\over 24}q_{\mu\nu}(w=0)(\tau^{\alpha}_{\phantom{\alpha}\alpha})^2~~.
\label{159}
\end{equation}   
Finally, on introducing a brane cosmological constant $\lambda$ and
decomposing the brane $\tau_{\mu\nu}$ as
\begin{equation}
\tau_{\mu \nu}=-\lambda q_{\mu\nu}(w=0)+S_{\mu\nu}
\label{160}
\end{equation}   
where $S_{\mu\nu}$ represents all other brane matter field sources, we
find that  Eq. (\ref{159}) can then be rewritten as 
\begin{equation}
^{(4)}G_{\mu \nu}=\Lambda_4q_{\mu\nu}(w=0)-\kappa_4^2S_{\mu\nu}
-\kappa_5^4 \pi_{\mu\nu}
-\bar{E}_{\mu\nu}(w=0)
\label{161}
\end{equation}   
where 
\begin{eqnarray}
\pi_{\mu \nu}&=&-{1\over 4}S_{\mu\alpha}S_{\nu}^{\phantom{\nu}\alpha}
 +{1\over 12}S^{\alpha}_{\phantom{\alpha}\alpha}S_{\mu \nu}
+{1\over 8}q_{\mu\nu}(w=0)S_{\alpha\beta}S^{\alpha\beta}
-{1\over 24}q_{\mu\nu}(w=0)(S^{\alpha}_{\phantom{\alpha}\alpha})^2~~,
\nonumber \\
\kappa_4^2&=&{\kappa_5^4\lambda \over 6}~~,~~
\Lambda_4={\kappa_5^2\over 12}(\kappa_5^2\lambda^2+6\Lambda_5)
=\frac{1}{2}\left(\kappa_4^2\lambda+\kappa_5^2\Lambda_5\right)~~.
\label{162}
\end{eqnarray}   

As constructed, Eq. (\ref{161}) thus represents the generalized
Einstein equation seen by an observer on the brane, with the
$\bar{E}_{\mu\nu}(w=0)$ and $\pi_{\mu \nu}$ terms heralding a departure
from a standard 4-dimensional gravity with an effective gravitational
constant $\kappa_4^2$ and an effective cosmological constant
$\Lambda_4$.\footnote{From the point of view of 4-dimensional physics, 
the massless 5-dimensional graviton of Eq. (\ref{152}) decomposes into
a massless 4-dimensional graviton and an infinite tower of massive
4-dimensional spin two modes \cite{Randall1999a,Randall1999b}, with
these  massive modes then contributing to Eq. (\ref{161}). As such,
these massive modes could also serve as potential dark matter
candidates.}   With regard to the effective cosmological constant, we
note that it is not
$\lambda$ which is to serve as the cosmological constant for cosmology
on the brane but rather
$\Lambda_4$, and since the anti-de Sitter bulk $\Lambda_5$ has to be
negative, the effective
$\Lambda_4$ has to be smaller than the $\lambda$ expected from ordinary
4-dimensional early universe phase transition physics. As we had noted
earlier, a possible way to quench the contribution of the cosmological
constant to cosmology is to have two cosmological constants, one being a
dynamical one coming from the energy-momentum tensor (viz. $\lambda$)
and the second being an a priori fundamental one. As we now see, what we
had thought of as a fundamental cosmological constant from the point of
view of 4-dimensional physics might actually be a
$\kappa_5^2\Lambda_5/2$ term coming from the embedding in a higher
dimensional bulk. While embedding thus does provide a natural rationale
for the existence of the
$\kappa_5^2\Lambda_5/2$ term, at the present time no rationale has
yet been found for why the $\kappa_5^2\Lambda_5/2$ term and the
$\kappa_4^2\lambda/2$ terms might be as close to each other as an
$\Omega_{\Lambda}(t_0)=0.7$ universe would require. Within this extra
dimension picture we note that it has also been suggested
\cite{Deffayet2002} that the acceleration of the universe might be
explainable not as a cosmological constant effect at all but rather
as a consequence of a leakage of gravitational signal into the extra
dimensions.

\subsection{The modified Friedmann equations on the brane}

To get a sense of the relevance of the $\bar{E}_{\mu\nu}(w=0)$
and $\pi_{\mu \nu}$ terms in Eq. (\ref{161}), it is convenient to
descend to the special case where $S_{\mu\nu}$ is taken to be the
perfect fluid 
\begin{equation}
S_{\mu \nu}=(\rho+p)U_{\mu}U_{\nu}+p q_{\mu\nu}(w=0)~~, 
\label{163}
\end{equation}   
since in this case $\pi_{\mu\nu}$ then also takes the form of the
perfect fluid
\begin{equation}
\pi_{\mu \nu}=(R+P)U_{\mu}U_{\nu}+P q_{\mu\nu}(w=0)
\label{164}
\end{equation}   
with energy density and pressure 
\begin{equation}
R={\rho^2 \over 12}~~,~~P={1\over 12}(\rho^2+2\rho p)~~. 
\label{165}
\end{equation}   
Considering now the cosmological case where $q_{\mu\nu}(w=0)$ is taken
to be a standard Robertson-Walker metric with scale factor $a(t)$, the
$(00)$ component of Eq. (\ref{161}) is then found to take the form
\begin{equation}
-{3(\dot{a}^2+k) \over a^2}=-\Lambda_4-\kappa_4^2\rho
-{\kappa_5^4 \rho^2 \over 12}
-\bar{E}_{00}(w=0)~~,
\label{166}
\end{equation}   
with it thus being Eq. (\ref{166}) which is to describe cosmic evolution
on the brane, to be compared with the standard (unembedded)
4-dimensional Friedmann evolution equation
\begin{equation}
-{3(\dot{a}^2+k) \over a^2}=-\Lambda_4-\kappa_4^2\rho
\label{167}
\end{equation}   
where the $\kappa_5^2\rho^2/12$ and $\bar{E}_{00}(w=0)$ terms are
of course absent. Now while it is geometrically possible to embed a
brane with a Robertson-Walker geometry into a 5-dimensional bulk with
an anti-de Sitter geometry, it is not obligatory, with the bulk
being able to have less than maximal 5-symmetry in general. Thus while
the $\bar{E}_{00}(w=0)$ term will vanish if the bulk is anti-de Sitter
(since the bulk Weyl tensor then vanishes), it should not in general be
expected to do so. Its non-vanishing would thus entail an unavoidable
modification of standard cosmology whose impact on the dark matter
and dark energy problems could be of interest. Moreover, even if the
bulk geometry is such that the $\bar{E}_{00}(w=0)$ term does in
fact vanish, Eq. (\ref{166}) will still differ from standard
cosmology due to the presence of a term which is quadratic in the
energy density $\rho$. Such terms will be of great importance when
$\rho$ is big, namely in the early universe, though in permanently
expanding cosmologies such terms would redshift away much faster than
the term which is linear in $\rho$ and thus be nicely suppressed
today. With the $\rho^2$ term leading to early universe modifications,
it would be of interest to see how they might modify the initial
condition fine-tuning flatness problem.

While it might be thought that a late time embedded
$\bar{E}_{00}(w=0)=0$ cosmology would always have to behave the same
way as an unembedded one, this is not necessarily the case. Thus in the
specific case where the effective $\Lambda_4$ is negative, the
Robertson-Walker 3-curvature $k$ is zero and $\bar{E}_{00}(w=0)$
is zero, Eq. (\ref{166}) is found to admit of the exact radiation era
($\rho=A/a^4$) solution
\begin{equation}
a^4(t)=-{\kappa_5^2Ab \over 6H^2}{\rm cos}(4Ht)+{\kappa_5^2A
(b^2-H^2)^{1/2}\over 6H^2}
\label{168}
\end{equation}   
where $b^2=-\kappa_5^2\Lambda_5/6$, $\Lambda_4=-3H^2$, and
$\kappa_4^2=\kappa_5^2(b^2-H^2)^{1/2}$. As we see from Eq. (\ref{168}), 
in this particular model $a(t)$ is not in fact late time expanding.
Rather, it only expands to a bounded maximum value given by
\begin{equation}
a^4_{\rm max}(t)={\kappa_5^2A \over 6H^2}\left[b+
(b^2-H^2)^{1/2}\right]~~.
\label{169}
\end{equation}   
With the ratio of the terms in Eq. (\ref{166}) which are quadratic 
and linear in $\rho$  being given by 
\begin{equation}
{\kappa_5^4\rho^2 \over 12 \kappa_4^2\rho}={ H^2 \over
2[b^2-H^2+b(b^2-H^2)^{1/2}]}
\label{170}
\end{equation}   
at this maximum, we see that when $H$ is of order $b$, the right-hand
side of Eq. (\ref{170}) is of then of order one. Since the $\rho^2/\rho
\sim 1/a^4$ ratio is at its minimum when $a(t)$ is at its maximum, we
thus see that in this particular cosmology at no time is the $\rho^2$
term ever negligible compared to the $\rho$ term, with this particular 
embedded cosmology never being able to approach the associated
unembedded one at all. With Eqs. (\ref{148}) and (\ref{151}) being
completely general purely geometric relations which will hold whenever
there is a higher-dimensional embedding of any sort whatsoever (they
do not even require the presence of a brane), despite the fact that
higher dimensional models do possess a massless graviton, we see
that such models could in principle yield modified 4-dimensional
physics, and thus each candidate such higher-dimensional model has to
be analyzed independently to see what kind of 4-dimensional physics it
might lead to.

\subsection{Relativistic MOND theory}

Through the use of additional fields it is also possible to generalize
Milgrom's non-relativistic MOND theory to the relativistic regime. The
first step towards doing this was taken in \cite{Bekenstein1984} where
a scalar field $\psi$ was introduced with action
\begin{equation}
I(\psi)=-\frac{1}{8\pi GL^2}\int
d^4x(-g)^{1/2}\tilde{f}\left(L^2g^{\alpha\beta}
\psi_{;\alpha}\psi_{;\beta}\right) 
\label{171}
\end{equation}   
($\tilde{f}$ is a scalar function and $L$ is a constant), with this
action then being added on to the standard Einstein-Hilbert one.
Additionally, the scalar field coupling to a test particle was taken to
be given by
\begin{equation}
I_m=-m\int e^{\psi}\left(-g_{\alpha\beta}
dx^{\alpha}dx^{\beta}\right)^{1/2}~~,
\label{172}
\end{equation}   
with the test particle thus obeying Eq. (\ref{12}) as written with
$S=e^{\psi}$. With the non-relativistic limit of $\tilde{f}$ being taken
to be the MOND function, the MOND formula of Eq. (\ref{96}) was then
found to ensue with the MOND acceleration parameter being identified
as $a_0=1/L$. While this treatment shows what is in principle needed in
order to recover MOND (or in general to produce a covariant theory whose
non-relativistic limit is not the standard second order Poisson
equation but a modified version of it), as formulated the above
specific treatment had two difficulties. One was that $\psi$ waves
would propagate faster than light (the requisite
$\tilde{f}$ not being quadratic), and the other being that $\psi$ would
have no effect  on the propagation of light waves (the light cone is
invariant under the conformal transformation
$g_{\alpha\beta}\rightarrow e^{2\psi}g_{\alpha\beta}$) leaving the
theory unable to explain gravitational lensing without dark matter. To
resolve the acausality problem a second scalar field $A$ was introduced
\cite{Bekenstein1988,Sanders1988} with the action of Eq. ({171}) being
replaced by 
\begin{equation}
I(\psi,~A)=-\frac{1}{2}\int
d^4x(-g)^{1/2}\left[
\frac{A^2}{\eta^2}g^{\alpha\beta}\psi_{;\alpha}\psi_{;\beta}
+g^{\alpha\beta}
A_{;\alpha}A_{;\beta}+\nu(A^2)\right]~~, 
\label{173}
\end{equation}   
where $\eta$ was a real parameter and $\nu$ a real valued function.
Building on this idea Bekenstein \cite{Bekenstein2004} then found a
a way to avoid the gravitational lensing problem as well in a
generalization which involved both scalar and vector fields. The theory
which Bekenstein constructed contains the metric $g_{\mu\nu}$,
a timelike 4-vector field $U_{\alpha}$ which obeys the constraint
$g^{\alpha\beta}U_{\alpha}U_{\beta}=-1$, a propagating scalar field
$\phi$ and a non-propagating scalar field $\sigma$. The action for the
metric is taken to be the standard Einstein-Hilbert action, to which is
added an action 
\begin{equation}
I(\phi,~\sigma)=-\frac{1}{2}\int
d^4x(-g)^{1/2}\left[
\sigma^2\left(g^{\alpha\beta}
-U^{\alpha}U^{\beta}\right)\phi_{;\alpha}\phi_{;\beta}
+\frac{G\sigma^4}{2\ell^2}F(kG\sigma^2)\right]
\label{174}
\end{equation}   
for the scalars and an action 
\begin{equation}
I(U_{\alpha})=-\frac{K}{32\pi G}\int
d^4x(-g)^{1/2}\left[
g^{\alpha\beta}g^{\mu\nu}\left(U_{\alpha;\mu}-U_{\mu;\alpha}\right)
\left(U_{\beta;\nu}-U_{\nu;\beta}\right)
-\frac{2\lambda}{K}\left(g^{\mu\nu}U_{\mu}U_{\nu}+1\right)\right]
\label{175}
\end{equation}   
for the vector. (Here $k$, $K$, and $\ell$ are constants, $F$ is a
general scalar function, and $\lambda$ is a Lagrange multiplier which
enforces the constraint $g^{\alpha\beta}U_{\alpha}U_{\beta}=-1$.) With
this formulation, MOND can now be regarded as being a fully-fledged,
fully consistent relativistic theory which retains all of MOND's
non-relativistic features, and it will be of interest to see how it
might address the dark energy and cosmological dark matter problems,
problems which we noted earlier fall right in the MOND regime. While the
above treatment does show how it is possible to embed the MOND formula
of Eq. (\ref{96}) in a fully relativistic setting, a drawback of the
approach is that the function $F$ (to which the MOND function
$\mu(a/a_0)$ of Eq. (\ref{96}) is related) is not specified by it, as
the only constraint on it is that $F(kG\sigma^2)$ be a general
coordinate scalar. It would thus be of interest to find a dynamics
which could constrain the function $F$ in a way that could naturally
lead to the behavior which is phenomenologically required of the MOND
function in its large and small argument limits. 

\subsection{Modifications of the nature of the spacetime geometry
itself}  

Beyond the above more or less conventional generalizations of Einstein
gravity (more Riemann tensor dependent terms, more fields, or more
spacetime dimensions),\footnote{Not that any modification of
Einstein gravity can really be described as being conventional.} there
are also some less orthodox ones which are also worthy of consideration,
generalizations which modify the role played by the metric and the
nature of the geometry itself. The first of these is to introduce
torsion, with the Christoffel symbols then no longer being symmetric in
their lower indices. Since such a modification would typically only be
expected to be of relevance to microscopic physics, it is not thought
to have any effect on the dark matter and dark energy problems. The
second modification is to allow the metric to be non-symmetric in its
indices, with $g_{\mu\nu}$ then being a 16-component tensor with ten
symmetric and six antisymmetric components. Such a proposal would not
affect the line element $ds^2=g_{\mu\nu}dx^{\mu}dx^{\nu}$ at all since
the symmetric $dx^{\mu}dx^{\nu}$ can only couple to the symmetric part
of $g_{\mu\nu}$. With the antisymmetric part of the metric having
precisely the same number of degrees of freedom as the 6-component
electromagnetic field strength ($\vec{E}$ and $\vec{B}$) this
generalization also opens up the possibility of unification of
electromagnetism with gravity, a program which has been developed in
detail by Kursunoglu \cite{Kursunoglu1952} and generalized by him to
encompass unification with the other fundamental forces as well
\cite{Kursunoglu1991}. Models based on a non-symmetric metric have
also been explored by Moffat, enabling him
\cite{Moffat1979,Moffat1995a,Moffat1995b,Moffat2004a,Moffat2004b} to
provide a reasonable explanation of galactic rotation curve systematics
and the accelerating universe without the need to invoke dark matter or
dark energy.

Perhaps the most radical departure from Einstein gravity was actually
the first generalization of it ever considered, when only a couple of
years after the very introduction of general relativity Weyl
\cite{Weyl1918a} actually generalized  Riemannian geometry itself. In
what was the very first attempt at unification of electromagnetism
with gravitation through metrication, Weyl introduced a local
transformation which he referred to as a \lq\lq gauge" transformation
under which the metric and the electromagnetic field would jointly
transform as 
\begin{equation}
g_{\mu\nu}(x) \rightarrow e^{2\alpha(x)} g_{\mu\nu}(x)~~,
\label{176}
\end{equation}
\begin{equation}
A_{\mu}(x) \rightarrow A_{\mu}(x)-e\partial_{\mu}\alpha(x)~~,
\label{177}
\end{equation}
with gravitation and electromagnetism thus unifying by sharing a common
$\alpha(x)$. Given such a joint transformation, Weyl then departed from
the Riemannian geometry of general relativity and replaced it with a new
geometry, \lq\lq Weyl Geometry", in which the Riemannian connection was
generalized to the $A_{\mu}$-dependent
\begin{equation}
\hat{\Gamma}_{\mu\nu}^{\lambda}={
1 \over 2}g^{\lambda\sigma}\left(\partial_{\mu}g_{\sigma\nu}
+\partial_{\nu}g_{\sigma\mu}-\partial_{\sigma}g_{\mu\nu}\right)
+{1 \over e}g^{\lambda\sigma}\left(g_{\sigma\nu}A_{\mu}
+g_{\sigma\mu}A_{\nu}-g_{\mu\nu}A_{\sigma}\right)~~,
\label{178}
\end{equation}
so that instead of the covariant derivative of the metric being zero, it
would instead be given by 
\begin{equation}
g^{\mu\nu}_{\phantom{\mu\nu};\nu}=\partial_{\nu}g^{\mu\nu}
+\hat{\Gamma}_{\nu\sigma}^{\mu}g^{\sigma\nu}
+\hat{\Gamma}_{\nu\sigma}^{\nu}g^{\sigma\mu}
={2A^{\mu}\over e}~~.
\label{179}
\end{equation}
The utility of this generalized definition of covariant derivative
is that Eq. (\ref{179}) has the remarkable property of being invariant
under the joint transformations of Eqs. ({\ref{176}) and (\ref{177})
(transformations under which $\hat{\Gamma}_{\mu\nu}^{\lambda}$
transforms into itself). With the covariant derivative of a tensor
transforming as a vector, Weyl was thus able to connect the tensor
$g^{\mu\nu}$ and the vector $A^{\mu}$ in an intricate geometrical
fashion, though at the price of departing from Riemannian
geometry.\footnote{In passing we note that there is no conflict with
general coordinate invariance in having a connection which is not simply
the standard pure metric based Christoffel symbol 
$\Gamma^{\lambda}_{\mu\nu}({\rm C.S.})
=(1/2)g^{\lambda\sigma}\left(\partial_{\mu}g_{\sigma\nu}
+\partial_{\nu}g_{\sigma\mu}-\partial_{\sigma}g_{\mu\nu}\right)$. 
In fact, in general the connection can even be completely independent of
the metric altogether and serve as a totally independent dynamical
degree of freedom. All that is required of a general connection
$\hat{\Gamma}^{\lambda}_{\mu\nu}$ is that it transform the same way under
a general coordinate transformation as the standard
$\Gamma^{\lambda}_{\mu\nu}({\rm C.S.})$, viz. as 
$\hat{\Gamma}^{\lambda}_{\mu\nu}\rightarrow (\partial x^{\prime\lambda}/
\partial x^{\rho}) (\partial x^{\tau}/\partial x^{\prime \mu})
(\partial x^{\sigma} /\partial x^{\prime \nu})
\hat{\Gamma}^{\rho}_{\tau\sigma}+ (\partial x^{\prime\lambda}/\partial
x^{\rho}) (\partial^2 x^{\rho} /\partial x^{\prime \mu}\partial x^{\prime
\nu})$, since then covariant derivatives constructed via its use will
still transform as true general coordinate tensors. (Because the 
$(1/e)g^{\lambda\sigma}\left(g_{\sigma\nu}A_{\mu}
+g_{\sigma\mu}A_{\nu}-g_{\mu\nu}A_{\sigma}\right)$ term in Eq.
(\ref{178}) is a true tensor, then given the way
$\Gamma^{\lambda}_{\mu\nu}({\rm C.S.})$ transforms, it follows that the
quantity $\hat{\Gamma}_{\mu\nu}^{\lambda}$ of Eq. (\ref{178}) does indeed
transform as a connection from which true tensor covariant derivatives
such as $g^{\mu\nu}_{\phantom{\mu\nu};\nu}$ of Eq. (\ref{179}) can be
constructed.) However, in geometries with a metric and an independent
connection, even though the metric will still give the invariant line
element and be responsible for raising and lowering indices, true tensor
quantities such as
$g^{\mu\nu}_{\phantom{\mu\nu};\tau} =\partial_{\tau}g^{\mu\nu}
+\hat{\Gamma}_{\tau\sigma}^{\mu}g^{\sigma\nu}
+\hat{\Gamma}_{\tau\sigma}^{\nu}g^{\sigma\mu}$ will not in general be
zero, with it then not being possible to move the metric in and out of
covariant derivatives in the usual Riemannian way. Additionally, in
terms of a general connection, the Riemann tensor as constructed as
$R^{\lambda}_{\mu\nu\kappa}=(\partial
\hat{\Gamma}^{\lambda}_{\mu\nu}/\partial x^{\kappa})
-(\partial \hat{\Gamma}^{\lambda}_{\mu\kappa}/
\partial x^{\nu})
+\hat{\Gamma}^{\eta}_{\mu\nu} \hat{\Gamma}^{\lambda}_{\kappa\eta} 
-\hat{\Gamma}^{\eta}_{\mu\kappa} \hat{\Gamma}^{\lambda}_{\nu\eta}$
will still transform as a true rank four tensor, with $g^{\mu\kappa}
R^{\lambda}_{\mu\lambda\kappa}$ still being a general covariant scalar.
Using this prescription for the Ricci scalar and forming covariant
derivatives with the general $\hat{\Gamma}^{\lambda}_{\mu\nu}$
connection will thus still enable us to build an action which is a true 
general coordinate scalar even when the connection is not taken to be
the Christoffel symbol or even not specified at all. And while the
variation of such a scalar action with respect to the metric will
automatically be a true rank two tensor, its independent (Palatini)
variation with respect to an independent connection (in general a
variation which is independent of that associated with the metric) will
still be a true tensor too, since even though the connection is not
itself a tensor, its variation is, as the $(\partial
x^{\prime\lambda}/\partial x^{\rho}) (\partial^2 x^{\rho} /\partial
x^{\prime \mu}\partial x^{\prime
\nu})$ term drops out in the transformation of the difference between
any two connections which both transform as 
$\hat{\Gamma}^{\lambda}_{\mu\nu}\rightarrow (\partial x^{\prime\lambda}/
\partial x^{\rho}) (\partial x^{\tau}/\partial x^{\prime \mu})
(\partial x^{\sigma} /\partial x^{\prime
\nu})\hat{\Gamma}^{\rho}_{\tau\sigma}+
(\partial x^{\prime\lambda}/\partial x^{\rho})
(\partial^2 x^{\rho} /\partial x^{\prime \mu}\partial x^{\prime
\nu})$. Consequently, dealing with a connection which is independent of
the metric is just as covariant as taking the connection to be the
standard Christoffel symbol, and doing so can thus generate departures
from standard Riemannian-based gravity which are every bit as covariant. 
(In such cases it should be noted that if the matter field is taken to
have an action which is independent of the connection altogether - such
as the free Maxwell action -- because of the matter field equations of
motion,  the covariant derivative of the matter field energy-momentum
tensor with respect to the Christoffel-symbol based connection will
vanish identically. However since this covariant derivative is
linear in $\Gamma^{\lambda}_{\mu\nu}({\rm C.S.})$, and since the
covariant derivative of the matter field energy-momentum tensor with
respect to $\hat{\Gamma}^{\lambda}_{\mu\nu}$ is necessarily a true
tensor, it follows that the covariant derivative of the matter field
energy-momentum tensor with respect to $\Gamma^{\lambda}_{\mu\nu}({\rm
C.S.})$ is a true tensor too, with its vanishing in one coordinate
system entailing its vanishing in all other coordinate systems too.)

While there is actually no difference between the Palatini approach and
the standard approach when the action is just the Ricci scalar (since
the stationary variation with respect to the
$\hat{\Gamma}^{\lambda}_{\mu\nu}$ connection then yields a condition 
whose solution requires the stationary $\hat{\Gamma}^{\lambda}_{\mu\nu}$
to actually be $\Gamma^{\lambda}_{\mu\nu}({\rm C.S.})$ after all), in
cases where the action contains additional fields and cross-terms
between these fields and the Ricci scalar, the Palatini approach can
lead to equations of motion for the metric and the connection which are
totally different from the ones in which the connection is taken to be
given by the standard $\Gamma^{\lambda}_{\mu\nu}({\rm C.S.})$ from the
outset. We thus note that taking advantage of this fact has enabled the
authors of \cite{Allemandi2005} to develop a Palatini-based scalar-tensor
generalization of the approach of \cite{Carroll2005} which also allows
for cosmic acceleration to emerge as a purely gravitational effect.}  It
was through the development of this theory that the notion of a gauge
transformation was  introduced into physics, with its usage in Eq.
(\ref{176}) entailing a change in the magnitude of
$g^{\mu\nu}$ and thus of its size (this being the meaning of the term
\lq\lq  gauge"). With the Einstein-Hilbert action not being invariant
under the transformation of Eq. (\ref{176}), Weyl's theory thus
possessed neither the Einstein equations or Riemannian geometry, and was
not so much a generalization of Einstein gravity as a rather substantial
departure from it. Moreover, following the development of quantum
mechanics with its complex-valued wave functions, Weyl was able to
redefine his gauge transformation so that instead of being accompanied
by Eq. (\ref{176}), the gauge transformation of Eq. (\ref{177}) would
instead be accompanied by a change in the complex phase of the wave
function ($\psi \rightarrow e^{i\alpha} \psi)$. With Weyl's original
invariance of Eq. (\ref{176})) entailing that all mass parameters would
be zero identically (something which at times prior to the development of
spontaneous symmetry breakdown was regarded as totally unacceptable),
and with the use of Eq. (\ref{177}) in accompaniment with $\psi
\rightarrow e^{i\alpha} \psi$ then becoming so fruitful, Weyl's
geometric theory has essentially been set aside\footnote{See however
\cite{Cheng1988} where it is suggested that Weyl's vector $A_{\mu}$
field might be become massive by a Higgs mechanism in which it
combines with the component of the complex doublet scalar
field whose vacuum expectation value spontaneously breaks the
$SU(2)\times U(1)$ electroweak symmetry.} -- so much so in fact
that the term \lq\lq  gauge invariance" is now taken to refer
exclusively to complex phase transformations, with the transformations
of Eq. (\ref{176}) now being known as conformal or scale
transformations instead.

\subsection{Conformal gravity}

In the course of developing his theory Weyl also discovered
\cite{Weyl1918b} a tensor with a remarkable geometric property, the
so-called conformal or Weyl tensor
\begin{equation}
C_{\lambda\mu\nu\kappa}= R_{\lambda\mu\nu\kappa}
-\frac{1}{2}\left(g_{\lambda\nu}R_{\mu\kappa}-
g_{\lambda\kappa}R_{\mu\nu}-
g_{\mu\nu}R_{\lambda\kappa}+
g_{\mu\kappa}R_{\lambda\nu}\right)
+\frac{1}{6}R^{\alpha}_{\phantom{\alpha}\alpha}\left(
g_{\lambda\nu}g_{\mu\kappa}-
g_{\lambda\kappa}g_{\mu\nu}\right)~~.
\label{180}
\end{equation}
As constructed, this particular combination of the Riemann and
Ricci tensors and the Ricci scalar has the property that under the local
transformation of Eq. (\ref{176}) it transforms as 
\begin{equation}
C^{\lambda}_{\phantom{\lambda}\mu\nu\kappa}(x)\rightarrow 
C^{\lambda}_{\phantom{\lambda}\mu\nu\kappa}(x)
\label{181}
\end{equation}
with all derivatives of $\alpha(x)$ dropping out identically.
As such the Weyl tensor $C^{\lambda}_{\phantom{\lambda}\mu\nu\kappa}$
thus bears the same relation to conformal transformations as the
Maxwell tensor $F_{\mu\kappa}$ does to gauge transformations,
with the kinematic relation $g^{\mu\kappa}F_{\mu\kappa}=0$ having as a
counterpart the kinematic
$g^{\mu\kappa}C^{\lambda}_{\phantom{\lambda}\mu\nu\kappa}=0$, with the
Weyl tensor being the traceless piece of the Riemann tensor.

Given this particular property of the Weyl tensor, a far less radical
version of Weyl's theory is suggested, one in which the invariance of
Eq. (\ref{176}) is retained but the unification with $A_{\mu}$ is
dropped, so that the geometry is then a standard strictly
Riemannian one in which the covariant derivative of the metric is zero
as usual. We thus retain the metric as the gravitational field,
have it couple covariantly in the usual way, but endow gravity with an
additional symmetry beyond coordinate invariance, viz. conformal
symmetry. As such, this theory is known as conformal gravity and has
been pursued by various authors, with (as we shall see below) the
interest of the present author being in applying it astrophysics and
cosmology with a view to solving the dark matter and dark energy
problems. The great appeal of this conformal symmetry is that its
imposition actually leads to a unique choice of gravitational action,
as there is one and only one action which is invariant under the local
conformal transformation of Eq. (\ref{176}), namely the Weyl action
\begin{eqnarray}
I_W&=&-\alpha_g\int d^4x (-g)^{1/2}C_{\lambda\mu\nu\kappa}
C^{\lambda\mu\nu\kappa}
\nonumber \\
&=&-\alpha_g\int d^4x (-g)^{1/2}\left[R_{\lambda\mu\nu\kappa}
R^{\lambda\mu\nu\kappa}-2R_{\mu\kappa}R^{\mu\kappa}+\frac{1}{3}
(R^{\alpha}_{\phantom{\alpha}\alpha})^2\right]
\label{182}
\end{eqnarray}
where $\alpha_g$ has to be dimensionless, with conformal gravity thus
possessing a dimensionless coupling constant.\footnote{The parallel
between $C^{\lambda}_{\phantom{\lambda}\mu\nu\kappa}$ and
$F_{\mu\kappa}$ also carries over to the form of the action with the
$-\alpha_g\int d^4x
(-g)^{1/2}C^{\lambda}_{\phantom{\lambda}\mu\nu\kappa}
C_{\lambda}^{\phantom{\lambda}\mu\nu\kappa}$ action being the conformal
analog of the
$-(1/4)\int d^4x (-g)^{1/2}F_{\mu\kappa} F^{\mu\kappa}$ action, an
action whose coefficient is also dimensionless.} Moreover, not only
does the conformal symmetry uniquely select out the Weyl action, it
expressly forbids the presence of any fundamental cosmological term,
and is thus a symmetry which is able to control the cosmological
constant;\footnote{It was the ability of conformal symmetry to control
the cosmological constant which first attracted the present author to
conformal gravity \cite{Mannheim1990}, with its application to the dark
matter and dark energy problems only coming later.} and as we shall see
below, even after the conformal symmetry is spontaneously broken (as is
needed to generate particle masses), the contribution of a then induced
cosmological constant to cosmology will still be under
control.\footnote{Conformal invariance Ward identities are not modified
by a change in vacuum, with the conformal energy-momentum tensor having
to remain traceless under a change of vacuum, with the induced
cosmological constant term then having to be neither bigger nor smaller
than the other terms in $T_{\mu\nu}$.}

However, in addition to forbidding a fundamental cosmological
constant, the same conformal symmetry also forbids a fundamental
Planck mass, and thereby excludes any fundamental Newton constant or
any fundamental Einstein-Hilbert action. Nonetheless, the
conformal theory cannot be excluded on these grounds, since, as we had
noted earlier, as long as a theory contains the Schwarzschild solution
as an exterior solution, that will suffice to recover the standard
solar system phenomenology. However, such an eventuality does not
immediately appear likely for the action $I_W$ since it involves the
Riemann tensor, a tensor which does not vanish in the Ricci flat
Schwarzschild solution. However, it turns out that the Lanczos
Lagrangian
\begin{equation}
L_L=(-g)^{1/2}\left[R_{\lambda\mu\nu\kappa}
R^{\lambda\mu\nu\kappa}-4R_{\mu\kappa}R^{\mu\kappa}+
(R^{\alpha}_{\phantom{\alpha}\alpha})^2\right]
\label{183}
\end{equation}
happens to be a total divergence \cite{Lanczos1938}, so that $I_W$ can
be rewritten as 
\begin{equation}
I_W=-2\alpha_g\int d^4x
(-g)^{1/2}\left[R_{\mu\kappa}R^{\mu\kappa}-\frac{1}{3}
(R^{\alpha}_{\phantom{\alpha}\alpha})^2\right]~~,
\label{184}
\end{equation}
to now only involve the Ricci tensor and scalar. Variation of the action
of Eq. (\ref{184}) with respect to the metric then yields
\begin{equation}
\frac{1}{(-g)^{1/2}}\frac{\delta I_{W}}{ \delta g_{\mu
\nu}}=-2\alpha_gW^{\mu \nu}=-2\alpha_g\left[W^{\mu
\nu}_{(2)}-\frac{1}{3}W^{\mu\nu}_{(1)}\right]
\label{185}
\end{equation}
where $W^{\mu\nu}_{(1)}$ and $W^{\mu\nu}_{(2)}$ were
respectively introduced in Eqs. (\ref{107}) and
(\ref{108}).\footnote{Alternatively, one can first
make the variation of the action
$I_{W_3}=\int d^4x(-g)^{1/2}R_{\lambda\mu\nu\kappa}
R^{\lambda\mu\nu\kappa}$ to obtain \cite{DeWitt1964}
$(-g)^{-1/2}\delta I_{W_3}/ \delta g_{\mu\nu}=W^{\mu \nu}_{(3)}
=4R^{\mu\nu;\sigma}_{\phantom{\mu\nu;\sigma};\sigma}
-2R^{\mu\sigma;\nu}_{\phantom{\mu\sigma;\nu};\sigma}                        
-2R^{\nu \sigma;\mu}_{\phantom{\nu \sigma;\mu};\sigma} 
-2R^{\mu}_{\phantom{\mu}\sigma\tau\rho}R^{\nu\sigma\tau\rho}
+(1/2)g^{\mu\nu}R_{\sigma\tau\rho\lambda}R^{\sigma\tau\rho\lambda}$,
and with the identical vanishing of variation $\delta
I_{L}/ \delta g_{\mu\nu}$ of the Lanczos action $I_L=\int d^4xL_L$
yielding \cite{DeWitt1964} the kinematic identity  
$-2R^{\mu}_{\phantom{\mu}\sigma\tau\rho}R^{\nu\sigma\tau\rho}
+(1/2)g^{\mu\nu}R_{\sigma\tau\rho\lambda}R^{\sigma\tau\rho\lambda}
+4R^{\mu\sigma\nu\tau}R_{\sigma\tau}
+4R^{\mu}_{\phantom{\mu}\sigma}R^{\nu\sigma}
-2g^{\mu\nu}R_{\sigma\tau}R^{\sigma\tau}
-2R^{\alpha}_{\phantom{\alpha}\alpha}R^{\mu\nu}
+(1/2)g^{\mu\nu}(R^{\alpha}_{\phantom{\alpha}\alpha})^2=0$, on using
Eq. (\ref{187}) one can then conclude that
$W^{\mu \nu}_{(3)}$ is not independent of $W^{\mu \nu}_{(1)}$ and
$W^{\mu \nu}_{(2)}$.} 
Then with the variation of the matter field action $I_M$  with respect
to the metric yielding a matter energy-momentum tensor $T^{\mu\nu}$, the
variation of the total action $I_W+I_M$ with respect to the metric 
yields as equation of motion
\begin{equation}
4\alpha_g W^{\mu\nu}=4\alpha_g\left[W^{\mu
\nu}_{(2)}-\frac{1}{3}W^{\mu\nu}_{(1)}\right]=T^{\mu\nu}~~,
\label{186}
\end{equation}
with use of the kinematic identity  
\begin{equation}
R^{\mu\sigma;\nu}_{\phantom{\mu\sigma;\nu};\sigma}=
\frac{1}{2}(R^{\alpha}_{\phantom{\alpha}\alpha})^{;\mu;\nu}
+R^{\mu\sigma\nu\tau}R_{\sigma\tau}
-R^{\mu}_{\phantom{\mu}\sigma}R^{\nu\sigma}
\label{187}
\end{equation}
allowing one to rewrite Eq. (\ref{186}) in the compact
form\footnote{Even though $C_{\lambda\mu\nu\kappa}(x)$ itself acquires
no derivatives of $\alpha(x)$ under a conformal transformation, because
of the way the Christoffel symbols transform its covariant derivatives
do, with it  being the so-called Bach tensor combination
$2C^{\mu\lambda\nu\kappa}_
{\phantom{\mu\lambda\nu\kappa};\lambda;\kappa}-
C^{\mu\lambda\nu\kappa}R_{\lambda\kappa}$ in which all derivatives of
$\alpha(x)$ drop out.}
\begin{equation}
4\alpha_g W^{\mu\nu}=4\alpha_g\left[
2C^{\mu\lambda\nu\kappa}_
{\phantom{\mu\lambda\nu\kappa};\lambda;\kappa}-
C^{\mu\lambda\nu\kappa}R_{\lambda\kappa}\right]=T^{\mu\nu}~~.
\label{188}
\end{equation}
Finally, as we see from the form of $W^{\mu\nu}_{(1)}$ and
$W^{\mu\nu}_{(2)}$ given in Eqs. (\ref{107}) and (\ref{108}), the
Schwarzschild solution is indeed an exact solution to the conformal
theory in the $T^{\mu\nu}=0$ region, just as required. 

Now we had noted earlier that
within the framework of gravitational theories which did not reduce to
the Einstein equations of motion in any limit whatsoever, as long as
those theories admit the Schwarzschild solution they would suffice
for solar system phenomenology. In principle such theories could include
gravitational actions based on arbitrarily high powers of the Ricci
tensor and Ricci scalar, with there thus being an infinite number of
such options. With there also being an infinite number of options for
theories which do contain the Einstein equations, the beauty of
conformal symmetry is that it serves as a principle which chooses one
and only one gravitational theory from amongst all possible
theories, regardless of whether they contain the Einstein
equations or not. Since the conformal symmetry sets to zero any
fundamental cosmological constant, it thus addresses the most severe
problem in physics head on, and is thus a reasonable starting point for
developing a gravitational theory; and so in the following we shall
look at its observational implications for astrophysics and cosmology.
However, we note now that the ability of the conformal theory to
address the cosmological constant problem so directly is that this
symmetry simultaneously excludes a fundamental Einstein-Hilbert term as
well; with the difficulties faced by the standard theory in addressing
the cosmological constant problem perhaps being due to the fact that in
any starting point which permits the Einstein-Hilbert action, there is
no reason to exclude a cosmological constant term at all, with it being
very hard to get rid of something which has no reason not to be there.

In the conformal theory Eq. (\ref{186}) is to replace the standard
Einstein equations of Eq. (\ref{1}), and with the rank two gravitational
tensor $W^{\mu\nu}$ being kinematically traceless, covariantly
conserved, and having to transform as  
\begin{equation}
W^{\mu\nu}(x)\rightarrow e^{-6\alpha(x)}W^{\mu\nu}(x)
\label{189}
\end{equation}
under the conformal transformation of Eq. (\ref{176}), it follows that
the consistency of the conformal theory requires its energy-momentum
tensor to be traceless, covariantly conserved and transform as 
\begin{equation}
T^{\mu\nu}(x)\rightarrow e^{-6\alpha(x)}T^{\mu\nu}(x)
\label{190}
\end{equation}
too. The consistency of the theory thus sharply constrains the
energy-momentum tensors that are allowable, with there being no freedom
to introduce new ad hoc sources (such as dark matter or dark energy) at
will, with a typical allowed energy-momentum tensor being the
generic one we gave earlier as Eq. (\ref{61}). Moreover, our use of
a conformal invariant action for gravity dovetails with our earlier
discussion of the constraints of an $SU(3)\times SU(2)\times U(1)$
invariant theory of massless fermions and gauge bosons on the structure
of the energy-momentum tensor. Specifically, even though the fermion
spin connection was only introduced into physics in order to make the
fermion kinetic energy term $-\int
d^4x(-g)^{1/2}i\bar{\psi}\gamma^{\mu}(x)
[\partial_\mu+\Gamma_\mu(x)]\psi$ be general coordinate invariant,
nonetheless under $\psi(x)\rightarrow e^{-3\alpha(x)/2}\psi(x)$,
$V^{\mu}_{a}(x)\rightarrow e^{-\alpha(x)}V^{\mu}_{a}(x)$
($V^{\mu}_{a}(x)$ is the vierbein from which
$\gamma^{\mu}(x)=V^{\mu}_{a}(x)\hat{\gamma}^{a}$ and
$\Gamma_{\mu}(x)=[\gamma^{\nu}(x),\partial_{\mu}\gamma_{\nu}(x)]/8
-[\gamma^{\nu}(x),\gamma_{\sigma}(x)]\Gamma^{\sigma}_{\mu\nu}/8$ 
are constructed) this term happens to
be locally conformal invariant as well. Similarly, the general coordinate
invariant Maxwell action is conformal invariant too. With the full
symmetry of the lightcone being the conformal group
$O(4,2)$, and with the conformal group being homomorphic to the Dirac
algebra $SU(2,2)$, we see that requiring massless gauge bosons and
massless fermions to couple covariantly to geometry forces them to
couple to it conformally as well. The only exception to this requirement
is the scalar field kinetic energy, since the minimally coupled scalar
field kinetic energy is not invariant under
$S(x)\rightarrow e^{-\alpha(x)}S(x)$, $g_{\mu\nu}(x)\rightarrow
e^{2\alpha(x)}g_{\mu\nu}(x)$. Rather it is only the combination 
$-\int d^4x(-g)^{1/2}\left[S^{;\mu}S_{;\mu}/2-S^2
R^\mu_{\phantom{\mu}\mu}/12\right]$ which is, an aspect of the
conformal theory which will prove central in the following. With
conformal gravity we thus resolve the conflict between standard gravity
and the tracelessness of $T^{\mu\nu}$ which is required of a theory
built out of massless fermions and gauge bosons, with a traceless
$T^{\mu\nu}$ being coupled in Eq. (\ref{186}) to a gravitational
rank two tensor $W^{\mu\nu}$ which is itself traceless too.
Interestingly, while conformal symmetry actually forces the
$SU(3)\times SU(2)\times U(1)$ theory to be second
order in the first place,\footnote{An action such as $\kappa\int
d^4x (-g)^{1/2}[F_{\mu\nu} F^{\mu\nu}]^2$ would still be
gauge invariant but would not be conformal invariant, and would have to
have a dimensionful coupling constant $\kappa$. The absence of terms
such as these is simply assumed by fiat in the standard $SU(3)\times
SU(2)\times U(1)$ theory, with it being conformal invariance which
actually provides a rationale for their absence, to thereby secure the
renormalizability of the standard model of strong, electromagnetic
and weak interactions.} this same symmetry forces gravity itself to be
fourth order. Thus while conformal invariance requires the
gravitational side of Eq. (\ref{186}) be fourth order, the same
symmetry obliges its matter side to be second order, so that with
conformal invariance the matter
$T^{\mu\nu}$ is of the same order (viz. second) as it is in the
standard gravitational theory. Having thus singled out and motivated
the use of conformal gravity, we turn now to see how it fares in
addressing the dark matter and dark energy problems, an exercise which
is of interest not just in and of itself, but also because it will
uncover some generic features which could serve as a guide for other
theoretical approaches which attempt to address these same problems.

\section{Alternatives to dark matter}

\subsection{Modifying Newtonian gravity at large distances or small
accelerations}

As it started to become clear that there were mass discrepancies in
galaxies, a few physicists started to entertain the possibility that
Newton's law of gravity might need modifying on large distance scales,
with some early ideas along these lines being made by Finzi
\cite{Finzi1963}, Tohline \cite{Tohline1982}, Kuhn and Kruglyak
\cite{Kuhn1987} and Sanders \cite{Sanders1990}, with some related
work being made by Fahr \cite{Fahr1990} and Gessner \cite{Gessner1992}.
Typical of these approaches was the introduction of an extra
potential which was to accompany the Newtonian one, and while such
approaches could in principle even be covariantized via the use of Eq.
(\ref{145}) and thereby be made compatible with the relativity principle,
by and large approaches such as these in the end tended not to fit
data well enough to make them compelling.\footnote{For a more detailed
discussion of candidate alternate theories of this type and of the
challenges they face see e.g. \cite{Aguirre2001}.} More compelling was
Milgrom's MOND approach because, without any need for dark matter,
it could encompass a large amount of data extremely well via the use of
a very compact formula which involved a very small number of
assumptions and a universal acceleration scale criterion which not only
explained why departures from the luminous Newtonian expectation are to
occur for galaxies, but also why none are to be expected for the solar
system. MOND fits to the data sample of Fig. 1 for instance are spot on
\cite{Begeman1991}, as are its fits to a large sample of low surface
brightness galaxies \cite{deBlok1998}, with reasonable fitting being
found for a quite wide range and variety of non-relativistic systems
from molecular clouds to clusters of galaxies \cite{Sanders2002}, systems
which in first approximation are found to obey the fundamental $v \sim
M^{1/4}$ relation between mass and velocity which the MOND theory
possesses.\footnote{In passing we note that a similar mass-velocity
relation was also found to hold \cite{Kazanas1990} in the alternate
conformal gravity theory which is described below.} And with MOND having
now been successfully covariantized, it should be taken seriously since,
as noted, it has captured a basic truth of nature, namely that it is a 
universal acceleration scale which determines when departures from the
luminous Newtonian expectation are to occur.

\subsection{Modifying Einstein gravity -- the conformal gravity approach}

While one can develop non-relativistic theories by starting with the
data themselves,\footnote{Such approaches while phenomenological should
not be thought of as being any more ad hoc than dark matter itself.} in
doing so it is not initially clear what covariantized relativistic
theory might ultimately ensue or whether such an approach would ever be
able to address problems such as the cosmological constant problem, a
problem which is intrinsically relativistic. On the other hand if one
starts with a relativistic theory which is chosen because it does
address the cosmological constant problem and then works down, one does
not know what non-relativistic limit one might then encounter. Encouraged
by the fact that conformal gravity could address the cosmological
constant problem, Mannheim and Kazanas \cite{Mannheim1989,Mannheim1994}
set out to determine the non-relativistic limit of the conformal theory,
and did so at that time with a quite limited objective, namely to see
whether a theory which did not contain the Einstein equations could
nonetheless still recover the standard Newton/Schwarzschild
phenomenology.

To this end they endeavored to find an exact conformal gravity analog of
the Schwarzschild exterior and interior solutions to standard gravity,
viz. they tried to solve the  equation $W^{\mu\nu}=T^{\mu\nu}/4\alpha_g$
for a static, spherically symmetric source. While this meant having to
handle a fourth order differential equation, in turned out that it was
possible to greatly simplify the problem through use of the underlying
conformal symmetry that the theory possessed. Specifically, they noted
that under the general coordinate transformation
\begin{equation}
\rho = p(r)~~,~~B(r) = {r^2 b(r)\over
p^2(r)}~~,~~                             
 A(r) = {r^2 a(r) p^{\prime 2}(r)\over p^2 (r)}
\label{191}
\end{equation}
with initially arbitrary function $p(r)$, the
general static, spherically symmetric line element
\begin{equation}
ds^2 = -b(\rho)dt^2 + a(\rho)d\rho ^2 + \rho ^2 d\Omega_2
\label{192}
\end{equation}
would be brought to the form
\begin{equation}
ds^2 = {p^2(r)\over r^2}\left[-B(r) dt^2 + A(r) dr^2 + r^2
d\Omega_2\right]~~.  
\label{193}
\end{equation}
On now choosing $p(r)$ according to 
\begin{equation}
-{1\over p(r)} = \int{dr\over r^2[a(r)b(r)]^{1/2}}~~,
\label{194}
\end{equation}
the function $A(r)$ would then be set equal to $1/B(r)$, with the 
line element then being brought to the convenient 
form                                               
\begin{equation}
ds^2= {p^2(r)\over r^2}\left[-B(r) dt^2+ {dr^2\over B(r)} + r^2
d\Omega_2\right]~~.
\label{195}
\end{equation}
While Eq. (\ref{195}) is coordinate equivalent to Eq. (\ref{192}) with the
two general functions $a(\rho)$ and $b(\rho)$ having been traded for two
equally general functions $p(r)$ and $B(r)$, its utility lies in the
fact that one of these two functions appears purely as an overall
multiplier in Eq. (\ref{195}). Consequently, through use of Eqs.
(\ref{189}) and (\ref{190}) the function $p(r)$ can be removed from the
theory altogether (i.e. gauged away) via a conformal transformation, with
the full kinematic content of the conformal theory then being contained in
the line element
\begin{equation}
ds^2 = -B(r)dt^2 + {dr^2\over B(r)} + r^2 d\Omega_2
\label{196}
\end{equation}
in the static, spherically symmetric case. Comparing with the standard
theory, we see that instead of having deal with coupled second order
equations for two independent functions, we instead have to deal with a
fourth order equation for one alone, an equitable enough trade-off.

Evaluating the form that $W^{\mu\nu}$ takes in the line element of Eq.
(\ref{196}) is straightforward though lengthy, and leads to       
\begin{eqnarray}
\frac{W^{rr}}{B(r)}&=& 
\frac{B^{\prime}B^{\prime \prime \prime}}{6}
-\frac{B^{\prime \prime 2}}{12}
-\frac{1}{3r} (B B^{\prime \prime\prime}-B^{\prime}B^{\prime\prime})
\nonumber \\ 
&&                                                        
- \frac{1}{3r^2}(B B^{\prime \prime} 
+B^{\prime 2})
+\frac{2B B^{\prime}}{3r^3} 
-\frac{B^2}{3 r^4}
+ \frac{1}{3 r^4}~~,
\label{197}
\end{eqnarray}
\begin{eqnarray}                                                                              
W^{00}&=&
-\frac{B^{\prime \prime \prime \prime}}{3}
+\frac{ {B^{\prime
\prime}}^2}{12B}        
-\frac{B^{\prime \prime \prime} B^{\prime}}{6B}
-\frac{B^{\prime \prime \prime}}{r}             
-\frac{B^{\prime\prime} B^{\prime}}{3rB}
\nonumber \\
&&+\frac{B^{\prime \prime}}{3r^2}  
+ \frac{B^{\prime 2}}{3r^2B} 
- \frac{2B^{\prime}}{3r^3}             
-\frac{1}{3r^4B} 
+ \frac{B}{3r^4}
\label{198}
\end{eqnarray}                                                 
for its components of interest. Combining Eqs. (\ref{197}) and
(\ref{198}) then yields the remarkably simple                            
\begin{equation}                                                                              
\frac{3}{B}\left(W^0_{{\phantom 0} 0} - W^r_{{\phantom r} r}\right)
=B^{\prime \prime \prime \prime} + \frac{4
B^{\prime \prime \prime}}{r} =                
\frac{1}{r}(rB)^{\prime\prime\prime\prime}=\nabla^4B~~,
\label{199}
\end{equation}                     
so that in terms of the convenient source function $f(r)$
defined via                   
\begin{equation}                                                                              
f(r) = \frac{3}{4\alpha_g B(r)}\left(T^0_{{\phantom 0} 0} -
T^r_{{\phantom r} r}\right)                   
\label{200}
\end{equation}                                                  
the equation of motion of Eq.
(\ref{186}) can then be written in the 
extremely compact form                                     
\begin{equation}                                                                               
\nabla^4 B(r) = f(r) ~~.
\label{201}
\end{equation}      

The remarkably simple equation of motion of Eq. (\ref{201}) contains 
the full dynamical content of the conformal theory in the static,
spherically symmetric case, an equation of motion which is completely
exact and which has not required the use of any perturbative
approximations whatsoever. While we have not advocated a preference for
one theory over another on the grounds of calculational simplicity,
preferring to make assessments on the basis of conceptual simplicity
instead, nonetheless from a purely calculational standpoint, Eq.
(\ref{201}) can claim to be just as simple as (and perhaps even simpler
than) its standard gravity counterpart given in Eqs. (\ref{44}) and
(\ref{45}). The solution to Eq. (\ref{201}) is readily given as the
analog of the solution to Eq. (\ref{86}) given earlier, and can be
written as 
\begin{eqnarray}
B(r>R)&=& -\frac{r}{2}\int_0^R
dr^{\prime}r^{\prime 2}f(r^{\prime})
-\frac{1}{6r}\int_0^R
dr^{\prime}r^{\prime 4}f(r^{\prime})+w-kr^2~~,
\nonumber \\
B(r<R)&=&-\frac{r}{2}\int_0^r
dr^{\prime}r^{\prime 2}f(r^{\prime})
-\frac{1}{6r}\int_0^r
dr^{\prime}r^{\prime 4}f(r^{\prime})
\nonumber \\
&&
-\frac{1}{2}\int_r^R
dr^{\prime}r^{\prime 3}f(r^{\prime})
-\frac{r^2}{6}\int_r^R
dr^{\prime}r^{\prime }f(r^{\prime})+w-kr^2~~,
\label{202}
\end{eqnarray}                                 
where the $w-kr^2$ term is the general solution to the homogeneous
$\nabla^4 B(r)=0$ equation. Finally, on defining
\begin{equation}
\gamma= -\frac{1}{2}\int_0^R
dr^{\prime}r^{\prime 2}f(r^{\prime})~~,~~
2\beta=\frac{1}{6}\int_0^R
dr^{\prime}r^{\prime 4}f(r^{\prime})~~,
\label{203}
\end{equation}                                 
on dropping the $kr^2$ term (as it does not couple to the source) and
setting $w=1$, we see that up to conformal equivalence the conformal
gravity metric exterior to a static, spherically source is thus given
without any approximation at all as
\begin{equation}
B(r>R)=-g_{00}=\frac{1}{g_{rr}}=1-\frac{2\beta}{r}+\gamma r~~.
\label{204}
\end{equation}                                 
In the region where $2\beta/r \gg \gamma r$ we thus nicely recover the
Schwarzschild solution, while seeing that departures from it occur only
at large distances and not at small ones, with the standard solar
system Schwarzschild phenomenology thus being preserved.\footnote{While
the vanishing of the Ricci tensor entails the vanishing of its
derivatives, the vanishing of the derivatives can be obtained without the
vanishing of the Ricci tensor itself, with solutions other than the
Schwarzschild solution then being possible. The constructive approach
which leads to Eq. (\ref{204}) shows that not only is the metric of Eq.
(\ref{204}) with its the linear potential an exact exterior solution to
the conformal theory, such a metric categorizes all departures from
Schwarzschild, as every solution to the conformal theory is conformally
equivalent to it. In solving the exterior fourth order Laplace equation
$\nabla^4 B(r)=0$ one would obtain \cite{Riegert1984,Mannheim1989} the
solution of Eq. (\ref{204}) as expressed in terms of what would then be
two free integration constants
$\beta$ and $\gamma$, with it being the matching \cite{Mannheim1994} to
the interior region via the fourth order Poisson equation
$\nabla^4 B(r)=f(r)$ which allows one to express
these integration constants in terms of moments of the source and
establish their physical meaning.} 

\subsection{Impact of the global Hubble flow on galactic rotation
curves}

With both $\beta$ and $\gamma$ being related to moments of the
energy-momentum tensor of the source, and with neither moment
having any reason to vanish, a general conformal gravity source
would then furnish both terms, with a source such as a star then
producing a non-relativistic potential of the form 
\begin{equation}
V^*(r)= -\frac{\beta^*c^2}{r}+\frac{\gamma^* c^2r}{2}~~,
\label{205}
\end{equation}                                 
to nicely yield none other than the potential of Eq. (\ref{95}) whose
merits were discussed earlier. In an application of the potential
$V^*(r)$ to spiral galaxies, we need to integrate the $V^*(r)$ potential
over a disk of $N^*$ stars distributed as
$\Sigma(R)=\Sigma_0e^{-R/R_0}$, with use of Eq. (\ref{A15}) then
enabling us to generalize Eq. (\ref{115}) to
\begin{eqnarray}
\frac{v_{{\rm lum}}^2}{R}=g^{{\rm lum}}=g_{\beta}^{{\rm
lum}}+g_{\gamma}^{{\rm lum}}&=&
\frac{N^*\beta^*c^2 R}{2R_0^3}\left[I_0\left(\frac{R}{2R_0}
\right)K_0\left(\frac{R}{2R_0}\right)-
I_1\left(\frac{R}{2R_0}\right)
K_1\left(\frac{R}{2R_0}\right)\right]
\nonumber \\
&&+\frac{N^*\gamma^* c^2R}{2R_0}I_1\left(\frac{R}{2R_0}\right)
K_1\left(\frac{R}{2R_0}\right)~~.
\label{206}
\end{eqnarray} 
With the large $R$ limit of $v_{{\rm lum}}^2/R$ then being given by
\begin{equation}
\frac{v_{{\rm lum}}^2}{R} \rightarrow \frac{N^*\beta^*c^2}{R^2}+
\frac{N^*\gamma^*c^2}{2}~~,
\label{207}
\end{equation} 
we see that we precisely recover the two $N^*$-dependent terms present in
the formula of Eq. (\ref{125}) that was extracted purely
phenomenologically from a study of the centripetal accelerations of the
last data points in each of the galaxies in the eleven galaxy sample of
Fig. 1. The potential of the conformal theory thus automatically captures
much of the essence of Eq. (\ref{125}).

However, absent from Eq. ({207}) is the $N^*$-independent $\gamma_0c^2/2$
term which we saw was crucial to the validity of Eq. (\ref{125}), and
finding its origin is quite subtle and not at all easy. Specifically,
intrinsic to the use of Newtonian gravity and to the intuition one then
acquires through its repeated use is a theorem due to Newton, namely
that if one has a spherical distribution of matter, the net Newtonian
potential produced at a given point depends only on the matter interior
to its location, with the potentials due to all the points exterior to
it mutually cancelling. In fact so central is this notion to our
thinking, that when a mass discrepancy is found we immediately introduce
dark matter in precisely the region where the discrepancy is found, a
strictly local approach to gravity. However, this theorem of Newton only
holds for a $1/r$ potential, and does not hold in any other case.
Consequently, in any modification to Newton whatsoever, one cannot
exclude the matter exterior to the point of interest, viz. one cannot
exclude the effect of the potentials due to the rest of matter in the
universe. Moreover, in modifications to Newton which grow with distance,
these effects will be quite pronounced, with the biggest contributors to
the gravitational potential at any given point then being not the matter
sources which are the nearest but rather those which are most distant, a
thus global gravitational effect, one which is quite reminiscent of
Mach's principle. The contributions of such distant objects would then
not depend on the mass of the local galaxy of interest, and since those
distant objects form the Hubble flow, one would even expect their net
contribution to come with a cosmologically relevant scale. From the rest
of the universe then we can thus naturally expect to obtain some form of
$N^*$-independent $\gamma_0c^2/2$ type term with strength of order the
phenomenologically obtained value $\gamma_0=3.06\times 10^{-30} {\rm
cm}^{-1}$ cited earlier.\footnote{For a galaxy containing $N^*$ stars,
the distance scale at which the galactic potential which they produce
becomes of order one is given by $r\sim 1/N^*\gamma^*$, a scale which
for a typical $N^*=10^{11}$ galaxy is numerically precisely of order
$10^{-30}$cm$^{-1}$, with a  cosmological scale of order $1/\gamma_0$
thus being precisely the scale on which galactic potentials become
strong.} However, what does not follow from this discussion is why the
net effect of the entire rest of the universe would be to explicitly
yield a linear potential term of the form $V=\gamma_0c^2R/2$, as it is
precisely such a contribution and no other which leads to a centripetal
acceleration of the form $v^2/R=\gamma_0c^2/2$.\footnote{An 
application of the purely local galactic Eq. (\ref{206}) alone to the
eleven galaxy sample of Fig. (1) was actually found
\cite{Mannheim1996,Carlson1996} to fit the shapes of the rotation curves
extremely well, but with the data wanting the strength of the total
galactic $\gamma^{*}N^{*}$ to be universal rather than $\gamma^{*}$
itself. It took the present author quite some time to grasp that the
apparent need for a universal $\gamma^{*}N^{*}$ in the fits was actually
due to the fact that in fitting data, Eq. (\ref{206}) was trying to
simulate the contribution of a universal, $N^*$-independent,  $\gamma^0$
contribution coming from the rest of the universe.}

To explicitly obtain this global $\gamma_0c^2R/2$ potential, we need to
ask how the comoving Hubble flow would look to an observer who uses a
coordinate system in which the center of any chosen galaxy is taken to be
at rest. With a Robertson-Walker geometry being homogeneous and 
isotropic, every point in the geometry can serve as the origin of
coordinates, and so we can take the center of any given comoving galaxy
of interest to serve as that center, with it being the presence of the
galaxy of interest itself which provides this choice for the origin of
coordinates. Having fixed the origin of coordinates this way we now need
to rewrite the comoving Robertson-Walker metric in static Schwarzschild
geometry coordinates as referred to this same origin.\footnote{The
Robertson-Walker geometry is spherically symmetric about every point
while the Schwarzschild geometry is only spherically symmetric about a
single point, with the Robertson-Walker geometry being spherically
symmetric about that one particular point too.} To this end we note that
the general coordinate transformation 
\begin{equation}
r=\frac{\rho}{(1-\gamma_0\rho/4)^2}~~,~~t=\int\frac{d\tau}{R(\tau)}
\label{208}
\end{equation}                                 
effects the metric transformation
\begin{eqnarray}
&&-(1+\gamma_0r)c^2dt^2+\frac{dr^2}{(1+\gamma_0r)}+r^2d\Omega_2
\nonumber \\
&&~~\rightarrow\frac{(1+\gamma_0\rho/4)^2}{R^2(\tau)(1-\gamma_0\rho/4)^2}
\left[-c^2d\tau^2+\frac{R^2(\tau)}{(1-\rho^2\gamma_0^2/16)^2}
\left(d\rho^2+\rho^2d\Omega_2\right)\right]~~,
\label{209}
\end{eqnarray} 
to yield a metric which we recognize as being conformal to a
Robertson-Walker metric with spatial 3-curvature $k=-\gamma_0^2/4$ (the
spatial part of the Robertson-Walker metric being written in isotropic
coordinates here). Now since a Robertson-Walker metric happens to be
conformal to flat, the Weyl tensor associated with it would not only be
zero, but would even remain so under conformal transformations such as
\begin{equation}
g_{\mu\nu}\rightarrow
\frac{(1+\gamma_0\rho/4)^2}{R^2(\tau)(1-\gamma_0\rho/4)^2}g_{\mu\nu}~~,
\label{210}
\end{equation}                                 
transformations which we are free to make in the conformal theory; with a
comoving Robertson-Walker geometry with $k=-\gamma_0^2/4$ thus being
coordinate and conformal equivalent to a static Schwarzschild coordinate
system metric with none other than a linear potential. As seen by an
observer at rest then it follows \cite{Mannheim1997} that the comoving
Hubble flow acts precisely as a universal linear potential. Moreover,
not only do we see that this effect is expressly associated with a
negative spatial 3-curvature,\footnote{If $k$ is zero there is no
effect, while if $k$ is positive, the transformation of Eq. (\ref{208})
would lead to a pure imaginary value for $\gamma_0$.} we shall show below
that $k$ actually is negative (and necessarily so in fact) in the
cosmology associated with the same conformal theory. We thus see that
despite our familiarity with the local nature of Newtonian gravity,
global cosmology can have a local observable effect in galaxies, an
effect which is of a quite general nature, and which should thus be
present in every theory of gravity, and not just the conformal theory
being considered here.

With the linear potential of Eq. (\ref{209}) being centered at the
center of any given spiral galaxy,\footnote{In a coordinate system in
which any given comoving galaxy is at rest, no other comoving galaxy
could simultaneously be at rest also. Nonetheless, since every comoving
galaxy can serve as the origin of the Robertson-Walker coordinate system,
for the purposes of determining the motion of particles with respect to
the center of any particular galaxy, we can refer Eq. (\ref{209}) with
its universal $\gamma_0$ to the center of the galaxy of interest
each and every time.} in the weak gravity limit we can directly add it
on to the luminous contribution of Eq. (\ref{206}), with the total
centripetal acceleration seen by a particle in the galaxy being given by 
\begin{eqnarray}
\frac{v_{{\rm tot}}^2}{R}=g^{{\rm tot}}=g_{\beta}^{{\rm
lum}}+g_{\gamma}^{{\rm lum}}+\frac{\gamma_0c^2}{2}&=&
\frac{N^*\beta^*c^2 R}{2R_0^3}\left[I_0\left(\frac{R}{2R_0}
\right)K_0\left(\frac{R}{2R_0}\right)-
I_1\left(\frac{R}{2R_0}\right)
K_1\left(\frac{R}{2R_0}\right)\right]
\nonumber \\
&&+\frac{N^*\gamma^* c^2R}{2R_0}I_1\left(\frac{R}{2R_0}\right)
K_1\left(\frac{R}{2R_0}\right)+\frac{\gamma_0c^2}{2}~~,
\label{211}
\end{eqnarray} 
with the $g_{\beta}^{{\rm lum}}$ and $g_{\gamma}^{{\rm lum}}$ terms 
being due to the luminous material within the galaxy and the
$\gamma_0c^2/2$ term  being due to the luminous material in the rest of
the universe. With the large $R$ limit of $v_{{\rm tot}}^2/R$ being given
by
\begin{equation}
\frac{v_{{\rm tot}}^2}{R} \rightarrow \frac{N^*\beta^*c^2}{R^2}+
\frac{N^*\gamma^*c^2}{2}+\frac{\gamma_0c^2}{2}~~,
\label{212}
\end{equation} 
we see that from the above $g^{{\rm tot}}$ we precisely recover the
phenomenologically established Eq. (\ref{125}), with conformal gravity
thus providing a theoretical rationale for its validity. With the
$\gamma^*$ and $\gamma_0$ parameters already being fixed by the
application of Eq. (\ref{125}) to the furthest data point in each
galaxy, and with the mass to light ratios of the luminous matter
distributions in each galaxy already being fixed by the maximum luminous
disk prescription for the inner regions of the rotation curves, there is
no more freedom left in the theory. The application of Eq. (\ref{211}) to
the entire 280 or so data points in the eleven galaxy sample of Fig. 1
is thus fully prescribed, with it leading to the fits of
\cite{Mannheim1997} which are displayed in Fig. 1. As we see, the
conformal theory fully captures the systematics of the data, and just
like the MOND fits, shows that it is possible to fit galactic rotation
curve data without needing to use any galactic dark matter at
all.\footnote{As with the analogous MOND case, the numerical values
obtained for the conformal gravity parameters used in Eq. (\ref{211})
ensure that no modifications to the luminous Newtonian expectation
are to occur on solar system distance scales.}

\subsection{Comparison of the conformal gravity and MOND fits}

To understand why fits to those of the rotation curves which are flat
(viz. the bright spirals) can be obtained in a theory which is based on
rising potentials, we recall that the bright spirals are all Freeman
limit galaxies with a common central surface brightness $\Sigma_0^F$,
and thus a common value for $N^*/2\pi R_0^2$. In a theory with rising
potentials, such potentials make a quite small contribution in the inner
regions of the rotation curves, with the peak at $R=2.2R_0$ being
controlled almost completely by the luminous Newtonian contribution, to
thus yield 
\begin{equation}
\frac{v^2_{{\rm lum}}}{c^2}\sim \frac{0.4 N^*\beta^*}{R_0}=
0.8\pi\Sigma_0^F\beta^*R_0
\label{213}
\end{equation} 
at the peak. With reference to say the typical galaxy NGC 3198, we
match the inner region peak value at $R=2.2R_0$ to the value at
the furthest data point at $R=10R_0$ where Eq. (\ref{125}) holds, to
obtain
\begin{equation}
\frac{0.4 N^*\beta^*}{R_0}=
0.8\pi\Sigma_0^F\beta^*R_0=
\frac{\beta^*N^*}{10R_0}+5R_0\left(\gamma^*N^*+\gamma_0\right)~~,
\label{214}
\end{equation} 
and thus 
\begin{equation}
\gamma^*N^*+\gamma_0=\frac{0.06 N^*\beta^*}{R_0^2}=
0.12\pi\Sigma_0^F\beta^*~~.
\label{215}
\end{equation} 
With $\gamma^*N^*$ and $\gamma_0$ being of the same order for such
galaxies, the universality of $\gamma_0$ and $\Sigma_0^F$ are thus
correlated, to not only automatically enable all Freeman limit galaxies
to match their inner and outer regions (something which we recall is
completely contrived in dark matter fits), but also to suggest that
the very existence of the universal $\Sigma_0^F$ itself is of
cosmological origin, and that it might therefore emerge naturally in a
cosmologically based theory of galaxy formation. Additionally, we also
note from Fig. 1 that in the intermediate region of the NGC 3198
rotation curve at around $R=6R_0$ or so, the luminous Newtonian
contribution is at about half of its inner region peak value, while the
linear potential contribution is at about half of its outer region
value, with the sum of their contributions thus being equal to the
value of $v^2_{{\rm lum}}$ obtained in the inner and outer regions
themselves. (With one contribution falling and the other rising, there
has to be some intermediate point where they cross.) Matching the inner
and outer regions to each other via Eq. (\ref{214}) then automatically
takes care of matching the outer and inner regions to the intermediate
region as well, to thus make the rotation curve pretty flat over the
entire $2.2R_0 \leq R \leq 10R_0$ region (again we recall how completely
contrived this particular matching is in dark matter fits). While such
flatness is thus naturally achieved in the conformal theory, it is
important to stress that it is only achievable out to some maximum value
of $R$ before the rise due to the linear potential wins
out over the falling Newtonian contribution, an aspect of the theory
which is immediately evident in the fits to the lower luminosity
galaxies shown in Fig. 1. The view of the conformal theory then is that
the currently rising low luminosity rotation curves will continue to
rise while the currently flat high luminosity rotation curves will
eventually start to rise at large enough
$R$. Such an expectation stands in sharp contrast to both the dark
matter and MOND expectations where the rising low luminosity rotation
curves are expected to flatten off and the flat high luminosity rotation
curves are expected to stay flat. Extending galactic rotation curves to
larger distances from the centers of galaxies could therefore be quite
instructive.

The essence of the conformal gravity fits presented above lies in the
specific role played by cosmology, with the view of conformal gravity
being that as a test particle orbits a galaxy, it not only sees the local
field produced by the luminous material within that selfsame galaxy, the
test particle also sees the gravitational field produced by the rest of
the universe. Particles orbiting a galaxy thus probe the global
cosmology and thereby enable us to measure the global spatial curvature
of the universe, which from the measured value for $\gamma_0$ is thus
given by $k=-2.3\times 10^{-60}$cm$^{-2}$. In the conformal theory the
rest of the universe thus replaces galactic dark matter, with the view
of the conformal theory being that the presence of dark matter in
galaxies is nothing more than an artifact which is engendered by trying
to describe a global phenomenon in purely local terms.\footnote{This
same could of course equally be said of dark matter in even larger
systems such as clusters of galaxies where the natural interplay in the
conformal theory between local and global effects would be even more
pronounced. With a cluster of galaxies being a large local inhomogeneity
in a homogeneous global cosmological background, a full treatment of
clusters in the conformal theory has to await the development of a
treatment of the growth of fluctuations in conformal cosmology.}

With Eq. (\ref{211}) being writable as
\begin{equation}
a=g^{{\rm lum}}+\frac{\gamma_0c^2}{2}~~,
\label{216}
\end{equation} 
we additionally see that we can write it in the form of Eq. (\ref{99}),
viz.
\begin{equation}
a=\nu\left(\frac{g^{{\rm lum}}}{(\gamma_0c^2/2)}\right)g^{{\rm lum}}
\label{217}
\end{equation} 
where
\begin{equation}
\nu(x)=1+\frac{1}{x}~~.
\label{218}
\end{equation} 
We thus recognize Eq. (\ref{216}) as being in the form of a MOND type
equation with $\gamma_0c^2/2$ playing the role of a universal
acceleration, but with the local $g_{\beta}^{{\rm lum}}$ being replaced
by the entire local $g^{{\rm lum}}=g_{\beta}^{{\rm lum}}
+g_{\gamma}^{{\rm lum}}$ which is due to both the Newtonian and the
linear potentials of the luminous matter within the galaxy. Conformal
gravity thus provides a rationale for why there should be a universal
acceleration in the first place, and also for why it should be
associated with a cosmic scale.\footnote{While the MOND and conformal
gravity theories agree on the need for a universal scale in galaxies,
they depart on how the theories are to behave in the region where the
existence of the scale is of consequence. (In passing we note that the
numerical relation between the $a_0$ and $\gamma_0c^2$ parameters may be
obtained by looking at the behavior of the typical galaxy NGC 3198 in the
intermediate region near $R=6R_0$ where the Newtonian and linear
contributions cross each other in Fig. 1. With $\gamma^*N^*$ being of
order $\gamma_0$ for this galaxy, at the crossing point we have
(c.f. Eq. (\ref{212})) $\beta^*N^*/R^2\sim \gamma_0$ and thus $v^2/R\sim
2\gamma_0c^2$. Comparing this with the MOND value (c.f. Eq. (\ref{102}))
of
$v^2/R\sim (a_0)^{1/2}(\beta^*N^*c^2/R^2)^{1/2}$ then yields the
relation $a_0=4\gamma_0c^2$, a relation which numerically is 
closely obeyed.)} Moreover, while it had been noted earlier that the
universal acceleration of MOND had a magnitude of order
$cH_0$, this was quite puzzling since $H_0$ is epoch-dependent. However,
cosmology can actually supply not just one but two natural scales, viz.
both $H_0$ and $k$, and unlike $H_0$, the spatial 3-curvature $k$ is
a true general coordinate scalar which is not epoch-dependent at all. It
is thus $k$ which is the natural quantity with which to associate a
universal scale, with it being $k$ which produces a global imprint on
galactic rotation curves. The success of the conformal gravity fits thus
suggests a role for cosmology in the interpretation of galactic rotation
curves as well as a role for the spatial curvature of the universe. We
believe these to be general features which should be sought in any
theory which attempts to explain galactic rotation curves, with the
absence of a fundamental scale such as $k$ being a possible
shortcoming of the standard hierarchical $\Omega_k=0$, $\Lambda$CDM 
models of dark matter halos.\footnote{Whatever one's views on alternate
theories such as MOND or conformal gravity, if the answer is to be dark
matter, then dark matter theory really has to be able to reproduce the
regularities in the galactic rotation curve data which these alternate
theories so readily capture, and has to do so with the same easy
facility.}

As a final comment on the conformal gravity fits to galactic rotation
curves which we have presented here, we note that even more remarkable
than the fits themselves is that the conformal theory was never
constructed for this particular purpose. Rather, the theory was selected
for an entirely different purpose (controlling the cosmological
constant), and when Mannheim and Kazanas set out to determine the
non-relativistic limit of the conformal theory, they had no idea what
they would find (as indicated earlier, they had only wanted to see
whether the theory could recover the standard Newton/Schwarzschild
phenomenology), and had no inkling at all as to how the theory might then
impact on galactic rotation curves. The fact that the theory then does
impact so well on the rotation curves suggests that one should
therefore give the theory serious consideration. With the motivation for
the conformal theory being the cosmological constant problem, we turn
now to a discussion of how the theory treats that particular problem,
and in the fitting to the Hubble plot which we shall describe below, we
shall again find the theory working remarkably well on data it had
not at all been designed for, data which actually did not even exist at
the time \cite{Mannheim1990,Mannheim1992} conformal cosmology was first
developed.

\section{Alternatives to dark energy}

\subsection{The conformal gravity alternative to dark energy}

In order to identify possible alternatives to dark energy, we first
present the conformal gravity theory treatment of the issue, and then
extract some generic features from it. In applying conformal gravity to
cosmology we note that since the Weyl tensor vanishes in a
Robertson-Walker geometry, in conformal cosmology the entire left-hand
side of Eq. (\ref{188}) vanishes identically, with the equation of
motion for conformal cosmology thus reducing to the extremely simple
\begin{equation}
T^{\mu\nu}=0~~.
\label{219}
\end{equation} 
With this vanishing of the cosmological $T^{\mu\nu}$, we see immediately
that in the conformal theory the zero of energy is completely determined,
to thus give us control of the cosmological constant.\footnote{While the
trace of the energy-momentum tensor must in general always vanish in
the conformal theory to thereby already give us control of its
various components, for cosmology the energy-momentum tensor itself must
vanish too, to give us yet more control.} To appreciate the implications
of Eq. (\ref{219}), we consider as cosmological energy-momentum tensor
the typical matter field conformal energy-momentum tensor given in Eq.
(\ref{64}). With this energy-momentum tensor having to transform
according to Eq. (\ref{190}) under a conformal transformation, its
vanishing will persist after any conformal transformation is made. With
the scalar field transforming as
$S(x) \rightarrow e^{-\alpha(x)}S(x)$ under a conformal transformation,
in field configurations in which $S(x)$ is non-zero, the full content of
the theory can be obtained by working in the particular gauge in which
the scalar field takes the constant value $S_0$. In such a gauge, use of
the matter field equations of motion allows us to then write the
energy-momentum tensor in the convenient form 
\begin{equation}
T^{\mu \nu} = i \bar{\psi} \gamma^{\mu}(x)[
\partial^{\nu}                    
+\Gamma^\nu(x)]                                                                 
\psi
-\frac{1}{6}S_0^2\left(R^{\mu\nu}
-\frac{1}{2}g^{\mu\nu}R^\alpha_{\phantom{\alpha}\alpha}\right)         
-g^{\mu\nu}\lambda S_0^4~~.
\label{220}
\end{equation}                                 
An incoherent averaging of
$i\bar{\psi}\gamma^{\mu}(x)[\partial^{\nu}+\Gamma^\nu(x)] \psi$  over 
all the fermionic modes propagating in a Robertson-Walker background
will bring the  fermionic contribution to $T^{\mu\nu}$  to the form of a
kinematic perfect fluid
\begin{equation}
T_{{\rm kin}}^{\mu \nu} =
\frac{1}{c}\left[(\rho_m+p_m)U^{\mu}U^{\nu}+p_mg^{\mu\nu}\right]~~,
\label{221}
\end{equation}                                 
with the conformal cosmology equation
of motion then taking the form
\cite{Mannheim1992,Mannheim2000,Mannheim2001}
\begin{equation}
T^{\mu \nu} = T_{{\rm kin}}^{\mu \nu} 
-\frac{1}{6}S_0^2\left(R^{\mu\nu}
-\frac{1}{2}g^{\mu\nu}R^\alpha_{\phantom{\alpha}\alpha}\right)         
-g^{\mu\nu}\lambda S_0^4=0~~.
\label{222}
\end{equation}                                 
Despite its appearance, the $T^{\mu \nu}=0$ condition can actually be
satisfied non-trivially, since the vanishing of the full $T^{\mu\nu}$ can
be effected by a cancellation of the perfect fluid contribution
against the $-(1/6)S_0^2\left(R^{\mu\nu}
-g^{\mu\nu}R^\alpha_{\phantom{\alpha}\alpha}/2\right)$ term associated
with the back reaction of the scalar field on the geometry. Thus in the
conformal theory there is energy density not just in the matter fields
but in the geometry as well (conformal invariance requires the
presence of the $-S^2R^\alpha_{\phantom{\alpha}\alpha}/12$ term in Eq.
(\ref{61}) with its necessarily negative sign), with it being this
additional energy density which actually drives the theory. Finally, we
note that despite the fact that it is only the full $T^{\mu \nu}$ which
is covariantly conserved, because the Einstein tensor and metric tensor
terms in Eq. (\ref{222}) are independently covariantly conserved, it
follows that $T_{{\rm kin}}^{\mu \nu}$ is covariantly conserved also. In
the $S=S_0$ gauge then the perfect fluid does not exchange energy and
momentum with the gravitational field, and thus obeys precisely the same
covariant conservation condition as is used in the standard theory, with
fermionic particles on average moving geodesically in the
Robertson-Walker geometry.\footnote{The energy-momentum tensor of Eq.
(\ref{222}) provides us with an explicit example of a phenomenon we
discussed earlier, namely that covariant conservation of the
kinematic fluid energy-momentum tensor is achievable even if the full
energy-momentum tensor differs from the purely kinematic one.}

With a slight rewriting of Eq. (\ref{222}) as
\begin{equation}
\frac{1}{6}S_0^2\left(R^{\mu\nu}
-\frac{1}{2}g^{\mu\nu}R^\alpha_{\phantom{\alpha}\alpha}\right) = 
T_{{\rm kin}}^{\mu \nu} -g^{\mu\nu}\lambda S_0^4~~,
\label{223}
\end{equation}                                 
we recognize the conformal cosmology Eq. (\ref{223}) as being of the form
of none other than the cosmological evolution equation of the standard
theory, save only for the fact that the standard $G$ has been replaced by
an effective, dynamically induced one given by
\begin{equation}
G_{{\rm eff}}=-\frac{3c^3}{4\pi S_0^2}~~.
\label{224}
\end{equation}                                 
Thus as noted in \cite{Mannheim1992}, conformal cosmology is 
controlled by an effective gravitational coupling constant which is
repulsive rather than attractive, and which becomes smaller the larger
$S_0$ might be.\footnote{In the conformal theory local non-relativistic
solar system gravity is controlled by the parameter $\beta$ in the
metric of Eq. (\ref{204}). With the conformal coupling constant
$\alpha_g$ not participating in Eq. (\ref{188}) in homogeneous
geometries such as the cosmological one in which the Weyl tensor is
zero, while participating in the static source $f(r)$ of Eq. (\ref{200})
in geometries such as the inhomogeneous Schwarzschild one in which the
Weyl tensor is non-zero,
$G_{{\rm eff}}$ is completely decoupled from the local $G=\beta c^2/M$
associated with a source of mass $M$. With the sign of the local $G$
being fixed by the sign of
$\alpha_g$, a negative effective global cosmological $G_{{\rm eff}}$ is
not in conflict with the existence of a positive local $G$. The fact
that the dynamically induced $G_{{\rm eff}}$ is negative in the
conformal theory had been thought of as being a disadvantage since it
seemed to imply that the local $G$ would be given by the same negative
$G_{{\rm eff}}$ and then be repulsive too. However, as we see, a
repulsive global cosmological $G_{{\rm eff}}$ and an attractive local
$G$ can coexist in one and the same theory, an aspect of the theory
which can now actually be regarded as a plus given the recent discovery
of cosmic repulsion. (Recall that the best fit line of Eq. (\ref{141})
to the accelerating universe data would go through the negative
$\Omega_M(t_0)=-0.34$ if $\Lambda=0$.)} If now we define conformal
analogs of the standard
$\Omega_M(t)$ and
$\Omega_{\Lambda}(t)$ via
\begin{equation}
\bar{\Omega}_{M}(t)=\frac{8\pi G_{{\rm eff}}\rho_{m}(t)}{3c^2H^2(t)}~~,~~
\bar{\Omega}_{\Lambda}(t)=\frac{8\pi G_{{\rm
eff}}\Lambda}{3cH^2(t)}~~
\label{225}
\end{equation}                                 
where $\Lambda=\lambda S_0^4$, 
then in a Robertson-Walker geometry Eq. (\ref{223}) yields
\begin{eqnarray}
\dot{R}^2(t) +kc^2
&=&\dot{R}^2(t)\left(\bar{\Omega}_{M}(t)+
\bar{\Omega}_{\Lambda}(t)\right)~~,~~\bar{\Omega}_M(t)+
\bar{\Omega}_{\Lambda}(t)+\bar{\Omega}_k(t)=1~~,
\nonumber \\
~~q(t)&=&\frac{1}{2}\left(1+\frac{3p_m}{\rho_m}\right)\bar{\Omega}_M(t)
-\bar{\Omega}_{\Lambda}(t)
\label{226}
\end{eqnarray}
as the evolution equation of conformal
cosmology,\footnote{In Eq. (\ref{226}) the quantity $\bar{\Omega}_k(t)$
is the same as the previously defined $\Omega_k(t)=-kc^2/\dot{R}^2$.} an
evolution equation which only departs from the standard Friedmann
evolution equation of Eq. (\ref{138}) through the replacement of $G$ by
$G_{{\rm eff}}$.\footnote{With conformal invariance forcing the matter
field action to be second order, in geometries in which the Weyl tensor
vanishes the gravitational equations are only second order, and thus
coincide in form with those of the standard second order Friedmann
equations. Thus even for cosmology a preference for simplicity of the
equations would not favor standard gravity over the conformal
alternative.} With $G_{{\rm eff}}$ being smaller the larger
$S_0$ itself might be, we see that the larger the cosmological constant
of the theory (viz. the larger $\lambda S_0^4$), the less it
will contribute to cosmic evolution, with $\bar{\Omega}_{\Lambda}(t)$
having a self-quenching property not possessed by the standard 
$\Omega_{\Lambda}(t)$, as the latter comes with a fixed, pre-assigned
$G$. Consequently, in the conformal theory a large unquenched $\Lambda$
can still yield a small $\bar{\Omega}_{\Lambda}(t)$, with it thus being
possible to have a $\Lambda$ as big as particle physics suggests, and
yet, as we shall see in detail below, nonetheless not be in conflict
with observation. The essence of the conformal gravity treatment of the
cosmological constant then will be not to quench $\Lambda$ at all but to
instead quench the amount (viz. $G_{{\rm eff}}$) by which it gravitates.

In applying Eq. (\ref{226}) to cosmology we have to recognize three
distinct epochs. Based on our notion of a universe that goes through
grandunification and electroweak symmetry breaking phase transitions
as it expands and cools, and of each such phase transition
having its own scalar field order parameter $S_0$ and critical
temperature $T_V$, the universe will start at the highest temperatures
above all phase transitions (where every $S_0$ will be zero), cool
through a grand-unified phase transition temperature $T^{{\rm GUT}}_V$
with order parameter $S_0^{{\rm GUT}}$, and then subsequently go through
the electroweak phase transition temperature $T^{{\rm EW}}_V$ with order
parameter
$S_0^{{\rm EW}}$ at which fermion and intermediate vector boson masses
are generated. For temperatures above $T^{{\rm EW}}_V$ all
particles will be massless, and $T_{{\rm kin}}^{\mu \nu}$ will act as a
radiation fluid with
\begin{eqnarray}
\rho_m=3p_m=\frac{A}{R^4(t)}~~.
\label{227}
\end{eqnarray}
For temperatures below $T^{{\rm EW}}_V$ there will be a net induced
cosmological constant which we can represent by an equivalent 
black-body with necessarily negative $\Lambda=-\sigma T_V^4/c$ (the free
energy having been being lowered by the phase transition, not raised)
where $T_V$ represents some blended value of $T^{{\rm GUT}}_V$ and
$T^{{\rm EW}}_V$. Analogously, $S_0^{{\rm GUT}}$ and $S_0^{{\rm EW}}$
can be represented by some blended value $S_0$. Since both of the
$T^{{\rm GUT}}_V$ and $T^{{\rm EW}}_V$ temperatures are so huge, it will
not matter which particular blended $T_V$ we might use since compared
to current temperatures it will be huge also; and with $\sigma T_V^4$
then being overwhelmingly larger then the currently measured
$\rho_m(t_0)$, it will not matter if we ignore particle masses
altogether.\footnote{For $T_V$ of order say $10^{15}$$^{\circ}$K and
$\rho_m(t_0)$ of order the detected luminous matter density in the
universe  at the current time $t_0$, the ratio $c\Lambda/\rho_m(t_0)$
will be of order
$10^{60}$, with $\rho_m(t)$ only being competitive with 
$c\Lambda$ at early universe temperatures of order $T_V$ where 
particle masses are anyway unimportant.} Thus we can define the model as
being one whose evolution at temperatures less than $T_V$ obeys Eq.
(\ref{226}) with
$\rho_m=3p_m=\sigma T^4$ and
$\Lambda=-\sigma T_V^4/c$ ($T_V$ being huge), and which at temperatures
above $T_V$ obeys
\begin{eqnarray}
T^{\mu\nu}_{{\rm kin}}=0
\label{228}
\end{eqnarray}
with $S_0$ being zero.\footnote{We should perhaps clarify what
we mean by $S_0=0$ here. Our view of the field $S_0$ is that it is a
classical order parameter field which represents a spontaneously
broken vacuum expectation value of a fermion composite (typically
quadrilinear for grandunification and mass generating bilinear for
electroweak) which is only operative in the critical regions below
$T_V^{\rm GUT}$ and $T_V^{\rm EW}$, and thus absent altogether at
temperatures above $T_V^{\rm GUT}$. However, there could still be
fundamental scalar fields in the theory as well (as there indeed would
for instance be if conformal gravity is given a conformal supergravity
extension). While such scalar fields would have vanishing vacuum
expectation values above the highest critical temperature (the vacuum
then being normal rather than spontaneously broken), nonetheless in such
a vacuum the expectation value of $S^2$ would not vanish, since a quantum
scalar field $\hat{S}$ can still connect a normal vacuum $|\Omega
\rangle$ to the one-particle state $|n\rangle$ created out of it. In
such a case the energy-momentum tensor associated with the kinetic
energy of the scalar field would be given (c.f. Eq. (\ref{64})) as
$T^{\mu\nu}=(2/3)S^{;\mu} S^{;\nu} -(1/6)g^{\mu\nu}S^{;\alpha}
S_{;\alpha} -(1/3)SS^{;\mu;\nu}
+(1/3)g^{\mu\nu}SS^{;\alpha}_{\phantom{;\alpha};\alpha}                          
-(1/6)S^2\left(R^{\mu\nu}
-(1/2)g^{\mu\nu}R^\alpha_{\phantom{\alpha}\alpha}\right)$ as evaluated
in modes which obey 
$S^{;\mu}_{\phantom{\mu};\mu}+(1/6)SR^\mu_{\phantom{\mu}\mu}=0$, with
the classical $S$ now denoting the quantum scalar
field matrix element $\langle \Omega |\hat{S}|n\rangle$.
Then even though the vacuum is such that $\langle \Omega |\hat{S} |\Omega
\rangle$ is zero, an incoherent sum over all $|n\rangle$ would still
generate a non-zero contribution to the scalar field bilinear
$T^{\mu\nu}$ which would then be included as part of $T_{\rm
kin}^{\mu\nu}$.}

At first glance the condition $T^{\mu\nu}_{{\rm kin}}=0$ would appear to
possess no non-trivial solution. However, that is not the case, since
$T^{\mu\nu}_{{\rm kin}}$ is constructed here as an incoherent averaging
of all the matter field modes propagating not in a flat space (where
the solution to $T^{\mu\nu}_{{\rm kin}}=0$ would of course be trivial)
but in a curved one, with the equation $T^{\mu\nu}_{{\rm kin}}=0$ being
found \cite{Mannheim2000} to actually support a non-trivial solution if
the spatial 3-curvature of the Robertson-Walker geometry is expressly
negative, since there is then negative energy density present in the
gravitational field to effect the needed cancellation. Conformal
cosmology thus specifies the sign of $k$,\footnote{That $k$ would be
negative in the conformal theory was first noted in
\cite{Mannheim1992}.} leading to the same negative value that we
previously found would impact on galactic rotation curves. With 
conformal cosmology early universe dynamics thus fixing the sign of $k$
(something not the case with the standard Friedmann cosmology), and
with $k$ being a general coordinate scalar, the global topology of the
universe would then not change as the universe cools down, with $k$ then
still being negative at temperatures below $T_V$ where Eq. (\ref{226})
applies.

\subsection{Conformal gravity and the accelerating universe}

Given the above considerations, the signs of various
parameters of the theory are then fully specified, with $k$, $\Lambda$
and $G_{{\rm eff}}$ all being negative and $\rho_m(t)$ being positive.
Given such a pattern of signs, the temperature evolution of the
cosmology is then completely determined. Specifically, with the
deceleration parameter
$q(t)$ being given by 
\begin{eqnarray}
q(t)=\bar{\Omega}_M(t)-\bar{\Omega}_{\Lambda}(t)
\label{229}
\end{eqnarray}
when $p_m=\rho_m/3$, we see that $q(t)$ is automatically negative in
every epoch, with the universe thus being a permanently accelerating one
no matter what the magnitudes of the various parameters.\footnote{With
$G_{{\rm eff}}$ and $\Lambda$ being negative and $\rho_m(t)$ being
positive, the quantity $\bar{\Omega}_{\Lambda}(t)$ is then positive
while the quantity $\bar{\Omega}_{m}(t)$ is negative.} Given the various
signs of the parameters, conformal cosmology is found to
admit of the exact solution \cite{Mannheim2001}
\begin{equation}
R^2(t)= -\frac{k(\beta-1)}{2\alpha}
-\frac{k\beta{\rm sinh}^2 (\alpha^{1/2} ct)}{\alpha}
\label{230}
\end{equation}
where 
\begin{equation}
\alpha c^2=-2\lambda S_0^2=\frac{8\pi G_{eff}\Lambda}{3c}~~,~~
\beta=\left(1- \frac{16A\lambda}{k^2
c}\right)^{1/2}~~.
\label{231}
\end{equation}
(When $A=0$ this solution reduces to the $k<0$, $\alpha>0$ solution given
in Eq. (\ref{133}).) With Eq. (\ref{230}) entailing that
the initial time $\dot{R}(t=0)$ is finite, conformal cosmology is
singularity-free ($G_{{\rm eff}}$ being repulsive rather than
attractive). The cosmology thus expands from a finite rather than
zero minimum size
$R_{{\rm min}}^2= -k(\beta-1)/2\alpha$, and thus from a finite maximum
temperature $T_{\rm max} \sim 1/R_{\rm min}$, with the temperature
evolution being given by
\begin{eqnarray}
\frac{T_{{\rm max}}^2}{T^2}&=&
1+\frac{2\beta{\rm sinh}^2 (\alpha^{1/2}
ct)}{(\beta-1)}~~,~~\beta=\frac{(T_{{\rm max}}^4+T_V^4)}{(T_{{\rm
max}}^4-T_V^4)}~~,
\nonumber \\
{\rm tanh}^2 (\alpha^{1/2}ct)&=&\left(1-\frac{T^2}{T_{max}^2}\right)
\left(1+\frac{T^2T_{max}^2}{T_V^4}\right)^{-1}~~.
\label{232}
\end{eqnarray}

In the solution of Eq. (\ref{230}) the temperature evolution of the
quantities
$\bar{\Omega}_{\Lambda}(t)$ and
$\bar{\Omega}_M(t)$ can be written in closed form as 
\begin{eqnarray}
\bar{\Omega}_{\Lambda}(t)&=& 
\left(1-\frac{T^2}{T_{max}^2}\right)^{-1}
\left(1+\frac{T^2T_{max}^2}{T_V^4}\right)^{-1}=
\left(1-\frac{T^2}{T_{max}^2}\right)^{-2}{\rm tanh}^2
(\alpha^{1/2}ct)~~,
\nonumber \\ 
\bar{\Omega}_M(t)&=&-\frac{T^4}{T_V^4}\bar{\Omega}_{\Lambda}(t)~~,
\label{233}
\end{eqnarray}
with Eq. (\ref{233}) holding at any $T(t)$ without any approximation
at all. Quite remarkably, from Eq. (\ref{233}) we see that no matter
what the actual value of $T_{\rm max}$, at temperatures which are well
below it, the late time $\bar{\Omega}_{\Lambda}(t \gg 0)$ will be given
by
\begin{equation}
\bar{\Omega}_{\Lambda}(t \gg 0) = {\rm tanh}^2 (\alpha^{1/2}ct)~~,
\label{234}
\end{equation}
to thus have to exclusively lie between zero and one, ultimately
asymptoting to a bound of one from below. Thus no matter how big $T_V$
might be, i.e. no matter how big $\Lambda$ might be, in the conformal
theory the current era $\bar{\Omega}_{\Lambda}(t_0)$ is automatically
bounded from above. With $\bar{\Omega}_M(t)$ being completely negligible
in the current era ($\bar{\Omega}_M(t_0)/\bar{\Omega}_{\Lambda}(t_0)
=-T_0^4/T_V^4=O(10^{-60})$), it follows from Eq. (\ref{226}) that the
current era $\bar{\Omega}_k(t_0)$ is given by 
\begin{equation}
\bar{\Omega}_k(t_0)= {\rm
sech}^2 (\alpha^{1/2}ct_0)
\label{235}
\end{equation}
with curvature thus contributing to current era cosmic
expansion,\footnote{With $k$ being negative and $\bar{\Omega}_M(t \gg
0)$ being negligible, the effective late time sum rule
$\bar{\Omega}_{\Lambda}(t \gg 0)+\bar{\Omega}_k(t
\gg 0)=1$ requires that
$\bar{\Omega}_{\Lambda}(t \gg 0)$ has to lie below one, with it being
the negative 3-curvature of the universe which was fixed in the very
early universe thus forcing the current era
$\bar{\Omega}_{\Lambda}(t_0)$ to be bounded from above.} with the late
time deceleration parameter $q(t \gg 0)$ being given by the accelerating
\begin{equation}
q(t \gg 0) =-{\rm tanh}^2 (\alpha^{1/2}ct)~~,
\label{236}
\end{equation}
a deceleration parameter which is therefore automatically bounded between
minus one and zero.

Not only does conformal cosmology automatically quench the current era
$\bar{\Omega}_{\Lambda}(t_0)$ and thereby show that it actually is
possible to live with a huge cosmological constant, since the cosmology
has no initial singularity, it also solves the Friedmann early universe
fine-tuning flatness problem, with the absence of
any initial singularity not obliging the initial
$\bar{\Omega}_M(t=0)+\bar{\Omega}_{\Lambda}(t=0)$ to have to be
one \cite{Mannheim1992}. Further, the conformal cosmology also solves
the cosmic coincidence problem, since $\bar{\Omega}_M(t_0)$ is nowhere
near close in value to $\bar{\Omega}_{\Lambda}(t_0)$ at all; and with
$\bar{\Omega}_M(t_0)$ not needing to be of order one (rather it is of
order $10^{-60}$) the cosmology even has no need for any cosmological
dark matter. With the conformal cosmology also having been shown to have
no horizon problem or universe age problem
\cite{Mannheim2000,Mannheim2001}, it thus directly addresses many of the
major challenges that contemporary cosmology has had to face, and does
so without needing to appeal to an early universe inflationary
era.\footnote{Since a de Sitter geometry is conformal to flat, conformal
gravity also admits of a de Sitter based cosmology
\cite{Mannheim1990}, with it not being excluded that there could have
been an early universe inflationary de Sitter phase in the conformal
theory even if the flatness and horizon problems that it addresses can
be solved without it.}

With $R(t)$ being completely determined in Eq. (\ref{230}), it is then
straightforward to calculate the dependence of the luminosity function
on redshift, with the late time luminosity function being found
\cite{Mannheim2001,Mannheim2003a} to be given by none other than the
pure $\Lambda$ universe formula given earlier as Eq. (\ref{140}), viz.
\begin{equation} 
d_L=-\frac{c}{H(t_0)}\frac{(1+z)^2}{q_0}\left(1-\left[1+q_0-
\frac{q_0}{(1+z)^2}\right]^{1/2}\right)~~,
\label{237}
\end{equation}
as expressed in terms of the current era $H(t_0)$ and $q_0$. With $q_0$
being treated as a free parameter which has to lie between minus
one and zero in the conformal case, as described earlier the accelerating
universe data are then fitted extremely well with $q_0=-0.37$ in the fits
exhibited in Figs. 2 and 3, with no fine-tuning of parameters being
required. The success of these fits shows that having a particle physics
sized cosmological constant is not incompatible with the supernovae data
at all, it is only having a particle physics sized cosmological constant
coupled to standard gravity which is incompatible. Since the conformal
theory was not constructed after the fact in order to explain the
accelerating universe data, and since it then does fit the accelerating
universe data as naturally and as readily as it does, it should be given
serious consideration. And as we indicated earlier, a $q_0=-0.37$,
$\bar{\Omega}_{\Lambda}(t_0)=0.37$, $\bar{\Omega}_{M}(t_0)=O(10^{-60})$,
$\bar{\Omega}_{k}(t_0)=0.63$ theory is a fully falsifiable one
whose continuing acceleration predictions for higher redshift exhibited
in Fig. 3 can be directly tested.

Beyond our interest here in conformal gravity as a classical theory
which is to be used to fit astrophysical and cosmological data without
dark matter or dark energy, the theory is also of interest as a
quantum theory. Since it is based on an action $I_W$ with a
dimensionless coupling constant, as a quantum gravity theory the
conformal theory is power counting renormalizable, to thus have far
better behavior in the ultraviolet that the non-renormalizable standard
theory, a difficulty which one tries to get round in the standard theory
by generalizing it to a 10-dimensional string theory. Because it is
based on fourth order equations of motion, the unitarity of quantum
conformal gravity has been called into question since it is suspected
that the theory might possess on-shell negative norm ghost states.
However, while this is indeed the case for the theories which involve
both second and fourth order components, it has recently been shown
\cite{Mannheim2005b} not to be the case for pure fourth order theories
themselves, a point we elaborate on in Appendix D. Conformal gravity may
thus well be a completely consistent quantum gravitational theory, one
which is formulated purely in four spacetime dimensions alone.

\subsection{Quenching the contribution of the cosmological constant to
cosmic evolution}

 The essence of the conformal gravity approach to the
cosmological constant problem is to have a symmetry which forces
$\Lambda$ to be zero in the exact symmetry limit (viz. above all
cosmological phase transitions), and to then, as per particle physics,
have a huge one induced as the temperature drops below the various
critical points, but to have the huge induced one gravitate far less
than it would do in the standard theory. Thus instead of trying to
quench $\Lambda$, one can instead try to quench $G$. As such, a
quenching of the cosmological
$G$ can be thought of as being a generic way to approach the
cosmological constant problem, one which could apply in theories other
than just conformal gravity, theories which need not depart so far from
the Einstein-Hilbert action as the conformal theory does. Thus in the
standard theory it could be the case that rather than be fixed,
$G$ could be a running coupling constant whose magnitude would depend on
epoch or on distance, with a difference between the cosmological early
universe $G$ and the cosmological late universe $G$ possibly allowing
one to solve the early universe fine-tuning problem, and a difference
between the large distance cosmological $G$ and the short distance local
$G$ possibly allowing one to solve the cosmological constant
problem.\footnote{For a possible role for a running $G$ in solving the
dark matter problem see \cite{Reuter2004}.} Also, it might be possible to
generate an epoch-dependent
$G$ through an embedding in a higher dimensional space, since as noted
in Eq. (\ref{161}), the 4-dimensional $G$ is then not fundamental but is
instead induced by the embedding. In having an epoch-dependent $G$ which
could quench the contribution of the cosmological constant to cosmology,
such a quenching would ordinarily be expected to quench the contribution
of ordinary matter to cosmology as well. However, in order to solve the
cosmological constant problem per se, it is actually only necessary to
quench the amount by which $\Lambda$ itself gravitates, with there being
no need to modify the amount by which $\rho_m(t)$ gravitates; with models
which can do this (by taking advantage of the fact that unlike
$\rho_m(t)$ the constant $\Lambda$ carries zero 4-momentum) currently
being under consideration \cite{Dvali2003,Dvali2002,Dvali2004}.
Given the fact that attempts to quench the cosmological constant itself
have so far foundered, attempting to instead quench the amount by
which it gravitates appears to be an attractive alternative.

\section{Future prospects and challenges}

In this review we have examined standard gravity and some of its
alternatives, and have identified the extension of the standard
Newton-Einstein gravitational theory beyond its solar system origins to
be the root cause of the dark matter and dark energy problems. However,
attempting to critique the standard theory is not an easy enterprise,
in part because the definition of what constitutes the standard theory is
something of a moving target. While it is agreed upon that the
standard theory is to be based on the Einstein equations, such a
definition leaves a great deal undetermined, as it leaves $T^{\mu\nu}$
completely unspecified, and does not make clear whether or not the
gravitational side of the Einstein equations is to also include a
fundamental cosmological constant term (to cancel the one induced in
by phase transitions in $T^{\mu\nu}$).\footnote{The freedom standard
gravity has in choosing $T^{\mu\nu}$ is due to its not using the
traceless $SU(3)\times SU(2)\times U(1)$ based $T^{\mu\nu}$ provided by
fundamental physics. And we note that if we were to couple standard
cosmology to a traceless $T^{\mu\nu}$, for $k=0$ we would obtain $R(t)
\sim t^{1/2}$ and $d_L=cz/H(t_0)$, with a best fit to the 54 data points
of Fig. 2 yielding $\chi^2=138$.} During the period between the
development of the inflationary universe model in 1981 and the discovery
of the accelerating universe in 1998 the standard model was taken to be
one in which both a fundamental and an induced cosmological constant
were zero and the matter density was at the critical density (viz.
$\Omega_M(t)=1$, $\Omega_{\Lambda}(t)=0$). Then, following the discovery
of the accelerating universe this definition was revised to mean a
universe with $\Omega_M(t_0)=0.3$, and a net $\Omega_{\Lambda}(t_0)=0.7$.
Such a universe has a ready test since it predicts deceleration at
redshifts above one (as per Fig. 3). However, if acceleration were to be
established from analysis of the redshift greater than one Hubble plot,
the response would not necessarily be to declare the standard theory
wrong but to instead introduce, say, a quintessence fluid (or possibly
even two or more such fluids) with a parameter $w=p/\rho$ which would
then depend on redshift in whatever way was needed. The weakness of the
standard theory is not so much that one necessarily would do this, but
rather that one could do it, i.e. that even now the theory is still not
uniquely specified in a way that could enable it to make fully
falsifiable predictions from which the theory could not subsequently
back away.\footnote{The contrast with conformal gravity theory for
instance in this regard is that the conformal theory is a uniquely
defined theory (with gravitational action $I_W$ and $SU(3)\times
SU(2)\times U(1)$ invariant matter action) which unambiguously predicts
continuing acceleration above a redshift of one. Showing that the
universe is in fact decelerating above a redshift of one would thus rule
out the conformal theory, while showing that it was accelerating above a
redshift of one would not necessarily rule out the standard theory.
Alternate theories such as conformal gravity and MOND which endeavor to
describe the universe without dark matter or dark energy do not enjoy
the luxury of still having functions to adjust, and any major failure
would rule them out.}

Nonetheless, even if the road which led to the present
$\Omega_M(t_0)=0.3$, $\Omega_{\Lambda}(t_0)=0.7$ standard paradigm was 
circuitous, it is still the case that these particular numbers work
extremely well when the standard theory is applied to the anisotropy
structure of the cosmic microwave background. In fact so well do they
work that it is doubted that there could possibly be any second theory
which could do as well. However, while it is extremely unlikely for
lightning to strike in the same place twice, as far as the standard
theory is concerned one has to ask if lightning has even struck once --
namely do $\Omega_M(t_0)=0.3$ and $\Omega_{\Lambda}(t_0)=0.7$ really
describe the energy content of the universe, or is the true
$\rho_M(t_0)$ in the real world the one given by luminous matter alone
and the true $\Lambda$ the one given by a particle physics scale,
since if they are, the standard model fits would then be disastrous. 

However, if in the real world $\rho_M(t_0)$ and $\Lambda$ are such that
$\Omega_M(t_0)$ is equal to $0.01$ and
$\Omega_{\Lambda}(t_0)$ is of order $10^{60}$, one has to ask why
does $\Omega_M(t_0)=0.3$, $\Omega_{\Lambda}(t_0)=0.7$ then work. As far
as the supernovae Hubble plot data are concerned an answer can be
provided to this question. Specifically, as we noted earlier, the best
($\Omega_M(t_0)$, $\Omega_{\Lambda}(t_0$) plane fits to the
supernovae data lie along the
$\Omega_{\Lambda}(t_0)=1.1\Omega_M(t_0)+0.37$ curve given as Eq.
(\ref{141}). The best fit minimum is thus quite shallow, and if the
conformal gravity theory is correct and the real world lies close to the
$\bar{\Omega}_M(t_0)=0$, $\bar{\Omega}_{\Lambda}(t_0)=0.37$ solution to
Eq. (\ref{226}) (a solution which as far as purely phenomenological
fitting is concerned is equivalent to using the $\Omega_M(t_0)=0$,
$\Omega_{\Lambda}(t_0)=0.37$ solution to Eq. (\ref{138})), the
$\Omega_M(t_0)=0.3$, $\Omega_{\Lambda}(t_0)=0.7$ standard model solution
would then just happen to be close to the real solution. Of course, the
argument cuts both ways, and it could be that it is the conformal theory
fit which is the one which is accidentally correct. To resolve this issue
will require fitting the cosmic microwave background with
conformal gravity to see if it works too,\footnote{To apply the conformal theory to the cosmic
microwave background requires the development of a theory for the
growth of inhomogeneities in the model, to see what size standard
yardstick the fluctuations would impose on the recombination
sky, and what size they would then appear to us. Since the size of the
fluctuation yardstick of the conformal theory would be different from
the one used in the standard theory, the conformal gravity value
$\bar{\Omega}_{k}(t_0)=0.63$ need not be in conflict with the
standard model's $\Omega_{k}(t_0)=0$.}
and if it does, to then see whether the  success of either of the
two theories (or of any other candidate theory) could reveal
why the successes of the other theories were then only accidental. 

The challenge to alternate theories to fit the cosmic microwave
background data is a daunting one, not just calculationally, but also
because alternate programs are extremely manpower limited with few
people working on them.\footnote{For the conformal gravity theory for
instance, while it has been shown that the metric produced by the sun 
yields the standard solar system Schwarzschild phenomenology, PPN-type
corrections to this phenomenology have yet to worked out. While there is
(necessarily) gravitational radiation reaction in the conformal theory,
its effect on the decay of the orbit of a binary pulsar has yet to be
worked out. Application of Eq. (\ref{211}) to galactic rotation curves
has so far only been made for the typical eleven galaxy sample of Fig.
(1). Studies of clusters of galaxies, gravitational lensing by them and
anisotropies in the cosmic microwave background have yet to be made
and all of them await the development of a treatment of the growth of
inhomogeneities in the conformal theory. Such a treatment is also needed
for the nucleosynthesis of primordial deuterium as not enough of it can
survive \cite{Knox1993,Elizondo1994,Lohiya1999,Sethi1999} in a strictly 
homogeneous cosmology which expands as slowly as that of Eq. (\ref{230})
-- though unlike the standard cosmology, this very same slow expansion
does allow
\cite{Lohiya1999} the conformal cosmology to get past the barrier
associated with the absence of stable nuclei with $A=8$ and primordially
synthesize $^9{\rm Be}$ and beyond. For cosmology, apart from the fits
of Figs. (2) and (3) to the accelerating universe data, at the present
time all the other achievements  of the conformal theory are
theoretical, with the cosmology having been shown to possess no flatness
problem, no horizon problem, no universe age problem, no cosmic
coincidence fine-tuning problem, no dark matter problem, no dark energy
problem, and no cosmological constant problem.} However, this has to be
weighed against the fact that the cosmological constant problem is an
extremely challenging one for the standard theory, and here there are
many workers and yet little progress.\footnote{One would especially like
to see string theory come up with an explanation for why
$\Omega_{\Lambda}(t_0)$ is equal to 0.7, and by that we mean not just to
explain why it is not of order $10^{60}$ to $10^{120}$, but why it is
not even equal to 1.7 or -1.7.} However, while this is all
for the future, for the present three instructive guidelines have been
identified  -- namely that there is good evidence that it is a universal
acceleration scale which determines when a luminous Newtonian
expectation is to fail to fit data, that a good case can be made that
there is a global cosmological effect on local galactic motions which
galactic dark matter might be simulating, and that in attempting to
solve the cosmological constant problem one does not actually need to
quench the cosmological constant itself, but only the amount by which it
gravitates.

\appendix  

\setcounter{equation}{0}
\def\theequation{A\arabic{equation}}

\section{The potential of a thin disk}
\medskip

In order to determine the weak gravity potential of an extended object such 
as a disk of stars each with gravitational potential $V(r)=-\beta c^2/r+
\gamma c^2r/2 $, we follow an approach originally developed by Casertano
\cite{Casertano1983} for both thin and thick Newtonian disks and then
generalize it to disks with linear potentials. For the Newtonian potential
of an initially non-thin axially symmetric distribution of matter sources
with matter volume density function  $\rho(R^{\prime},z^{\prime})$ we need
to evaluate the quantity
\begin{equation}
V_{\beta}(R,z)=-\beta c^2 
\int_0^{\infty}dR^{\prime} \int_0^{2\pi}d\phi^{\prime}
\int_{-\infty}^{\infty}dz^{\prime}  {R^{\prime}\rho(R^{\prime},z^{\prime}) \over 
(R^2+R^{\prime 2}-2RR^{\prime}cos\phi^{\prime}+(z-z^{\prime})^2)^{1/2}} 
\label{A1}
\end{equation} 
where $R^{\prime},~\phi^{\prime},~z^{\prime}$ are cylindrical coordinates 
of the source and $R$ and $z$ are the only observation point coordinates
of relevance. To evaluate Eq. (\ref{A1}) it is convenient to make use of
the cylindrical  coordinate Green's function Bessel function expansion
\begin{equation}
{1 \over \vert \vec{r} -\vec{r^{\prime}} \vert }=\sum_{m=-\infty}^{\infty}
\int_0^\infty dk J_m(kr)J_m(kr^{\prime})
e^{im(\phi-\phi^{\prime})-k \vert  z -  z^{\prime} \vert }~~,
\label{A2}
\end{equation} 
with its insertion into Eq. (\ref{A1}) yielding 
\begin{equation}
V_{\beta}(R,z)=-2\pi\beta c^2\int_{0}^{\infty}
dk\int_{0}^{\infty}dR^{\prime}
\int_{-\infty}^{\infty}dz^{\prime}
R^{\prime} \rho(R^{\prime},z^{\prime}) J_0(kR)J_0(kR^{\prime}) e^{-k\vert z
-  z^{\prime}\vert}~~.
\label{A3}
\end{equation} 
For the case of infinitesimally thin disks with
$\rho(R^{\prime},z^{\prime})=\Sigma(R^{\prime})\delta(z^{\prime})$ Eq. (\ref{A3}) then
simplifies to 
\begin{equation}
V_{\beta}(R)=-2\pi\beta c^2\int_{0}^{\infty} dk\int_{0}^{\infty}dR^{\prime}
R^{\prime} \Sigma(R^{\prime})J_0(kR)J_0(kR^{\prime})
\label{A4}
\end{equation} 
for observation points in the $z=0$ plane of the disk.
Hence, for a disk with an exponential matter distribution
$\Sigma(R^{\prime})=\Sigma_0e^{-\alpha R^{\prime}}$ and total number of
stars
$N=2\pi\Sigma_0/\alpha^2$ ($R_0=1/\alpha$ is the scale length of the disk),
use of the standard Bessel function integral formulae
\begin{equation}
\int_0^\infty dR^{\prime}
R^{\prime}J_0(kR^{\prime})e^{-\alpha R^{\prime}}= {\alpha \over
(\alpha^2+k^2)^{3/2}}~~,
\label{A5}
\end{equation} 
\begin{equation}
\int_0^\infty dk {J_0(kR) \over (\alpha^2+k^2)^{3/2}}
=\frac{R}{2\alpha}\left[I_0\left(\frac{R\alpha}{2}\right)
K_1\left(\frac{R\alpha}{2}\right)-
I_1\left(\frac{R\alpha}{2}\right)
K_0\left(\frac{R\alpha}{2}\right)\right]~~,
\label{A6}
\end{equation} 
then leads directly to 
\begin{eqnarray}
V_{\beta}(R)&=&-2\pi\beta c^2\Sigma_0 \int_{0}^{\infty}dk
{\alpha J_0(kR) \over (\alpha^2
+k^2)^{3/2}}
\nonumber \\
&=& -\pi\beta c^2\Sigma_0
R\left[I_0\left(\frac{R\alpha}{2}\right)
K_1\left(\frac{R\alpha}{2}\right)-
I_1\left(\frac{R\alpha}{2}\right)
K_0\left(\frac{R\alpha}{2}\right)\right]~~.
\label{A7}
\end{eqnarray} 
Finally, on differentiating Eq. (\ref{A7}) and using the modified Bessel
function relations
\begin{equation}
I_0^{\prime}(z)=I_1(z)~~,~~I_1^{\prime}(z)=I_0(z)-\frac{I_1(z)}{z} ~~,
~~K_0^{\prime}(z)=-K_1(z)~~,~~K_1^{\prime}(z)=-K_0(z)-\frac{K_1(z)}{z}~~,
\label{A8}
\end{equation} 
we obtain the relation 
\begin{equation}
RV^{\prime}(R)=
\frac{N\beta c^2 R^2\alpha^3}{2}
\left[I_0\left(\frac{R\alpha}{2}\right)
K_0\left(\frac{R\alpha}{2}\right)-
I_1\left(\frac{R\alpha}{2}\right)
K_1\left(\frac{R\alpha}{2}\right)\right]
\label{A9}
\end{equation} 
presented in the text.

The utility of the above formalism is that it immediately generalizes to 
the linear potential case (and by extension to the cubic potential and so
on), since on setting
$\vert
\vec{ r} -\vec{ r^{\prime}}
\vert = (\vec{ r} -\vec{ r^{\prime}})^2/
\vert \vec{r} -\vec{r^{\prime}} \vert$, we immediately obtain the potential
\begin{eqnarray}
V_{\gamma}(R,z)&=&{\gamma c^2\over 2}
\int_0^{\infty}dR^{\prime} \int_0^{2\pi}d\phi^{\prime}
\int_{-\infty}^{\infty}dz^{\prime} R^{\prime}\rho(R^{\prime},z^{\prime})  
[R^2+R^{\prime 2}-2RR^{\prime}cos\phi^{\prime}+(z-z^{\prime})^2]^{1/2}
\nonumber \\
&=&\pi\gamma c^2
\int_0^{\infty}dk \int_0^{\infty}dR^{\prime} \int_{-\infty}^{\infty}dz^{\prime} 
R^{\prime}\rho(R^{\prime},z^{\prime})
\bigg{[}(R^2+R^{\prime 2}+(z-z^{\prime})^2)J_0(kR)J_0(kR^{\prime})
\nonumber \\
&&~~~~~~~~~~~~~~~~~~~~~~~~~~~~~~~~~~~~~~~~~~~~~~~~~~~-2RR^{\prime}
J_1(kR)J_1(kR^{\prime})\bigg{]} e^{-k\vert z -  z^{\prime}\vert}~~.
\label{A10}
\end{eqnarray} 
For an infinitesimally thin disk Eq. (\ref{A10}) then reduces at $z=0$ to
\begin{equation}
V_{\gamma}(R)=\pi\gamma c^2\int_{0}^{\infty}
dk\int_{0}^{\infty}dR^{\prime}R^{\prime}
\Sigma(R^{\prime})\left[ (R^2+R^{\prime 2})J_0(kR)J_0(kR^{\prime}) 
-2RR^{\prime}
J_1(kR)J_1(kR^{\prime})\right]~~.
\label{A11}
\end{equation} 
The use of Eqs. (\ref{A5}) and (\ref{A6}) and the additional integral 
formula
\begin{equation}
\int_0^\infty dR^{\prime} R^{\prime
2}J_1(kR^{\prime})e^{-\alpha R^{\prime}} ={3\alpha k\over
(\alpha^2+k^2)^{5/2}}
\label{A12}
\end{equation} 
then yields 
\begin{eqnarray}
V_{\gamma}(R)&=&
\pi \gamma c^2\Sigma_0
\int_{0}^{\infty}dk 
\left( {\alpha R^2 J_0(kR) \over (\alpha^2+k^2)^{3/2}}
-{9\alpha J_0(kR) \over (\alpha^2+k^2)^{5/2}}
+{15\alpha^3J_0(kR) \over (\alpha^2+k^2)^{7/2}}
-{6\alpha kRJ_1(kR) \over (\alpha^2+k^2)^{5/2}} \right)
\nonumber \\
&=&\pi\gamma c^2\Sigma_0
\int_{0}^{\infty}dk J_0(kR)
\left( {\alpha R^2 \over (\alpha^2+k^2)^{3/2}}
+{15\alpha \over (\alpha^2+k^2)^{5/2}}
-{15\alpha^3 \over (\alpha^2+k^2)^{7/2}} \right)~~.
\label{A13}
\end{eqnarray} 
Through use of the modified Bessel function relations given above
taken in conjunction with Eq. (\ref{A6}) and its derivatives, Eq. (\ref{A13}) readily
evaluates to
\begin{eqnarray}
V_{\gamma}(R)=~\frac{\pi\gamma c^2\Sigma_0
R}{\alpha^2}\left[I_0\left(\frac{R\alpha}{2}\right)
K_1\left(\frac{R\alpha}{2}\right)-
I_1\left(\frac{R\alpha}{2}\right)
K_0\left(\frac{R\alpha}{2}\right)\right]&&
\nonumber \\
+ \frac{\pi\gamma
c^2\Sigma_0R^2}{2\alpha}\left[I_0\left(\frac{R\alpha}{2}\right)
K_0\left(\frac{R\alpha}{2}\right)+
I_1\left(\frac{R\alpha}{2}\right)
K_1\left(\frac{R\alpha}{2}\right)\right]&&. 
\label{A14}
\end{eqnarray} 
Finally, differentiation with respect to $R$ and repeated use of the
recurrence relations of Eq. (\ref{A8}) then yields the expression
\begin{equation}
RV^{\prime}(R)=
\frac{N\gamma c^2R^2\alpha}{2}I_1\left(\frac{R\alpha}{2}\right)
K_1\left(\frac{R\alpha}{2}\right)
\label{A15}
\end{equation} 
presented in the text.
\medskip

\section{The potential of a separable thick disk}
\medskip

For non-thin disks simplification of Eqs. (\ref{A1}) and
(\ref{A10}) can be achieved for disks with a separable matter distribution
$\rho(R^{\prime},z^{\prime})=\Sigma(R^{\prime})f(z^{\prime})$ where the
symmetric thickness function $f(z^{\prime})=f(-z^{\prime})$ is
normalized according to
\begin{equation}
\int_{-\infty}^\infty dz^{\prime}f(z^{\prime})
=2\int_0^\infty dz^{\prime}f(z^{\prime})=1~~.
\label{A16}
\end{equation} 
\noindent 
Recalling that 
\begin{equation}
e^{-k\vert z -  z^{\prime}\vert}=\theta(z -  z^{\prime})e^{-k(z -  z^{\prime})}+
\theta(z^{\prime}-z)e^{+k(z -  z^{\prime})}~~,
\label{A17}
\end{equation} 
we find that Eqs. (\ref{A1}) and (\ref{A10}) then respectively yield for
points with
$z=0$
\begin{equation}
V_{\beta}(R)=-4\pi\beta c^2\int_{0}^{\infty} dk\int_{0}^{\infty}dR^{\prime}
\int_0^{\infty}dz^{\prime}R^{\prime} \Sigma(R^{\prime})f(z^{\prime})
J_0(kR)J_0(kR^{\prime}) e^{-kz^{\prime}}
\label{A18}
\end{equation} 
and
\begin{eqnarray}
V_{\gamma}(R)&=&2\pi\gamma c^2\int_{0}^{\infty}
dk\int_{0}^{\infty}dR^{\prime}
\int_0^{\infty}dz^{\prime}R^{\prime}
\Sigma(R^{\prime})f(z^{\prime})e^{-kz^{\prime}}
\nonumber \\
&&\times\bigg{[}(R^2+R^{\prime 2}+z^{\prime 2})J_0(kR)J_0(kR^{\prime}) 
-2RR^{\prime} J_1(kR)J_1(kR^{\prime})\bigg{]}
\label{A19}
\end{eqnarray} 
in the separable case. Further simplification is possible if
$\Sigma(R^{\prime})$  is again the exponential
$\Sigma_0e^{-\alpha R^{\prime}}$, with use of
the recurrence relation 
$J_1^{\prime}(z)=J_0(z)-J_1(z)/z$ then yielding 
\begin{equation}
RV_{\beta}^{\prime}(R)=2N\beta c^2\alpha^3 R \int_{0}^{\infty}dk
\int_0^{\infty}dz^{\prime}{ f(z^{\prime})e^{-kz^{\prime}}kJ_1(kR)
\over (\alpha^2+k^2)^{3/2}}
\label{A20}
\end{equation} 
and
\begin{eqnarray}
RV_{\gamma}^{\prime}(R)&=&N \gamma c^2\alpha^3 R \int_{0}^{\infty}dk
\int_0^{\infty}dz^{\prime} f(z^{\prime})e^{-kz^{\prime}}
\nonumber\\
&&\times\left( -{4RJ_0(kR) \over (\alpha^2+k^2)^{3/2}}
+{6\alpha^2RJ_0(kR) \over (\alpha^2+k^2)^{5/2}}
-{(R^2+z^{\prime 2})kJ_1(kR) \over (\alpha^2+k^2)^{3/2}}
+{9kJ_1(kR) \over (\alpha^2+k^2)^{5/2}}
-{15\alpha^2kJ_1(kR) \over (\alpha^2+k^2)^{7/2}} \right).
\nonumber \\
\label{A21}
\end{eqnarray} 

Further simplification is possible if a specific form for $f(z^{\prime})$
is specified, with two commonly considered ones being of the form
\begin{equation}
f(z^{\prime})=\frac{1}{2z_0}{\rm
sech}^2\left(\frac{z^{\prime}}{z_0}\right)~~,
\label{A22}
\end{equation} 
and
\begin{equation}
f(z^{\prime})=\frac{1}{\pi
z_0}{\rm sech}\left(\frac{z^{\prime}}{z_0}\right)~~,
\label{A23}
\end{equation} 
with both of them falling off very rapidly once $z$ is much greater than
the scale height $z_0$. 
The thickness function of Eq. (\ref{A22}) is found
to lead to rotational velocities of the form
\begin{eqnarray}
RV^{\prime}_{\beta}(R)&=&
\frac{N\beta c^2\alpha^3 R^2}{2}\left[I_0\left(\frac{\alpha R}{2}\right)
K_0\left(\frac{\alpha R}{2}\right)- 
I_1\left(\frac{\alpha R}{2}\right)
K_1\left(\frac{\alpha R}{2}\right)\right]
\nonumber\\
&&-N \beta c^2\alpha^3 R\int_0^\infty dk {k^2J_1(kR)z_0
\over 
(\alpha^2+k^2)^{3/2}}\beta\left(1+{kz_0 \over 2}\right)
\label{A24}
\end{eqnarray} 
and
\begin{eqnarray}
RV_{\gamma}^{\prime}(R)&=&N\gamma c^2\alpha^3 R \int_{0}^{\infty}dk
\left[1-kz_0\beta\left(1+{kz_0 \over 2}\right)\right]
\nonumber\\
&&\times~ \left( -{2RJ_0(kR) \over (\alpha^2+k^2)^{3/2}}
+{3\alpha^2RJ_0(kR) \over (\alpha^2+k^2)^{5/2}}
-{R^2kJ_1(kR) \over 2(\alpha^2+k^2)^{3/2}}
+{9kJ_1(kR) \over 2(\alpha^2+k^2)^{5/2}}
-{15\alpha^2kJ_1(kR) \over 2(\alpha^2+k^2)^{7/2}} \right)
\nonumber\\
&&~~+N\gamma c^2\alpha^3 R \int_{0}^{\infty}dk{kJ_1(kR)
\over 2(\alpha^2+k^2)^{3/2}} 
{d^2 \over dk^2}\left[ kz_0\beta\left(1+{kz_0 \over 2}\right)\right]~~,
\label{A25}
\end{eqnarray} 
where
\begin{equation}
\beta(x)=\int_0^1dt{t^{x-1} \over (1+t)}~~. 
\label{A26}
\end{equation} 
Similarly, the thickness function of Eq. (\ref{A23}) leads to
\begin{equation}
RV^{\prime}_{\beta}(R)={2N\beta c^2\alpha^3 R \over \pi}
\int_0^\infty dk {kJ_1(kR) \over (\alpha^2+k^2)^{3/2}}\beta\left({1+kz_0
\over 2}\right)
\label{A27}
\end{equation} 
and
\begin{eqnarray}
RV_{\gamma}^{\prime}(R)&=&{N\gamma c^2\alpha^3 R \over \pi}
\int_{0}^{\infty}dk
\beta\left({1+kz_0 \over 2}\right)
\nonumber\\
&&\times~ \left( -{4RJ_0(kR) \over (\alpha^2+k^2)^{3/2}}
+{6\alpha^2RJ_0(kR) \over (\alpha^2+k^2)^{5/2}}
-{R^2kJ_1(kR) \over (\alpha^2+k^2)^{3/2}}
+{9kJ_1(kR) \over (\alpha^2+k^2)^{5/2}}
-{15\alpha^2kJ_1(kR) \over (\alpha^2+k^2)^{7/2}} \right)
\nonumber\\
&&-{N\gamma c^2\alpha^3 R \over \pi} \int_{0}^{\infty}dk{kJ_1(kR)
\over (\alpha^2+k^2)^{3/2}}
{d^2 \over dk^2}\left[\beta\left({1+kz_0 \over 2}\right) \right] ~~.
\label{A28}
\end{eqnarray} 
\noindent
In all of these expressions the needed functions of $\beta(x)$ 
and its derivatives all converge very rapidly to their asymptotic
values as their arguments increase. Consequently the relevant $k$ 
integrations all converge very rapidly. As a practical matter, galactic
scale heights $z_0$ are usually much smaller than galactic disk scale
lengths $R_0$. Consequently, the thickness corrections are only of
significance in the inner galactic region, and thus have
essentially no effect on the linear potential contribution.  For the
Newtonian term the thickness corrections of Eqs. (\ref{A24}) and
(\ref{A27}) both tend to slightly reduce the overall 
Newtonian contribution, and serve to help ensure that the inner region
rotation curves of Fig. 1 are well described by the luminous Newtonian
contribution alone.

\section{The potential of a spherical bulge}

For a spherically symmetric matter distribution such as the central 
bulge region of a galaxy with radial matter density $\sigma(r)$
and $N=4\pi \int  dr^{\prime}r^{\prime 2}\sigma(r^{\prime})$ stars, 
the potential is readily found to take the form
\begin{equation}
rV_{\beta}^{\prime}(r)={4\pi\beta c^2\over r}\int_0^r
dr^{\prime}\sigma(r^{\prime}) r^{\prime 2}
\label{A29}
\end{equation} 
for a Newtonian potential, and take the form 
\begin{equation}
rV_{\gamma}^{\prime}(r)={2\pi\gamma c^2\over 3r}\int_0^r
dr^{\prime}\sigma(r^{\prime}) (3r^2r^{\prime 2}-r^{\prime 4})
+{4\pi\gamma c^2r^2\over 3}\int_r^{\infty} dr^{\prime}\sigma(r^{\prime})
r^{\prime } 
\label{A30}
\end{equation} 
for a linear potential. Both of these expressions can readily be
integrated once a specific form for $\sigma(r)$ is specified, and it is
from Eq. (\ref{A29}) that dark matter halo contributions
to galactic rotation curves are calculated. Despite the simplicity of
these expressions, because of projection effects,
it is unfortunately not the 3-dimensional $\sigma(r)$ which is directly
measured in spherical astronomical systems. Rather, it is only the
two-dimensional surface matter distribution $I(R)$ which is measured,
with $\sigma(r)$ having to be extracted from it via an Abel transform
\begin{equation}
\sigma(r)=-{1 \over \pi} \int _r^{\infty} dR{ I^{\prime}(R) 
\over (R^2-r^2)^{1/2}}~~,~~
I(R)=2\int _R^{\infty} dr {\sigma(r) r \over (r^2-R^2)^{1/2}}~~,
\label{A31}
\end{equation} 
to thus initially lead to double integrals in Eqs. (\ref{A29}) and
(\ref{A30}). 

Reduction of these integrals to one-dimensional ones
which only involve  the measured $I(R)$ is however possible since
on introducing the strip brightness $S(x)$ which obeys
\begin{equation}
S(x)=2\int_x^{\infty}dR {RI(R) \over (R^2-x^2)^{1/2}}~~,
~~\sigma(x)=-{S^{\prime}(x) \over 2\pi x}~~,
\label{A32}
\end{equation} 
we can
rewrite Eqs. (\ref{A29}) and (\ref{A30}) as
\begin{equation}
rV_{\beta}^{\prime}(r)=-2\beta c^2S(r) +  {2\beta c^2\over
r}\int_0^rdr^{\prime}S(r^{\prime})
\label{A33}
\end{equation} 
for the Newtonian potential and 
\begin{equation}
rV_{\gamma}^{\prime}(r)=\gamma  c^2r 
\int_0^rdr^{\prime}S(r^{\prime})
-{\gamma c^2\over r}\int_0^rdr^{\prime }r^{\prime 2}
S(r^{\prime})
\label{A34}
\end{equation} 
for the linear one. 
Then, use of the relations
\begin{eqnarray}
{d \over dr}\left\{ 2\int _r^{\infty}dRRI(R) {\rm arcsin} \left({r \over
R} 
\right)
 \right\}
&=&-\pi rI(r)+S(r)
\nonumber \\
{d \over dr}\left\{ 2\int _r^{\infty}dRRI(R) \left[ R^2 {\rm arcsin}
\left({r 
\over  R} \right) -r(R^2-r^2)^{1/2} \right] \right\}
&=&-\pi r^3I(r)+2r^2S(r)
\label{A35}
\end{eqnarray} 
enables us to conveniently reexpress Eqs. (\ref{A33}) and (\ref{A34})
entirely in terms of $I(R)$, to yield
\begin{equation}
rV_{\beta}^{\prime}(r)={4 \beta c^2\over r}\int_r^{\infty}dRRI(R)
\left[ {\rm arcsin} \left({r \over R} \right) -{r \over
(R^2-r^2)^{1/2}}\right]
\label{A36}
\end{equation} 
for the  Newtonian potential \cite{Kent1986} and 
\begin{eqnarray}
rV_{\gamma}^{\prime}(r)&=&{2 \pi \beta c^2\over r} \int_0^rdRRI(R) +
{\gamma c^2 \pi \over 2r} \int_0^rdRRI(R)(2r^2-R^2)
\nonumber\\
&&+{\gamma c^2  \over r}\int_r^{\infty}dRRI(R)
\left[ (2r^2-R^2){\rm arcsin} \left({r \over R} \right)
+r(R^2-r^2)^{1/2}\right]
\label{A37}
\end{eqnarray} 
for the linear one.

\section{The ghost problem in fourth order theories}

With the fourth order propagator 
\begin{equation}
D(k^2,M^2)=
\frac{1}{(k_0^2-\vec{k}^2)(k_0^2-\vec{k}^2-M^2)}
\label{A38}
\end{equation}
associated with the prototype action
\begin{equation}
I=-\frac{1}{2}\int d^4x
\left(M^2\partial_{\mu}S\partial^{\mu}S
+\partial_{\mu}\partial_{\nu}S\partial^{\mu}\partial^{\nu}S\right)
\label{A39}
\end{equation}
and equation of motion 
\begin{equation}
(-\partial_0^2+\nabla^2)(-\partial_0^2+\nabla^2-M^2)S=0
\label{A40}
\end{equation}
being writable as a sum of two opposite signatured second order
propagators, viz.
\begin{equation}
D(k^2,M^2)=
\frac{1}{M^2(k_0^2-\vec{k}^2-M^2)}
-\frac{1}{M^2(k_0^2-\vec{k}^2)}~~,
\label{A41}
\end{equation}
it is quite widely thought that when quantized, fourth order theories
such as conformal gravity would possess ghost states.
However, the equation of motion which is given in Eq. (\ref{A40}) is not
actually the equation of motion of a pure fourth order theory per se,
rather it is that of a second plus fourth order theory. With the pure
fourth order action being the one obtained by setting $M^2$ to zero in
Eq. (\ref{A39}), and with the separate positive and negative signatured
second order propagators in Eq. (\ref{A41}) becoming singular in this
limit, no conclusion about the particle content of the pure fourth order
theory can be immediately drawn. 

To investigate what the particle content of the fourth order theory does
look like, it is sufficient to specialize to field configurations of the
form $S(\bar{x},t)=q(t)e^{i\vec{k}\cdot\vec{x}}$, configurations in
which Eq. (\ref{A40}) then reduces to
\begin{equation}
\frac{d^4q}{dt^4} +(\omega_1^2+\omega_2^2)\frac{d^2q}{dt^2} 
+\omega_1^2\omega_2^2 q=0~~,
\label{A42}
\end{equation}
where
\begin{equation}
\omega_1^2+\omega_2^2=2\vec{k}^2+M^2~~,~~\omega_1^2\omega_2^2
=\vec{k}^4+\vec{k}^2M^2~~,
\label{A43}
\end{equation}
with Eq. (\ref{A42}) itself reducing to the equal frequency
$\omega_1=\omega_2$ when $M^2=0$. The equation of motion given in Eq.
(\ref{A42}) can be derived by variation of the
acceleration-dependent action first introduced by Pais and Uhlenbeck
\cite{Pais1950}
\begin{equation}
I_{PU}=\frac{\gamma}{2}\int dt\left[\ddot{q}^2
-(\omega_1^2+\omega_2^2)\dot{q}^2+\omega_1^2\omega_2^2q^2\right]
\label{A44}
\end{equation}
where $\gamma$ is a constant, and we thus seek to quantize the theory
based on $I_{PU}$ to see what happens when we take the $\omega_1
\rightarrow \omega_2$ limit.

For the theory associated with $I_{PU}$ considered first as a
classical theory, we cannot treat $q$ and 
$\dot{q}$ as independent coordinates, since if $\dot{q}$ is to be an
independent coordinate with canonical conjugate
$\partial L/\partial\ddot{q}$, we could not then use $\partial L/\partial
\dot{q}$ as the canonical conjugate of $q$. In order to reexpress
this theory in terms of an unconstrained set of variables, we 
introduce a new variable $x$ to replace
$\dot{q}$, with the method of Dirac constraints then converting the
classical theory associated with $I_{PU}$ into a theory of two coupled
classical oscillators $(q,p_q)$ and $(x,p_x)$, with the
unconstrained classical Hamiltonian 
\begin{equation}
H=\frac{p_x^2}{2\gamma}+p_qx+\frac{\gamma}{2}(\omega_1^2+\omega_2^2)x^2
-\frac{\gamma}{2}\omega_1^2\omega_2^2q^2
\label{A45}
\end{equation}
then resulting \cite{Mannheim2005b}.
With this Hamiltonian generating the closed set of Poisson bracket
relations 
\begin{eqnarray}
\{x,p_x\}&=&1~~,~~\{q,p_q\}=1~~,
\nonumber \\
\{x,H\}=\frac{p_x}{\gamma}~~,~~\{q,H\}=x~~,~~\{p_x,H\}&=&
-p_q-\gamma(\omega_1^2+\omega_2^2)x~~,
~~\{p_q,H\}=\gamma\omega_1^2\omega_2^2q~~~~
\label{A46}
\end{eqnarray}
for Poisson brackets defined by 
\begin{equation}
\{A,B\}=\frac{\partial A}{\partial x}\frac{\partial B}{\partial p_x}
-\frac{\partial A}{\partial p_x}\frac{\partial B}{\partial x}
+\frac{\partial A}{\partial q}\frac{\partial B}{\partial p_q}
-\frac{\partial A}{\partial p_q}\frac{\partial B}{\partial q}~~,
\label{A47}
\end{equation}
this Hamiltonian is indeed the correct classical one, and thus 
the appropriate one for quantization.

At the classical level, Eq. (\ref{A46}) yields equations of
motion of the form 
\begin{equation}
\dot{x}=\frac{p_x}{\gamma}~~,~~\dot{q}=x~~,~~\dot{p}_x
=-p_q-\gamma(\omega_1^2
+\omega_2^2)x~~,~~\dot{p}_q=\gamma\omega_1^2\omega_2^2q~~,
\label{A48}
\end{equation}
to nicely recover Eq. (\ref{A42}), with the substitution of
this solution into the Hamiltonian of Eq. (\ref{A45}) yielding a
stationary Hamiltonian 
\begin{equation}
H_{\rm
STAT}=\frac{\gamma}{2}\ddot{q}^2-\frac{\gamma}{2}(\omega_1^2
+\omega_2^2)\dot{q}^2 -\frac{\gamma}{2}\omega_1^2\omega_2^2 q^2-\gamma
\dot{q}
\frac{d^3q}{dt^3}
\label{A49}
\end{equation}
whose time independence in solutions to Eq. (\ref{A42}) can readily be
confirmed. The unequal frequency theory has explicit two-oscillator
solution
\begin{equation}
q(t)=a_1e^{-i\omega_1t}+a_2e^{-i\omega_2t}+{\rm c.c.}
\label{A50}
\end{equation}
with energy
\begin{equation}
H_{\rm
STAT}(\omega_1\neq\omega_2)=2\gamma(\omega_1^2-\omega_2^2)
(a_1^*a_1\omega_1^2-a_2^*a_2\omega_2^2)~~, 
\label{A51}
\end{equation}
while the $\omega_1=\omega_2=\omega$ equal frequency theory has explicit
solution
\begin{equation}
q(t)=c_1e^{-i\omega t}+c_2te^{-i\omega t}+{\rm c.c.}
\label{A52}
\end{equation}
with energy
\begin{equation}
H_{\rm
STAT}(\omega_1=\omega_2)=4\gamma
\omega^2\left(2c_2^*c_2+i\omega c_1^*c_2-i\omega c_2^*c_1\right)~~. 
\label{A53}
\end{equation}
As we see, despite the fact that the equal frequency solution involves a
temporal runaway, the appropriately defined energy is still time
independent, with temporal runaways not being the problem for fourth
order theories they are often thought to be.

Before proceeding to quantize the fourth order theory, we note that a
comparison of Eqs. (\ref{A51}) and (\ref{A53}) reveals a difference
between the unequal and equal frequency theories already at the
classical level, one that will prove to be of great significance in the
following. Specifically, while the unequal frequency theory is diagonal
in the $(a_1,a_2)$ basis, with one of the oscillators having negative
energy (the disease which translates into negative norm ghost states in
the quantum theory), the equal frequency theory is not diagonal in the
$(c_1,c_2)$ basis. Moreover, if we set $c_2=0$ we obtain 
\begin{equation}
H_{\rm
STAT}(\omega_1=\omega_2)=0~~, 
\label{A54}
\end{equation}
while if we set $c_1=0$ we obtain
\begin{equation}
H_{\rm
STAT}(\omega_1=\omega_2)=8\gamma
\omega^2c_2^*c_2~~. 
\label{A55}
\end{equation}
Of these two modes, we see that only the $c_2$ one carries energy.
The zero energy result of Eq. (\ref{A54}) is reminiscent of the conformal
gravity zero energy theorem, with Boulware, Horowitz and Strominger
\cite{Boulware1983} having shown that at the classical level the fourth
order gravity theory energy would vanish identically if the
gravitational field solutions were required to be asymptotically flat.
With the restriction to asymptotic flatness in the gravitational case
being  equivalent in the Pais-Uhlenbeck case to requiring no temporal
runaway, the equivalence of Eq. (\ref{A54}) to the zero energy theorem of
\cite{Boulware1983} is manifest. However, in our analysis of the
conformal gravity analog of the Schwarzschild solution, we found that
conformal gravity also possessed the non-asymptotically flat linear
potential term of Eq. (\ref{204}). We now see from Eq. (\ref{A55}) that
when the asymptotic flatness requirement is dropped, the temporal
runaway solution would then be allowed, and would (for an appropriate
choice of the sign of $\gamma$) actually have a perfectly acceptable
positive energy, with the constraints of the zero energy theorem (a
theorem long considered to be a shortcoming of conformal gravity) then
being evaded. With this is mind, we shall now show that the quantization
of the equal frequency Pais-Uhlenbeck theory will lead to one oscillator
which propagates with positive energy and to a second oscillator which
does not propagate at all, with the would be ghost mode not being an
eigenstate of the Hamiltonian. 

To quantize the theory based on the Hamiltonian of Eq. (\ref{A45}),
the introduction of the four Fock space operators $a_1$, $a_1^{\dagger}$,
$a_2$ and $a_2^{\dagger}$ defined via 
\begin{eqnarray}
q(t)&=&a_1e^{-i\omega_1t}+a_2e^{-i\omega_2t}+{\rm H.c.}~~,~~
p_q(t)=i\gamma \omega_1\omega_2^2a_1e^{-i\omega_1t}+
i\gamma \omega_1^2\omega_2 a_2e^{-i\omega_2t}+{\rm H.c.}~~,
\nonumber \\
x(t)&=&-i\omega_1a_1e^{-i\omega_1t}-i\omega_2 a_2e^{-i\omega_2t}+{\rm
H.c.}~~,~~
p_x(t)=-\gamma\omega_1^2a_1e^{-i\omega_1t}-
\gamma \omega_2^2a_2e^{-i\omega_2t}+{\rm H.c.}~~
\nonumber \\
\label{A56}
\end{eqnarray}
then furnishes us with a Fock space representation of the
quantum-mechanical commutation relations
\begin{equation}
[x,p_x]=[q,p_q]=i~~,~~[x,q]=[x,p_q]=[q,p_x]=[p_x,p_q]=0
\label{A57}
\end{equation}
provided the Fock space operators obey 
\begin{equation}
[a_1,a_1^{\dagger}]=\frac{1}{2\gamma\omega_1
(\omega_1^2-\omega_2^2)}~~,~~
[a_2,a_2^{\dagger}]=\frac{1}{2\gamma\omega_2
(\omega_2^2-\omega_1^2)]}~~,~~
[a_1,a_2^{\dagger}]=0~~,~~[a_1,a_2]=0~~.~~~~
\label{A58}
\end{equation}
In terms of the Fock space operators the quantum-mechanical
Hamiltonian takes the form
\begin{equation}
H=2\gamma(\omega_1^2-\omega_2^2)(\omega_1^2a_1^{\dagger}a_1-
\omega_2^2a_2^{\dagger}a_2)
+\frac{1}{2}(\omega_1+\omega_2)
\label{A59}
\end{equation}
with its associated commutators as inferred from Eq. (\ref{A46}) then
automatically being satisfied. With $\omega_1$, $\omega_2$ and
$\gamma(\omega_1^2-\omega_2^2)$ all being taken to be positive for
definitiveness, we see that the $[a_1,a_1^{\dagger}]$ commutator is
positive definite while the $[a_2,a_2^{\dagger}]$ commutator is
negative definite. And with the Hamiltonian being diagonal in the
$a_1^{\dagger}a_1$, $a_2^{\dagger}a_2$ occupation number operator basis,
we see that the state defined  by
\begin{equation}
a_1 |\Omega\rangle=0~~,~~a_2|\Omega\rangle =0
\label{A60}
\end{equation}
is its ground state, that the one-particle states
\begin{equation}
|+1\rangle =[2\gamma\omega_1(\omega_1^2-\omega_2^2)]^{1/2}
a_1^{\dagger}|\Omega\rangle~~,~~
|-1\rangle =[2\gamma\omega_2(\omega_1^2-\omega_2^2)]^{1/2}
a_2^{\dagger}|\Omega\rangle
\label{A61}
\end{equation}
are both positive energy eigenstates with respective energies $\omega_1$
and $\omega_2$ above the ground state, that the state $|+1\rangle$ has a
norm equal to plus one, but that the state $|-1\rangle$ has norm minus
one, a ghost state. Thus, as anticipated from Eq. (\ref{A41}), we find
that the Hamiltonian of the unequal frequency theory can indeed be
diagonalized in a basis of positive and negative norm
states.\footnote{Despite the presence of ghost states, the propagator of
the theory is still causal, with the relative minus sign in Eq.
(\ref{A41}) not affecting the fact that each of the two second order
propagators which appear in $D(k^2,M^2)$ is a standard causal second
order propagator.} In the presence of interactions that would
be added on to the Pais-Uhlenbeck action
$I_{PU}$, the asymptotic states for the S-matrix that such interactions
would then generate would be the eigenstates of the free Hamiltonian of
Eq. (\ref{A59}). While there are negative norm asymptotic states in the
eigenspectrum of the free Hamiltonian, in and of itself that would only
be a disaster if interactions would connect in and out states of
opposite signature, and it is thus of interest to note that Hawking and
Hertog
\cite{Hawking2002} have argued via a path integral treatment of the
constraints of $I_{PU}$ that this does not in fact occur, with the
unequal frequency theory then being viable. Hence even when $M^2$ is
non-zero, assessing fourth order theories on the basis of the structure
of Eq. (\ref{A41}) is too hasty, and a careful quantization of the
theory which takes constraints into account first needs to be made. 

With the commutation relations of Eq. (\ref{A58}) being singular in the
$\omega_1 \rightarrow \omega_2$ limit, we see that we cannot infer
anything about the structure of the equal frequency Fock space from a
study of the unequal frequency one. To take the limit we need to find
some new set of Fock operators whose commutation relations would
instead be non-singular. To this end we thus introduce the basis
\begin{eqnarray}
&&a_1=\frac{1}{2}\left(a-b+\frac{2b\omega}{\epsilon}\right)~~,~~
a_2=\frac{1}{2}\left(a-b-\frac{2b\omega}{\epsilon}\right)~~,
\nonumber \\
&&a=a_1\left(1+\frac{\epsilon}{2\omega}\right)
+a_2\left(1-\frac{\epsilon}{2\omega}\right)~~,~~b=
\frac{\epsilon}{2\omega}(a_1-a_2)~~.
\label{A62}
\end{eqnarray}
for the unequal frequency theory
where we have set
\begin{equation}
\omega=\frac{(\omega_1+\omega_2)}{2}~~,~~\epsilon=\frac{(\omega_1
-\omega_2)}{2}~~.
\label{A63}
\end{equation}
These new variables are found to obey commutation relations of the form
\begin{equation}
[a,a^{\dagger}]=\lambda~~,~~
[a,b^{\dagger}]=\mu~~,~~[b,a^{\dagger}]=\mu~~,~~
[b,b^{\dagger}]=\nu~~,~~[a,b]=0~~,
\label{A64}
\end{equation}
where
\begin{equation}
\lambda=\nu=-\frac{\epsilon^2}{16\gamma(\omega^2-\epsilon^2)\omega^3}~~,~~
\mu=\frac{(2\omega^2-\epsilon^2)}{
16\gamma(\omega^2-\epsilon^2)\omega^3}~~.
\label{A65}
\end{equation}
In terms of these new variables the coordinate $q(t)$ of Eq.
(\ref{A56}) gets rewritten as
\begin{equation}
q(t)=e^{-i\omega t}\left[(a-b){\rm cos}~\epsilon t-
\frac{2ib\omega}{\epsilon} {\rm sin}~\epsilon t\right] +{\rm H.c.}~~,
\label{A66}
\end{equation}
and thus has a well defined $\epsilon \rightarrow 0$ limit, viz.
\begin{equation}
q(t,\epsilon =0)=e^{-i\omega t}(a-b-2ib \omega t) +{\rm H.c.}~~.
\label{A67}
\end{equation}
Similarly, in the new variables the Hamiltonian of Eq. (\ref{A59}) gets
rewritten as    
\begin{equation}
H=8\gamma \omega^2\epsilon^2(a^{\dagger}a-b^{\dagger}b)+
8\gamma \omega^4 (2b^{\dagger}b+a^{\dagger}b
+ b^{\dagger}a)+\omega~~,
\label{A68}
\end{equation}
and it too has a well-defined limit, viz.
\begin{equation}
H(\epsilon=0)=8\gamma \omega^4 (2b^{\dagger}b+a^{\dagger}b
+ b^{\dagger}a)+\omega~~,
\label{A69}
\end{equation}
with the following commutators of interest having limiting 
form 
\begin{eqnarray}
&&[H(\epsilon=0),a^{\dagger}]
=\omega (a^{\dagger}+2b^{\dagger})~~,~~
[H(\epsilon=0),a]=-\omega (a+2b)~~,
\nonumber \\
&&[H(\epsilon=0),b^{\dagger}]=\omega b^{\dagger}~~,~~
[H(\epsilon=0),b]=-\omega b~~,
\nonumber \\
&&[a+b,a^{\dagger}+b^{\dagger}]=2\hat{\mu}~~,~~
[a-b,a^{\dagger}-b^{\dagger}]=-2\hat{\mu}~~,
~~[a+b,a^{\dagger}-b^{\dagger}]=0~~,
\label{A70}
\end{eqnarray}
where $\hat{\mu} =\mu(\epsilon=0)=1/(8\gamma \omega^3)$.

With the equal frequency theory now being well-defined, we find that
for the Fock vacuum $|\Omega\rangle$ defined by
$a|\Omega\rangle=b|\Omega\rangle=0$,
$H(\epsilon=0)|\Omega\rangle=\omega |\Omega\rangle$, the
$H(\epsilon=0)$ Hamiltonian possesses only one and not two one-particle
states. The state
$b^{\dagger}|\Omega\rangle$ is an eigenstate with expressly positive
energy $2\omega$, while the state $a^{\dagger}|\Omega\rangle$ is not an
eigenstate at all, a quantum-mechanical one-particle spectrum which thus
reflects the features of the classical spectrum exhibited in Eqs.
(\ref{A54}) and (\ref{A55}). With this same pattern repeating for
the multi-particle states, the equal frequency theory is thus a very
unusual one in which the full Fock space has the same number of basis
states as a two-dimensional harmonic oscillator, while the Hamiltonian
itself only has the number of basis states associated with a
one-dimensional one. The $a^{\dagger}|\Omega\rangle$ states thus only
couple off shell (where they can serve to regulate the ultraviolet
behavior of theory) but do not materialize as on-shell asymptotic states.
While the disappearance of these modes from the eigenspectrum of the
Hamiltonian solves the ghost problem in the theory, this disappearance 
is still quite perplexing and requires further explanation.

The reason for this highly unusual outcome derives from the fact that
while the normal situation for square matrices is that the number of
independent eigenvectors of a square matrix is the same as the
dimensionality of the matrix, there are certain matrices, known as
defective matrices, for which this is not in fact the case. A typical
example of such a defective matrix is the non-Hermitian two-dimensional
matrix
\begin{eqnarray}
M= \pmatrix{1&c \cr 0&1}
\label{A71}
\end{eqnarray}
with $c$ non-zero but otherwise arbitrary, since even though this matrix
has two eigenvalues both of which are real (despite the lack of
hermiticity) and equal to $1$ (no matter what the value of $c$),
solving the equation 
\begin{eqnarray}
\pmatrix{1&c \cr 0&1}\pmatrix{p \cr q}=\pmatrix{p+cq\cr q}=\pmatrix{p
\cr q}
\label{A72}
\end{eqnarray}
leads only to $q=0$ when $c \neq 0$, and thus to only one eigenvector
despite the two-fold degeneracy of the eigenvalue, with the space on
which the matrix $M$ acts not being complete. The defective
matrix $M$ given above is in the form of a Jordan block matrix, with
Jordan having shown that under a similarity transform an arbitrary square
matrix can be brought to either a diagonal form, or to a triangular form
such as that of Eq. (\ref{A71}) in which every matrix element on one
side of the diagonal is zero. Jordan block form matrices have the
property that no matter what values the matrix elements take on the
other side of the diagonal, the secular equation for the eigenvalues
only involves the elements on the diagonal itself. Then if these
diagonal elements are all real, the matrix will have real eigenvalues
despite not being Hermitian. (While Hermitian matrices must have real
eigenvalues, there is no converse theorem which would oblige the
eigenvalues of non-Hermitian matrices to necessarily be complex, with
non-Hermitian matrices still being able to possess real eigenvalues in
certain cases.) As such, a Jordan block form matrix can be thought of as
being a diagonal matrix $M_1$ to which has been added a second matrix
$M_2$ with non-zero elements on only one side of the diagonal. With
this second matrix being a divisor of zero, while its addition to the
diagonal matrix does not affect eigenvalues, its being a divisor of
zero does affect eigenvectors, to thereby cause the full Jordan
block matrix $M_1+M_2$ to have a smaller number of eigenvectors than
eigenvalues. 

To see how these considerations apply in the case of interest to us
here, we need to track the $\epsilon \rightarrow 0$ limit carefully. As
regards first the unequal frequency Hamiltonian of Eq. (\ref{A68}), we
note that its action on the one-particle states
$a^{\dagger}|\Omega\rangle$,
$b^{\dagger}|\Omega\rangle$ yields
\begin{eqnarray}
Ha^{\dagger}|\Omega\rangle&=&
\frac{1}{2\omega}\left[(4\omega^2+\epsilon^2)a^{\dagger} |\Omega\rangle 
+(4\omega^2-\epsilon^2)b^{\dagger}|\Omega\rangle\right]
\nonumber \\
Hb^{\dagger}|\Omega\rangle
&=&\frac{1}{2\omega}\left[\epsilon^2a^{\dagger} |\Omega\rangle 
+(4\omega^2-\epsilon^2)b^{\dagger}|\Omega\rangle\right]~~.
\label{A73}
\end{eqnarray}
In this sector we can define a matrix
\begin{eqnarray}
M(\epsilon)=
\frac{1}{2\omega}\pmatrix{4\omega^2+\epsilon^2&4\omega^2-\epsilon^2
\cr \epsilon^2&4\omega^2-\epsilon^2}
\label{A74}
\end{eqnarray}
whose eigenvalues are given as $2\omega+\epsilon$ and
$2\omega -\epsilon$. For such eigenvalues, 
energy eigenvectors which obey  
\begin{eqnarray}
H|2\omega \pm \epsilon\rangle =(2\omega \pm \epsilon)|2\omega \pm
\epsilon\rangle 
\label{A75}
\end{eqnarray}
are then readily constructed as
\begin{equation}
|2\omega \pm \epsilon\rangle =\left[\pm\epsilon a^{\dagger}+(2\omega
\mp \epsilon)b^{\dagger}\right]|\Omega \rangle~~.
\label{A76}
\end{equation}
As we see, as long as $\epsilon \neq 0$, the two one-particle sector 
eigenvectors of the Hamiltonian $H$ are distinct, and $H$ has two
eigenvectors to go with its two eigenvalues. However when we now let
$\epsilon$ go to zero, the two eigenvectors in Eq. (\ref{A76}) collapse
onto a single eigenvector, viz. the vector $b^{\dagger}|\Omega \rangle$,
the two eigenvalues collapse onto a common eigenvalue, viz. $2\omega$,
and the matrix $M(\epsilon=0)$ of Eq. (\ref{A74}) becomes defective,
with the $H(\epsilon=0)$ Hamiltonian then being free of negative norm
eigenstates. The equal frequency theory is thus fully acceptable, to
thereby resolve what had been thought of as being one of the biggest
difficulties for fourth order theories. 

As regards the conformal gravity theory itself, we note that
in a $g_{\mu\nu}=\eta_{\mu\nu} +h_{\mu\nu}$ linearization of the theory
around flat spacetime, the conformal gravity rank two 
tensor of Eq. (\ref{185}) reduces to 
\begin{equation}
W^{\mu\nu}=\frac{1}{2}\Pi^{\mu\rho}
\Pi ^{\nu\sigma}K_{\rho \sigma}- \frac{1}{6}\Pi^{\mu \nu} \Pi ^{\rho
\sigma}K_{\rho\sigma}
\label{A77}
\end{equation}
where
\begin{equation}
K^{\mu
\nu}=h^{\mu \nu}-
\frac{1}{4}\eta^{\mu \nu} h^{\alpha}_{\phantom{\alpha}\alpha}~~,~~
\Pi^{\mu \nu}
=\eta^{\mu \nu} \partial^{\alpha}\partial_{\alpha}-
\partial^{\mu}\partial^{\nu}~~.
\label{A78}
\end{equation}
In the
conformal gauge 
\begin{equation}
\partial_{\nu}g^{\mu \nu}
-\frac{1}{4}g^{\mu\sigma}g_{\nu\rho}\partial_{\sigma}g^{\nu 
\rho}=0
\label{A79}
\end{equation}
(viz. the gauge condition which is left invariant under 
$g_{\mu \nu}(x)\rightarrow e^{2\alpha(x)}g_{\mu \nu}(x)$)  the
source-free region gravitational fluctuation equation
$W^{\mu\nu}=0$ then reduces to 
\begin{equation}
(-\partial_0^2+\nabla^2)^2K^{\mu\nu}=0~~.
\label{A80}
\end{equation}
With this equation of motion being decoupled in its tensor indices, we
see that each tensor component precisely obeys none other than the
$M^2=0$ limit of Eq. (\ref{A40}). Consequently, it would be of interest
to see if the structure we have found for the prototype equal frequency
Pais-Uhlenbeck theory can carry over to the conformal gravity theory
once its gauge and tensor structure is taken into consideration, since
that would then permit the construction of a fully renormalizable, fully
unitary gravitational theory in four spacetime dimensions, one which
despite its fourth order equation of motion, would nonetheless only
possess one on-shell graviton and not two.\footnote{It would also be of
interest to see whether the typical linear in time growth given in Eq.
(\ref{A52}) might enable conformal gravity early universe tensor
fluctuations to grow to macroscopic size by recombination.}

\end{document}